\documentclass[12pt]{uhthesis}

\usepackage{uhaas}
\usepackage{rotating}
\usepackage[T1]{fontenc}
\usepackage[polish,main=english]{babel}

\renewcommand{\baselinestretch}{1.5}

\usepackage{savesym}
\usepackage{threeparttable}
\savesymbol{tablenum}
\usepackage{siunitx}
\restoresymbol{TXF}{tablenum}

\usepackage{multirow}
\savesymbol{tablehead}
\savesymbol{tabletail}

\usepackage{subfig}
\usepackage{amsmath}	
\usepackage{breqn}
\usepackage{pifont} 
\usepackage{float}
\usepackage{tablefootnote}
\newcommand{\angstrom}{\textup{\AA}}
\newcommand{\mbh}{$M_{\mathrm{BH}}$}
\newcommand{\dotm}{$\dot{m}$}
\newcommand{\dotM}{$\dot{M}$}
\newcommand{\as}{$a_{\ast}$}
\newcommand{\incl}{$i$}
\newcommand{\mbhlit}{$M_{\mathrm{BH}}^{\mathrm{lit}}$}
\newcommand{\mbhbay}{$M_{\mathrm{BH}}^{\mathrm{Bay}}$}
\newcommand{\dotmlit}{$\dot{m}_{\mathrm{lit}}$}

\newcommand{\Hb}{H{\small $\beta$}}
\newcommand{\MgII}{\ion{Mg}{2}}
\newcommand{\CIV}{\ion{C}{4}}
\newcommand{\FeII}{\ion{Fe}{2}}
\newcommand{\NV}{\ion{N}{5}}

\newcommand{\mmbh}{M_{\mathrm{BH}}}
\newcommand{\mdotm}{\dot m}
\newcommand{\mdotM}{\dot M}
\newcommand{\mas}{a_{\ast}}
\newcommand{\mmbhlit}{M_{\mathrm{BH}}^{\mathrm{lit}}}
\newcommand{\mdotmlit}{\dot{m}_{\mathrm{lit}}}

\newcommand{\Ks}{K$_{\rm S}$}
\newcommand{\Msun}{${\rm M_{\odot}}$}
\newcommand{\mMsun}{{\rm M_{\odot}}}
\newcommand{\FWHM}{\textit{FWHM}}
\newcommand{\mFWHM}{\mathit{FWHM}}

\newcommand{\mRblr}{R_{\mathrm{BLR}}}

\newcommand{\mLsun}{{\rm L_{\odot}}}

\newcommand{\mHbeta}{H{\small \beta}}
\newcommand{\mMgII}{\mathrm{Mg} \ \text{\small{\( \textrm{II} \)}}}

\newcommand{\ergs}{erg s$^{-1}$}
\newcommand{\kms}{km s$^{-1}$}
\newcommand{\mergs}{\mathrm{erg\, s}^{-1}}
\newcommand{\mkms}{\mathrm{km\, s}^{-1}}

\newcommand{\ok}{\ding{52}}

\usepackage[acronym,toc,nonumberlist]{glossaries}

\makeglossaries

\newacronym{AGN}{AGN}{Active Galactic Nucleus}
\newacronym{AD}{AD}{Accretion Disk}
\newacronym{WLQ}{WLQ}{Weak emission-line Quasar}
\newacronym{BH}{BH}{Black Hole}
\newacronym{SMBH}{SMBH}{Supermassive black hole}
\newacronym{QSO}{QSO}{Quasar}
\newacronym{BLR}{BLR}{Broad Line Region}
\newacronym{NLR}{NLR}{Narrow Line Region}
\newacronym{UV}{UV}{Ultraviolet}
\newacronym{IR}{IR}{Infrared}
\newacronym{SF}{SF}{Starformation}
\newacronym{IGM}{IGM}{Intergalactic medium}
\newacronym{EW}{EW}{Equivalent Width}
\newacronym{FWHM}{FWHM}{Full Width at Half Maximum}
\newacronym{SED}{SED }{Spectral Energy Distribution}
\newacronym{NT}{NT}{Novikov-Thorne}
\newacronym{SS}{SS}{Shakura-Sunayaev}
\newacronym{f}{f}{virial factor}

\begin{document}


\frontmatter




\begin{titlepage}
    \begin{center}
        \vspace*{1cm}
            
        \large
        {\textbf{GLOBAL PARAMETERS OF QUASARS WITH ANOMALOUS ELECTROMAGNETIC SPECTRUM}}
            
        \vspace{2.0cm}
            
        \normalsize{DOCTOR OF PHILOSOPHY\\
        IN\\
        PHYSICS}
            
        \vspace{1.5cm}
        
        \large{
        \textbf{Marcin Marculewicz}}
        \vspace{2.0cm}
        
        \normalsize{Faculty of Physics\\
        University of Białystok\\
        \vspace{1.0cm}
        PhD thesis supervisor:\\
        dr hab. Marek Nikołajuk, prof. UwB\\
        \vspace{3.5cm}
        Białystok, June 2021}
        
        \makecopyright{2021}{Marcin Marculewicz}   
    \end{center}
\end{titlepage}


\makeacknowledgements
I would like to thank my advisor, dr hab Marek Nikolajuk, prof. UwB, for guiding and supporting me over the years.


\printglossary[type=\acronymtype,nonumberlist]  

\tableofcontents
\listoftables
\listoffigures

\mainmatter
\chapter*{Preface}
\addcontentsline{toc}{chapter}{Preface}

The thesis concerns the analysis of the Active Galactic Nuclei. These are galaxies with an active core. The most luminous type of Active Galactic Nuclei is Quasar. It contains the supermassive black hole at the center. One of the least known subtype of Quasars are: \textit{Weak emission-Line Quasars.} Their recognizable feature are weak emission-lines.

The primary goal of PhD thesis is to evaluate \textbf{the global parameters} such as: the black hole mass, the accretion rate, spin of the black hole, and the inclination of weak emission-line quasars based on \textit{the continuum fit method}. This method apart from the literature black hole masses estimation methods \textbf{does not depend} on the observed Full Width at Half Maximum of emission line, which could be biased due to the weakness or lack of the emission lines in these quasars. Using the Spectral Energy Distribution of quasars, I have fitted the geometrically thin and optically thick accretion disk model described by Novikov \& Thorne equations. I have obtained the model of the continuum of the accretion disk for the 10 weak emission-line quasars.

The second project concerned the description of abnormal, deep absorption of SDSS J110511.15+530806.5 quasar. I checked the correctness of the thesis posed that corona and warm skin concept above/around an accretion disk explain this phenomenon.

\newpage
My main motivation during my thesis is: 
\begin{itemize}
    \item Examining the validity of hypotheses of the nature of the weak emission-line quasars,
    \item Analysis of the Broad Line Region in the vicinity of the host of weak emission-line quasars,
    \item Development of the virial factor for the weak emission-line quasars,
    
    \item Development of the new method of the \mbh\ estimation of weak emission-line quasars,
    \item Comparison of my results with the previous literature reports,
    \item Modeling the corona and warm skin for SDSS J110511.15+530806.5
    \item Validate the hypothesis of the phenomena in SDSS J110511.15+530806.5
\end{itemize}

The following work is divided into Chapter 1: Introduction of the Active Galactic Nuclei and Weak emission-line Quasars description. In Chapter~2, I have described the sample selection and data reduction. I have collected photometric points and spectra of 10 Weak emission-line Quasars in the broad range (Infrared/Optic/Ultraviolet) using the WISE, 2MASS, SDSS, Galex. In Chapter~3, I have focused on the model and results description. Chapter 4 presents a discussion of results and potential explanation of the Weak emission-line Quasars. Chapter 5 contains the description of the SDSS J110511.15+530806.5. Additionally, I have described the modeling and the results of the fitting of the corona and the warm skin of SDSS J110511.15+530806.5. I have concluded my thesis in Chapter 6.

\newpage
The doctoral dissertation was created on the basis of:
\begin{itemize}
   \item{\underline{M. Marculewicz} and M. Nikolajuk - {"Black Hole Masses of Weak Emission Line Quasars Based on the Continuum Fit Method"}, the Astrophysical Journal, Volume 897, Number 2 (2020)}

\item{\underline{M. Marculewicz} and M. Nikolajuk - {"Models of Continuum of Weak Emission-Line Quasars"}, Proceedings of the Polish Astronomical Society, Volume 10, 243-247, ISBN: 978-83-950430-8-6 (2020)}
\end{itemize}

and unpublished material described in Chapter 5 (in preparation).

My additional papers not included into this PhD thesis are:

\begin{itemize}
\item{H. Abdalla et al. (Cherenkov Telescope Array Consortium) - {"Sensitivity of the Cherenkov Telescope Array for probing cosmology and fundamental physics with gamma-ray propagation"}, Journal of Cosmology and Astroparticle Physics, Issue 02, article id. 048 (2021)},
\item{A. Acharyya et al. (Cherenkov Telescope Array Consortium) - {"Sensitivity of the Cherenkov Telescope Array to a dark matter signal from the Galactic centre"}, Journal of Cosmology and Astroparticle Physics, Issue 01, article id. 057 (2021)},
\item{A. Acharyya et al. (Cherenkov Telescope Array Consortium) - {"Monte Carlo studies for the optimisation of the Cherenkov Telescope Array layout"}, Astroparticle Physics, 111, 35-53 (2019)},
\item{M. Ostrowski et al. - {"Progress of the Cherenkov Telescope Array project in Poland"}, Proceedings of the Polish Astronomical Society, vol. 7, 343-348 (2018)},
\item{M. Giedyk et al. - {"Photoinduced Vitamin B12-Catalysis for Deprotection of (Allyloxy)arenes"}, Organic Letters, 19, 10, 2670-2673, (2017)}
\end{itemize}






\chapter{Introduction}

\section*{Abstract \label{abstract_chapter1}}

An \acrshort{AGN} (Active Galactic Nucleus) provides a unique test of fundamental physics in regimes inaccessible on Earth and plays an important role in our understanding of the evolution of galaxies. We describe these systems through the "Unified Model" in which the AGN contains a central \acrshort{BH} (Black Hole), an \acrshort{AD} (Accretion Disk), jets, magnetic fields, and exhibit chaotic inflow/outflow processes. Modeling these components, therefore, requires understanding physical processes across a wide range of scales. While the Unified Model explains many of the peculiar behaviors of different types of AGN, they still fail to explain some classes such as \acrshort{WLQ}s (Weak emission-Line Quasars). 



\section{Active Galactic Nuclei}

\begin{figure}[!htb]

 \centering
   \includegraphics[width=0.85\textwidth]{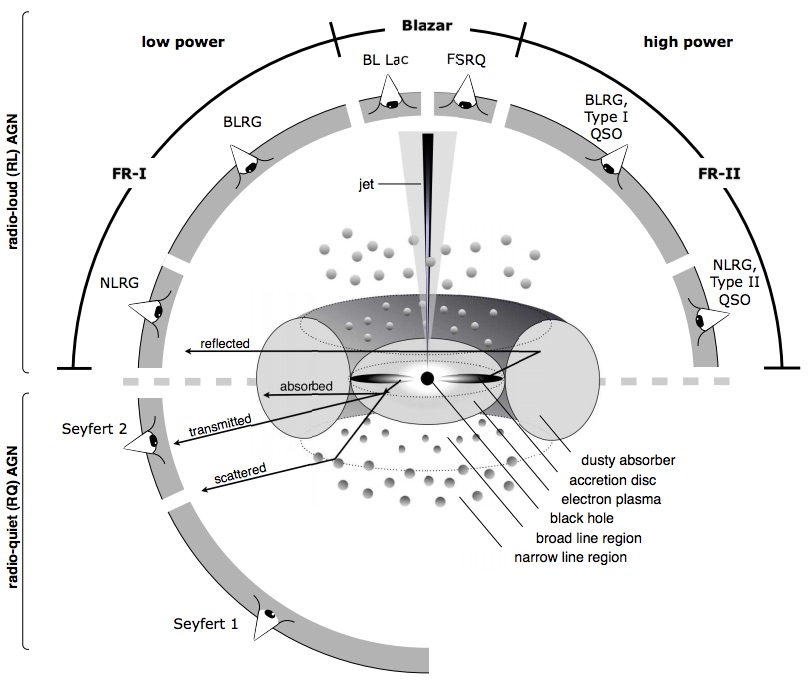}
    
    \caption[{Types of AGNs}]{Types of AGNs including radio, inclination, and power dependency \citep{Antonucci2002}.}
    \label{fig:type_agn}
\end{figure}

\begin{figure}[!htb]
\centering
    \includegraphics[width=1.0\textwidth]{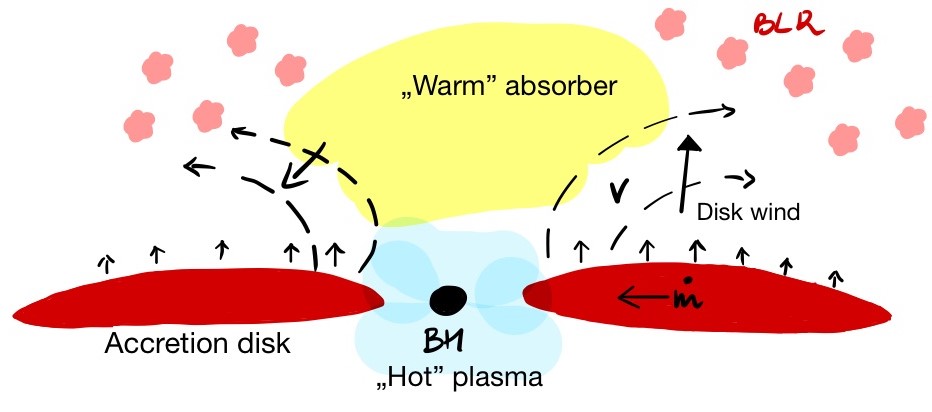}
    \caption {The zoom picture of the vicinity of SMBH}
    \label{fig:agn_my}
\end{figure}

For hundreds of years, we have been looking at the sky and observing the astonishing nature of the cosmos. Recent decades have shown many discoveries about the nature of \acrshort{AGN}s. They are powered by accretion onto \acrshort{SMBH} and we are able to observe their features across the full electromagnetic spectrum. From the zoo collection of all of the types of AGN, thanks to a recent study we can constrain them well. One of the main discrepancies/differences is the angle at which the observer sees the object. Currently, we are certain that the line of sight is crucial. The type of AGN is determined among others by the inclination angle (see Fig.\ref{fig:type_agn}). If we could look closer, the vicinity of the \acrshort{SMBH} (Supermassive black hole) would look like in Fig.\ref{fig:agn_my} which represents the SMBH at the center. The blue cloud is the 'hot' plasma commonly named a 'corona'. Above it is the 'warm' absorber which might be a part of the corona. This region is mostly dominated by X-ray emissions. A sub-parsec further one would see the AD. Looking at the geometry and optical depth, we can distinguish a few ADs: the geometrically thin (\acrshort{SS} - Shakura-Sunayaev) or thick (slim) AD \citep{Abramowicz1988}. An optically thin AD is called radiation inefficient accretion flow (RIAF, \citeauthor{Yuan04}, 2004) or advection dominated accretion flow (ADAF, \citeauthor{Narayan_Yi_ADAF1994}, 1994). Gaseous, high density clouds \acrshort{BLR} (Broad Line Region) ($\sim 10^{10}$ cm$^{-3}$) are present roughly at the distance of 0.01 - 1 parsec \footnote{1 pc = 3.0857$\times10^{16}$ m} from the central engine \citep{Netzer13}. Subsequently, at a distance of 0.1 - 10 parsec is a dusty torus, often called the central torus. Moving further one would observe a low density ($\sim 10^{10}$ cm$^{-3}$), low velocity ionized gas, the \acrshort{NLR} (Narrow Line Region). This extends from the outside of the torus to a hundred or a thousand parsec. A few of the AGNs include a central radio jet associated with $\gamma$-ray emission \citep{Netzer15_agn_unified}. The jet can be stretched to a few hundred kpc \citep{Blandford2019jets}.

\clearpage

\clearpage

In the late 1970s, studies of AGNs started, long after the first discovery of quasi-stellar objects (QSOs) in the 1960s. Three decades before that Carl Seyfert observed one of the first galaxies. In honor of him, there are two types of Seyfert galaxies: type 1 and type 2. Nowadays, all objects containing active SMBHs are referred to as AGNs. The main discrepancy between these two is in the optical-ultraviolet spectra. Seyfert 1 galaxies have strong and broad emission lines (2000 - 10 000 $\mkms$), whereas Seyfert 2 galaxies do not exceed a width of 1200 $\mkms$ \citep{Peterson97_book}. Differences in the broadness are assigned the discrepancy in the viewing angle to the central source. Tab. \ref{tab:AGN_unification_Peterson} presents the AGN Unification (according to Table 7.1 \citeauthor{Peterson97_book}, 1997). The types of AGNs are LIRG -- luminous infrared galaxies; BL Lac -- BL Lacertae (rapid and large-amplitude flux variability and significant optical polarization); BLRG -- broad-line radio galaxies; NLRG -- narrow-line radio galaxies; OVV -- optically violently variable \acrshort{QSO}; FR (Fanaroff–Riley) class I and II. 

\begin{table}[ht]
  \centering
\begin{tabular}{lll}
\hline
\multicolumn{2}{c}{Orientation} \\
Face-On & Edge-On & Radio properties\\
\hline
Seyfert 1 & Seyfert 2 & Radio Quiet\\
QSO  & LIRG galaxy & \\
\hline
 BL Lac & FR I & Radio Loud\\
 BLRG & NLRG & \\
 Quasar/OVV & FR II & \\
 \hline
\end{tabular}
\caption{AGN Unification}
\label{tab:AGN_unification_Peterson}
\begin{quote}
    Note: R - radio loudness parameter; the ratio between radio 5GHz to optical (B-band = 4400 \AA\ based on Johnson-Morgan system) monochromatic luminosity. L (5GHz)/L (4400\AA). The dividing line between radio-loud and radio-quiet AGNs is usually set at R = 10 \citep{Netzer13}
\end{quote}
\end{table}

AGNs are a stronger emitter than the nuclei of normal galaxies due to a very energetic event -- accretion onto a central SMBH. They have very high luminosities (up to $L_{bol} \approx 10^{45} - 10^{48}\mergs$) and can emit radiation across the whole electromagnetic spectrum. These features provide an excellent field for the astrophysical research of their nature.

The multi-wavelength study of AGN provides many windows for observing their properties. The infrared (\acrshort{IR}) band is sensitive to obscuring matter and dust. The optical and \acrshort{UV} bands are related to the emission from the AD. The X-ray emission is assigned mostly to the corona. According to \cite{Padovani17} the non-thermal emission is attributed to radio and $\gamma$ emission. Table 1 in \cite{Padovani17} presents a list of the AGN classes.

The AGNs are some of the biggest objects in the whole Universe, although in the description of their physical phenomena they are simpler than their size would indicate. In summary, the previously mentioned features -- orientation; accretion rate; the absence or the presence of jets; the environment within the host galaxy is enough to describe those behemoths well.

\begin{figure}[ht]
\centering
\includegraphics[width=1.0\textwidth]{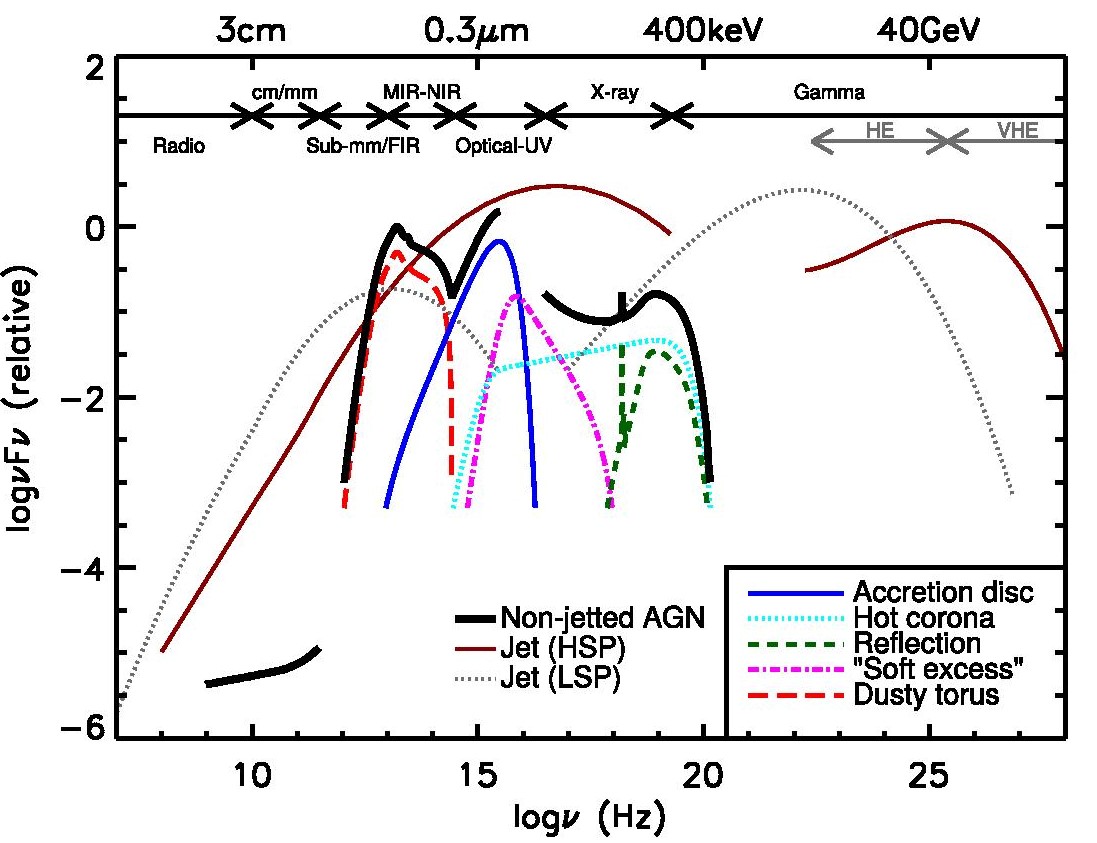}

    \caption [AGN spectral energy distribution]{AGN spectral energy distribution (SED). HSP - high synchrotron peaked blazar, LSP - low synchrotron peaked blazar \citep{Padovani17}. Each component is represented by different color curves (for better clarity the authors have shifted it down).}
    \label{fig:padovani_sed}
\end{figure}

Fig. \ref{fig:padovani_sed} represents a schematic \acrshort{SED} (Spectral Energy Distribution) of an AGN. Total emission is represented by the black solid curve. The intrinsic shape of the SED in the mm-far infrared (FIR) regime is uncertain; however, it is widely believed to have a minimal contribution (to an overall galaxy SED) compared to \acrshort{SF} (Starformation), except in the most intrinsically luminous quasars and powerful jetted AGN. The primary emission from the AGN accretion disk peaks in the UV region. The jet SED is also shown for a high synchrotron peaked blazar (e.g. Mrk 421) and a low synchrotron peaked blazar (e.g 3C 454.3).


\newpage
\section{The spectrum of AGN}


The AGN continuum in the optical-UV range is dominated by accretion processes. In the simplest case, it is spherically-symmetrical accretion \citep{Bondi1952}. The geometrically thin and optically thick AD is described by \cite{SS73} equations, where matter is accreting onto a non-rotating BH. An extension of this approach for a rotating BH is the \cite{NT73} approach. 

\begin{figure}[ht]
\centering
     \includegraphics[width=0.7\textwidth]{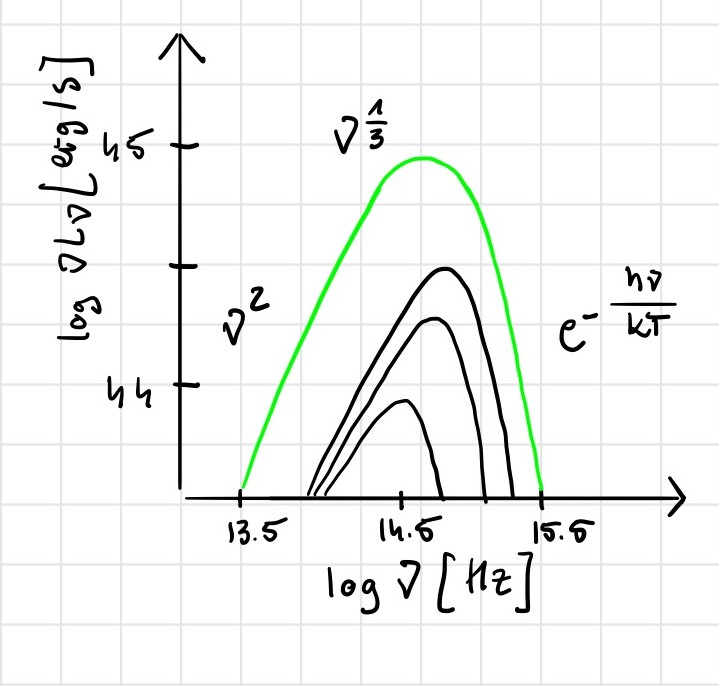}

    \caption [{Accretion disk as a blackbody}]{The continuum of the accretion disk. The green line is the sum of the black lines. Each black line represents the one temperature blackbody.}
    \label{fig:AD_logniu_logniuLniu}
\end{figure}

The low-order approximation of the SED of AGNs can be described as a power law of the form $F_{\nu} \propto \nu^{-\alpha}$, where $\alpha$ called a spectrum index. 
The AD is described as a perfect blackbody emission with multi-temperature components. Each of the components is assigned as a single ring with a specific temperature. 
The single ring radiates as a blackbody (see Fig.\ref{fig:AD_logniu_logniuLniu}).

The short-band (optics/UV) SED of a quasar continuum approximately also takes the form of power dependency expressed mainly with $0 \leq \alpha \leq 1$ from the observer's point of view. Dependency of the SED may have the form: $L_{\lambda} = \lambda^{-\beta}$, where $\beta$ is the wavelength spectral index, $\beta = 2 -\alpha$ \citep{Netzer13}.
For example, the observed 1200-6000 \angstrom\ continuum of many luminous AGNs is described by 
$F_{\nu} \approx \nu^{-0.5}$. This single power law approximation fails for wavelengths below 1200 or above 6000 \angstrom.

In the AGN, bolometric luminosity can be seen mainly from the UV part of the maximum peak emission called the big blue bump. The inner part of it is generally agreed to be thermal in origin, although it is not clear whether it is optically thick (blackbody) or optically thin (free-free) emission \citep{Peterson97_book}. Many proponents of the optically thick interpretation ascribe the big blue bump to the accretion disk. 

\begin{figure}[ht]
\centering
    \includegraphics[width=0.85\textwidth]{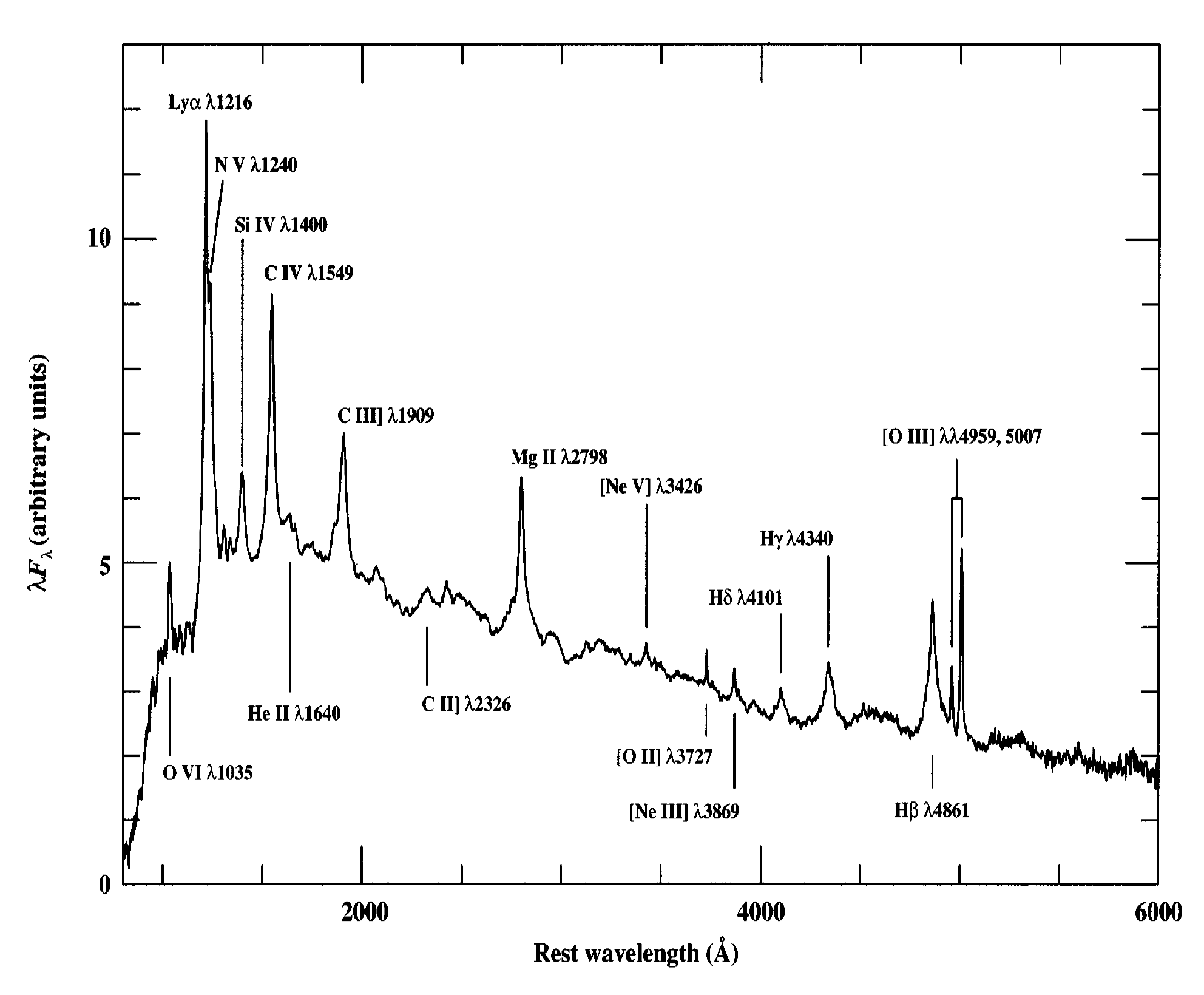}
    \caption [{The mean QSO spectrum}]{The mean QSO spectrum of 700 objects from the Large Bright Quasar Survey. Source: Fig 2.\cite{Francis1991}, Fig 2.2 \cite{Peterson97_book}.}
    \label{fig:mean_spec_qso}
\end{figure}

The continuum of the object is enriched by the existence of emission and absorption lines. The mean quasar spectrum of the Large Bright Quasar Survey (Fig. \ref{fig:mean_spec_qso}) presents prominent emission lines such as Lyman $\alpha$, Si IV, C IV, Mg II, and Balmer lines. There is a weaker feature superimposed on the big blue bump between $\sim$ 2000-4000 \angstrom. This feature is attributable to a combination of Balmer continuum emission and blends of Fe II emission lines arising in the broad-line region \citep{Peterson97_book}. The AGN spectrum contains thermal and non-thermal emissions. Non-thermal emission comes from particles whose velocities are not described by a Maxwell-Boltzmann distribution (e.g. particles which give rise to Compton scattering (corona) or synchrotron power-law spectra (jets)).


The secondary instance is the emission by gas which receives its energy from the previous processes and re-radiates the emission. This is the free-free emission from photoionized or collisionally ionized gas i.e. reprocessing and reflection \citep{Garcia_xillver2013}. Isotropy plays an important role in determining the type of AGN. Let's focus on the non-thermal emission of jets. If a line of sight is close to the axis of the jet, the spectrum might be dominated by it since the emission is highly beamed in the direction of the electron motion. However, the total amount of energy in the beam may be far less than that emitted by thermal processes, which is radiated isotropically.

The influence of the new direction on the differentiation of AGN types can be seen in the example of Seyfert galaxies
The main observational difference between Seyfert 1 and 2 galaxies is their different optical-ultraviolet spectra \citep{Krolik1999book, Netzer13}. Seyfert 1 shows strong, very broad (2000 - 10000 \kms) permitted and semiforbidden emission lines, whereas the broadest lines in Seyfert 2 have widths that do not exceed 1200 \kms. Such differences are now interpreted as arising from different viewing angles to the centers of such sources and from a large amount of obscuration along the line of sight. Type 2 are heavily obscured along the line of sight that extinguishes basically all the optical-UV radiation from the inner parsec. Sources with unobscured lines of sight to their centers are called Type 1. \cite{Veron_Cetty_catalogue2001} adopted classification introduced by \cite{Winkler1992} accordingly, where $R_{sey}:= \frac{\mHbeta}{[OIII]}$:

\begin{table}[H]
    \centering
    \begin{tabular}{|c|c|}
    \hline
    S 1.0 & 5.0 < $R_{sey}$ \\
    \hline
    S 1.2 & 2.0 < $R_{sey}$ < 5.0 \\
    \hline
    S 1.5 & 0.333 < $R_{sey}$ < 2.0 \\
    \hline
    S 1.8 & $R_{sey}$ < 0.333, broad components visible in $H \alpha$ and \Hb \\
    \hline
    S 1.9 & broad components visible in $H \alpha$ but not \Hb \\
    \hline
    S 2.0 & no broad component visible \\
    \hline
    \end{tabular}
\end{table}

\noindent For example, the broad components are very weak in Seyfert 1.8 galaxies, but detectable at \Hb\ as well as $H \alpha$. Indeed, the AGN continuum is usually so weak in Seyfert 2 galaxies that it is very difficult to unambiguously isolate it from the stellar continuum.





\section{Weak Emission-lines Quasars (WLQs)}\label{sec:intro_WLQ}


WLQs are an unsolved puzzle in the model of active galactic nuclei (AGN). There are $\sim$ 200 confirmed sources or candidates \citep{DS09, Shemmer10, Hryniewicz10, Wu11, Plotkin15, Ni18, Tim20}. The typical \acrshort{EW} (Equivalent Width) of the \CIV\ emission line is extremely weak ($\la 10{\angstrom}$) compared to normal quasars and very weak or absent in Ly$\alpha$ emission \citep{Fan99,DS09}. WLQs are type 1, radio-quiet quasars with weak or no emission lines; X-ray emission is also suppressed. 

\citet{DS09} note that WLQs have optical continuum properties similar to normal quasars, although Ly$\alpha$+\NV\ line luminosities are significantly weaker, by a factor of 4. An explanation for the weak or absent emission lines has not been found so far. The results of \cite{DS09} support the idea of WLQs as intrinsically weak UV emission-lines quasars.


\begin{figure}[ht]
\centering
    \includegraphics[width=0.85\textwidth]{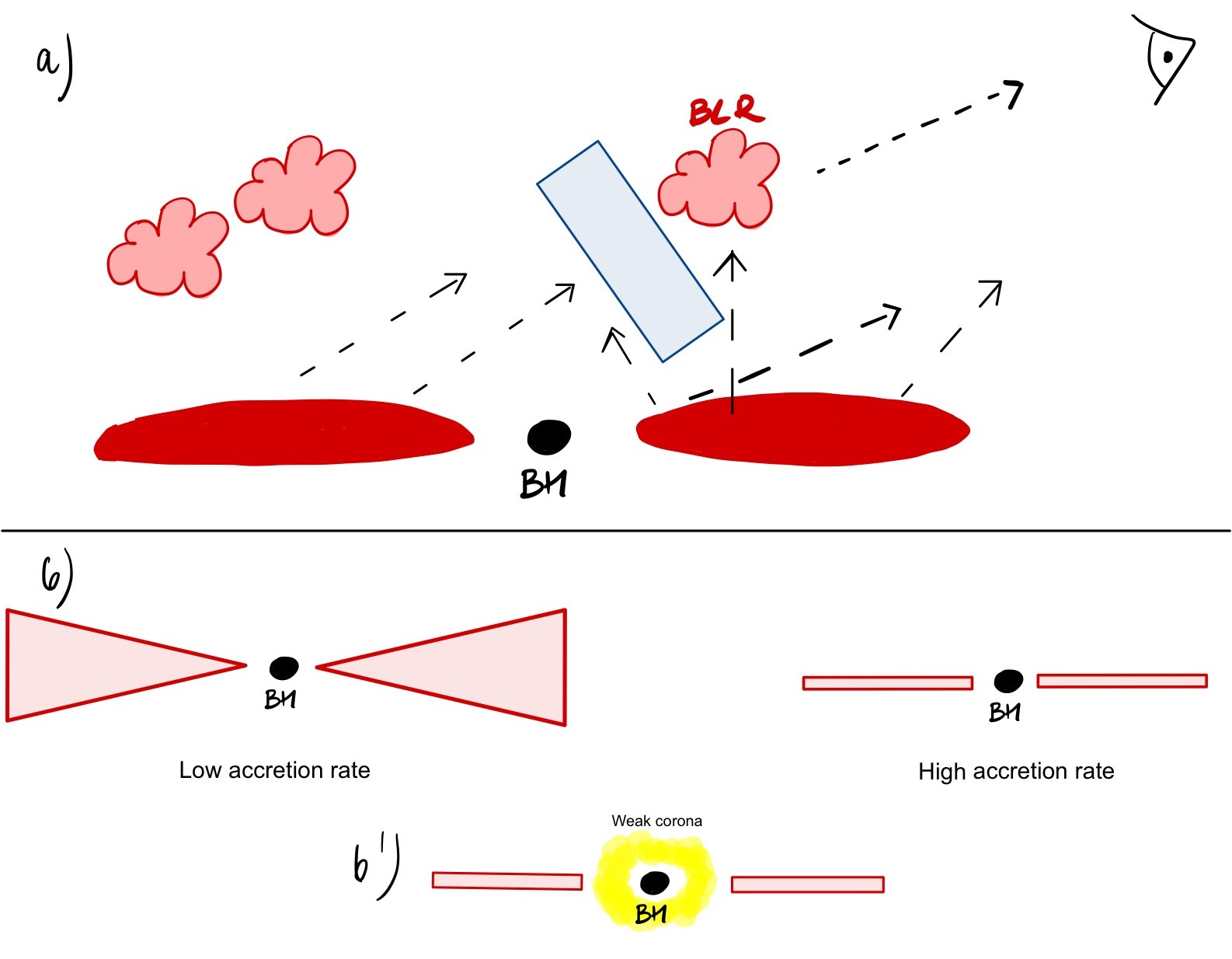}

    \caption [Potential vicinity of the WLQ]{Potential vicinity of the WLQ (cigar-shaped area - AD, red clouds - the BLR, and blue rectangle - 'shielding gas').}
    \label{fig:WLQ_vicinity}
\end{figure}

Fig. \ref{fig:WLQ_vicinity} represents the potential image of the vicinity of the WLQ. On the upper panel of Fig. \ref{fig:WLQ_vicinity} the cigar-shaped area represents AD, red clouds refer to the BLR, and the blue rectangle is the 'shielding gas'. Below I present hypotheses that describe approaches to explain the nature of quasars. 

The explanation of the nature of the WLQs can be divided into two main divisions. First, something is happening with the photoionized photons illuminating the BLR. Photoionized photons (soft X-ray/hard UV) are produced by the close vicinity of SMBH. Second idea, something is happening with the BLR itself. 
\newpage
\noindent Each of the explanations can still be divided into subsections.

\textbf{Photoionized photons illuminating the BLR:}
\vspace{-0.2in}
\begin{itemize}
\itemsep-0.2em 
\item[\ok] an extremely high accretion rate where the continuum is too soft and the photoionized flux is inefficient \citep{Leighly07a,Leighly07b} (Fig. \ref{fig:WLQ_vicinity} the \textit{b} right panel)
\item[\ok] a cold accretion disk with a small accretion rate \citep{Laor11}(Fig. \ref{fig:WLQ_vicinity} the \textit{b} left and \textit{b'} lower panels)
\item[\ok] \citet{Wu11} postulate the presence of a shielding gas between the accretion disk and the broad line region (BLR), which could absorb the high-energy ionizing photons from the accretion disk (Fig. \ref{fig:WLQ_vicinity} the \textit{a} panel).  
\item[\ok] a radiatively inefficient accretion flow \citep{Yuan04}
\end{itemize}

\textbf{The BLR itself}
\vspace{-0.2in}
\begin{itemize}
\itemsep-0.2em 
\item[\ok] an unusual BLR structure i.e., anemic in construction and/or weak gas abundance \citep{Shemmer10, Nikolajuk12} 
\item[\ok] the idea that WLQs can also be in an early stage of AGN evolution \citep{Hryniewicz10, Liu11, Banados14, Meusinger14}
\end{itemize}

\citet{Leighly07a,Leighly07b} suggest that the combination of large \mbh\ and a high accretion rate might be at plausible explanation for this weakness, which could be due to the relative deficiency in high-energy photons in the SED. It is worth pointing out that the photoionized flux is too soft in energy and not efficient enough to produce strong emission lines. 

In this approach the prominent high-ionization emission line, e.g. \CIV\, is suppressed to the low-ionization line \Hb\ \citep{Shemmer10}. This effect is a plausible explanation of the nature of PHL 1811. This is a high accretion rate quasar with EW(\CIV) = 6.6 \angstrom, whereas \Hb\ is more typical, but still weak with EW~=~50 \angstrom\ \citep{Leighly07a, DS09}. Similar to PHL 1811 is PG1407+265 described by \cite{Mcdowell95}. 

\cite{Laor11} postulate that the cold AD provides a good explanation for SDSS J094533.99+100950.1. The ionizing continuum in such an object may originate in the X-ray power-law component, which produces an extended partially ionized region in the illuminated gas, which cools mostly through low-ionization lines. Additionally, I would like to note that this approach allows them to place an upper limit on the BH spin ($a_* \leq 0.3$).


The analogs of PHL 1811 from \cite{Wu11} confirm previous assumptions. They have a blue UV/optical continua without detectable broad absorption lines or dust reddening. PHL 1811 analogs seem to be an X-ray weak subset of WLQs with strong UV Fe emission and \CIV\ blueshift. \cite{Wu11} claim that unusual UV properties might be explained by the intrinsically X-ray weak SED. However, a confirming hypothesis is still missing. It cannot be ruled out that WLQs are heavy X-ray absorption objects. The radio-quiet study of PHL 1811 analogs suggests the connection of very high-ionization absorbers (e.g. O VI) to the UV and X-ray properties of WLQs. 

\begin{figure}[ht]
\centering
    \plotone{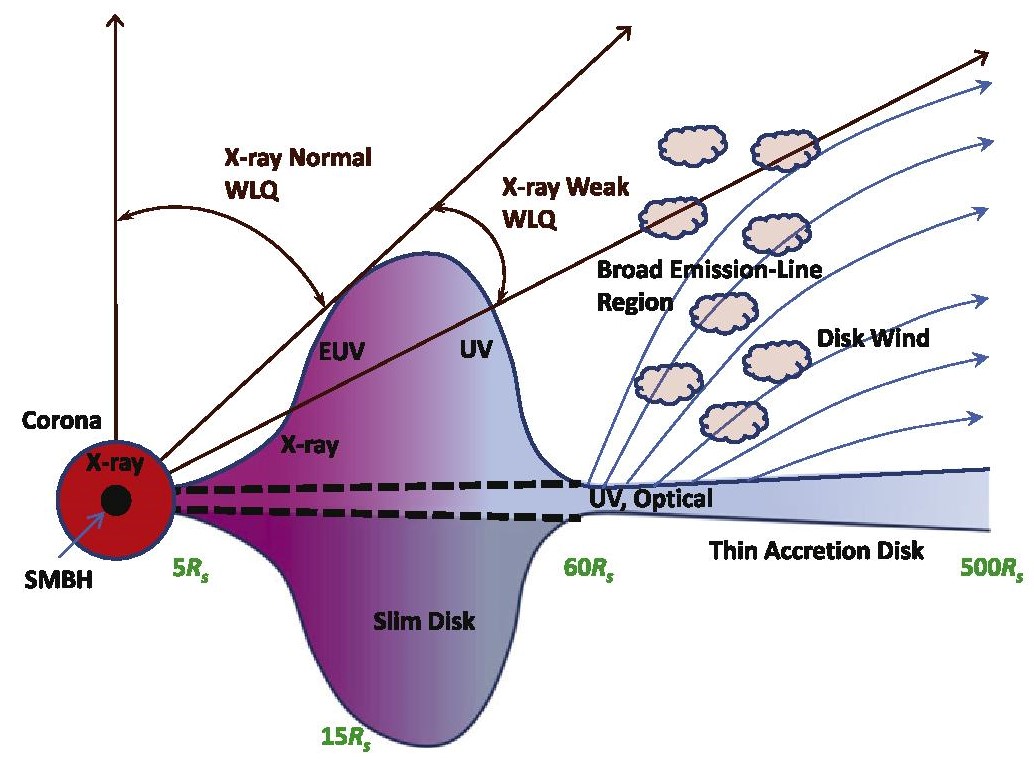}

    \caption [{The 'shielding gas scenario'}]{A schematic diagram of the geometrically thick disk scenario for WLQ. PHL 1811 analogs are denoted as X-ray weak WLQ \citep{Luo15}.}
    \label{fig:shielding_gas}
\end{figure}

\cite{Luo15} measured a relatively hard $\Gamma = {1.16 ^{+0.32}_{-0.37}}$ effective power-law photon index\footnote{It is worth noting $N(E) \propto E^{-\Gamma}$, where N(E) is the amount of photons and E is the energy. For example $\Gamma \geq 2.0$ indicates soft X-ray source \citep[e.g.][]{Haardt_Maraschi1991, Cao2009}} for a stack of the X-ray weak subsample, suggests X-ray absorption. If WLQs and PHL 1811 analogs have very high Eddington ratios, the inner disk could be significantly puffed up to resemble a slim disk. This suggests that the shielding gas could be described as a geometrically thick inner accretion disk that shields the broad-line region from the ionizing continuum (Fig. \ref{fig:shielding_gas}). Shielding of the broad emission-line region by a geometrically thick disk may have a significant role in setting the broad distributions of C IV EW and blueshift for quasars more generally \citep{Luo15}.

Several works about X-ray properties of WLQs have recently appeared \citep[e.g.][]{Wu11,Wu12,Ni18, Marlar18}. The conclusion arising from these works is that WLQs are more likely to be X-ray weaker (about half of them) than normal quasars. For example, \citet{Ni18} mention that 7 of the 16 WLQs in their sample are X-ray weak. \citet{Luo15} suggests that this weakness may be caused by the shielding gas which prevents the observer from seeing the central X-ray emitting region. 




Now, let's move to the second type, which is the BLR itself. \citet{Shemmer09, Nikolajuk12} claim that abnormally broad emission line region properties -- e.g. a significant deficit of line-emitting gas in BLR -- impact the weakness. \cite{Nikolajuk12} postulate that the weakness or absence of emission lines in WLQs does not seem to be caused by their extremely soft ionizing continuum but rather by the low covering factor ($\Omega/4\pi$) of their BLR, their low-ionization emission lines are weak. \cite{Hryniewicz10} hypothesize an early stage of formation of QSO. WLQ would be a quasar where BLR is just forming. In this case, where the BLR is just created the covering factor might be lower than in normal AGNs. According to \cite{Nikolajuk12} the low value of $\Omega$ means that the BLR in the WLQ could have fewer clouds.  However, they found that the $\Omega$ of regions responsible for producing CIV and Ly $\alpha$ are similar in WLQs and QSOs, whereas the covering factor of high-ionization lines to low-ionization lines are lower in WLQ than in QSOs (this result was observed only in four sources of their sample).



As I mentioned above, WLQs have stunning X-ray properties. An X-ray to optical power-law slope parameter ($\alpha_{OX}$)\footnote{$\alpha_{OX} \equiv$ log$(L_{2keV}/L_{2500})$/log$(\nu_{2keV}/\nu_{2500})$} correlates with the luminosities at 2500 \angstrom\ in the typical radio-quiet quasar without broad absorption lines (BALs) \cite[e.g.][]{Shen11}. However, about half of the WLQs have notably lower X-ray luminosities compared to the expectation of $\alpha_{OX}-L_{2500\angstrom}$ relation \citep{Nikolajuk12, Luo15}. These WLQs populations have a high apparent level of intrinsic X-ray absorption, Compton reflection, and scattering \citep{Ni18,Ni20}. It is worth noting that the other half of the WLQ population that is not X-ray weak indicates high Eddington ratios \citep{Luo15}. The shielding mechanism represents the thick, inner accretion disk that prevents radiation to reach the BLR. WLQs generally have not been associated with extreme X-ray variability before. However, recently \cite{Ni20} reported a dramatic increase in the X-ray of SDSS J1539+3954 by a factor of > 20. At the same time, they found that the overall UV continuum and emission-line properties do not show significant changes compared with previous observations.
\newpage
\section{Estimation of Black hole mass}
Knowledge about the values of BH masses and accretion rates is crucial in understanding the phenomena of accretion flows. The most robust technique is the reverberation mapping method \citep[RM,][]{Blandford82, Peterson93, Peterson14, Fausnaugh17, Bentz18, Shen19}. This method is based on the study of the dynamics surrounding the black hole gas. In this way, we are able to determine
the supermassive black hole (SMBH) mass:
\begin{equation}
    \mmbh = 
    \frac{v^2_{\mathrm{BLR}} R_{\mathrm{BLR}}}{G} =
    f \frac{\mathrm{\FWHM}^2 R_{\mathrm{BLR}}}{G}
\label{eq:blr}    
\end{equation}
where \mbh\ is the black hole mass, G is the gravitational constant, $R_{\mathrm{BLR}}$ is the distance between the SMBH and a cloud in the BLR, and $v_{\mathrm{BLR}}$ is the velocity of the cloud around the SMBH. This speed is unknown and we express our lack of knowledge in the form of the \acrshort{FWHM} of an emission-line and $f$ -- the virial factor, which describes the distribution of BLR clouds. In the RM the $\mRblr$\ is determined as the time delay between the continuum change and the BLR response. This technique requires a significant number of observations. 

A modification of this method is the single-epoch virial BH mass estimator \citep[see][for review]{Shen13}. The correlation between $\mRblr$ and the continuum luminosity ($\nu L_\nu$) is observed \citep{Kaspi05, Bentz09} and incorporated in Eq.~(\ref{eq:blr}): $R_{BLR} = $coeff.$ \times (\nu L_\nu)^{0.5}$, where the coefficient may differ depending on the geometry of the BLR (e.g $32.9^{+2.0}_{-1.9}$ in \citeauthor{Kaspi2000} 2000). Thus, the method is powerful and eagerly used because of its simplicity \citep{Kaspi2000, Peterson04, Vestergaard06, Shen13, Plotkin15, Wang19}. 

The non-dynamic method, which is the spectra disk-fitting method (see chapter 2), is based on a well-grounded model of emission from an AD surrounded black hole \citep[e.g.][]{SS73, NT73}. The most important parameter in such models is the mass of the black hole and the accretion rate. The spin of the black hole and the viewing angle are also taken into account. In this technique, the observed SED of an AGN is fitted to the theoretical model and one can constrain these four parameters \citep{Marculewicz2020}. More advanced disk spectra models, which take into account the irradiation effect, limb-darkening/brightening effects, the departure from a blackbody due to radiative transfer in the disk atmosphere, and the ray-tracing method to incorporate general relativity effects in light propagation, can be fitted \citep[e.g.][]{Hubeny2000, Loska04, Sadowski09, Czerny11HNS, Laor11,Czerny19BLRscaling}.

\section{Accretion disk model}\label{sec:AD_model}


The spectrum of the Keplerian disk is a perfect blackbody approximation. The first approximation of the AD is the disk in the Newtonian gravitational potential. Energy generated in accretion is transported vertically to the plane of the symmetry of the disk and radiated outside. The disk flux radiated at radius $r$ is expressed as:
\begin{equation}
    F(r) = \frac{3GM \mdotM}{8 \pi r^3} (1-\frac{R_{in}}{r}),
    \label{eq:Newtonian}
\end{equation}
where \dotM\ is the accretion rate, $R_{in}$ the beginning of the disk. Inner boundary condition for the Schwarzschild ($a_* = 0$, non-rotating) BH is $R_{in} = 3R_{Schw}$ and for Kerr ($a_* \neq 0$, rotating) BH is $\frac{1}{2} R_{Schw} \leq R_{in} \leq 3R_{Schw}$.
\newpage
The part in parentheses specifies the internal boundary condition in the Newtonian approximation. When we consider a case involving a more real situation, namely the Kerr instance, we consider the local flux Eq. (\ref{eq:Newtonian}) with \cite{NT73} correction:

\begin{equation}
    F(r) = \frac{3GM \mdotM}{8 \pi r^3} \frac{L}{BC^{1/2}},
    \label{eq:NT_flux}
\end{equation}
where:

\begin{equation}
    \begin{aligned}
    B = 1 + \frac{a_{*}}{x^3} \\
    C = 1 - \frac{3}{x^2} + 2 \frac{a_{*}}{x^3} \\
    x = \sqrt{\frac{r}{GM/c^2}},
    \label{eq:Bcx_equation}
    \end{aligned}
\end{equation}

The \textit{L} is an integral function that depends on the radius of the circular orbit where the accreting matter is. For a stationary disk the \textit{L} function takes the form \citep{Page_Thorne1974}:

\begin{dmath}
    L = \frac{1+a_*x^{-3}}{\sqrt{1-3x^{-2}+2a_{*}x^{-3}}}\frac{1}{x} \left[x-x_{0}-\frac{3}{2}a_*\ln\left(\frac{x}{x_0}\right)
    -\frac{3(x_1-a_*)^2}{x_1(x_1-x_2)(x_1-x_3)}\ln\left(\frac{x-x_1}{x_0-x_1}\right)-
    \frac{3(x_2-a_*)^2}{x_2(x_2-x_1)(x_2-x_3)} \\ 
    \ln\left(\frac{x-x_2}{x_0-x_2}\right)-\frac{3(x_3-a_*)^2}{x_3(x_3-x_1)(x_3-x_2)}
    \ln\left(\frac{x-x_3}{x_0-x_3}\right)
    \right],
   \label{eq:L_expresion}
\end{dmath}
where $x_1, x_2,$ and $x_3$ are the three zeros of the equation $x^3 - 3x + 2a_* = 0$.

\begin{equation}
    \begin{aligned}
    x_1 = 2 \cos \left( \frac{\arccos(a_*) - \pi}{3} \right)\\
    x_2 = 2 \cos \left( \frac{\arccos(a_*) + \pi}{3} \right)\\
    x_3 = - 2 \cos \left( \frac{\arccos(a_*)}{3} \right),
    \end{aligned}
    \label{eq:three_zero_of_eq}
\end{equation}

In this approach, we assume that the disk radiated as a perfect blackbody. Using the temperature relation:

\begin{equation}
    \sigma T_{eff}^4(r) = F(r), 
    \label{eq:eff_temp}
\end{equation}
where $\sigma$ is the Stefan-Boltzmann constant,
we are able to describe flux and temperature dependency.

In the case of a disk as a black-body, the Stefan-Boltzmann law holds, hence the intensity of the radiation as a function of frequency is expressed as the Planck spectrum:

\begin{equation}
    B_{\nu}(T) = \frac{2h\nu^3}{c^2}\frac{1}{e^{\frac{h\nu}{kT_{eff}(r)}}-1},
    \label{eq:Planck_eq}
\end{equation}
Putting Eq. (\ref{eq:eff_temp}) into Eq. (\ref{eq:Planck_eq}) we are able to calculate the $F_{\nu}$ (see Eq. \ref{eq:f_nu_2pirdr}). For the AD and the radiation flux that reaches the observer in the distance ($D$) and the disc observed inclined at an angle \textit{i} is equal:

\begin{equation}
    F_{\nu} = \frac{\cos i}{D^2} \int_{R_{out}}^{R_{in}} B_{\nu} \left [ T(r)\right] \times 2\pi RdR,
    \label{eq:f_nu_2pirdr}
\end{equation}
where $F_{\nu}$ is a monochromatic flux observed on the frequency $\nu$; $R_{out}$ and $R_{in}$ are the end radius of a disk and the beginning of the radiated disk respectively. $2 \pi r$ because we are integrated rings.

Based on Eq. (\ref{eq:f_nu_2pirdr}) we are able to estimate \mbh\, $\mdotM$, $a_*$, and $i$ parameters. For example, in the case $a_* = 0$ after solving Eq. (\ref{eq:f_nu_2pirdr}) we get:
\begin{equation}
    F_{\nu} \sim \nu^{\frac{1}{3}} \times \frac{\cos i}{D^2} \times (M\mdotM)^{\frac{2}{3}},
    \label{eq:F_nu_mdot}
 \end{equation}

\noindent Based on it and using the observations we are able to constrain the $M$, $\mdotM$, and $cos$ $i$.

\chapter{Sample selection and data reduction}


\section*{Abstract}

An unexpected population of quasars display exceptionally weak or completely missing broad emission lines
in the ultraviolet (UV) rest-frame \citep{Plotkin15}. There are above 200 objects known as WLQs or WLQs candidates \citep{DS09, Shemmer10, Hryniewicz10, Wu11, Plotkin15, Ni18, Tim20}. Only over a dozen have the estimated masses of a black hole by the single-epoch method. Many contributing factors were taken into consideration. In this chapter, I describe the influence of the most impactful ones on the data.

In this thesis I compute luminosity distances using the standard cosmological model ($H_{0}$ = 70 \kms\ Mpc$^{-1}$, $\Omega_{\Lambda}$ = 0.7, and $\Omega_{\rm M}$ = 0.3 \citealt{Spergel}).


\section{Sample selection and data preparation}
\label{sec:2}
\subsection{Sample selection}
\label{sec:2.1}
In Sec. \ref{sec:intro_WLQ} I described the properties of WLQs and their history. From nearly 200 known WLQ candidates I have chosen 10. The selection was dictated by the calculation of their masses by the single-epoch virial BH mass method. According to \citet{Shemmer10, Hryniewicz10, Wu11, Plotkin15}, only SMBH in 10 of the WLQs were well estimated by this method.

The sample contains WLQs whose positions cover a wide range of redshift from 0.2 to 3.5 (see Tab.~\ref{tab:coordinates}). Four objects, namely SDSS J083650.86+142539.0 (hereafter J0836), SDSS J141141.96+140233.9 (J1411), SDSS J141730.92+073320.7 (J1417), and SDSS J144741.76-020339.1 (J1447) were analysed by \citet{Shen11} and \citet{Plotkin15}. Three other sources -- SDSS J114153.34+021924.3 (J1141) and SDSS J123743.08+630144.9 (J1237) were studied by \citet{DS09}, and SDSS J094533.98+100950.1 (J0945) by \citet{Hryniewicz10}. The quasar SDSS J152156.48+520238.5 (J1521) was inspected by \citet{Just07} and \citet{Wu11}. The number of the above objects in the sample taken from the SDSS campaign \citep{SDSSDR7} has been increased by the two next WLQs: PG 1407+265 and PHL 1811. The first object named PG1407+265 (hereafter PG1407) is the first observed WLQ in history and was intensively examined by \citet{Mcdowell95}. PHL 1811 is the low redshift source classified also as the NLS1 galaxy \citep{Leighly07a, Leighly07b}. 
\subsection{Observed data}
\label{sec:2.2}
Observational data was collected from various catalogs. The infrared range was provided by the Wide-field Infrared Survey Explorer (WISE) and the Extended Source Catalog of the Two Micron All Sky Survey (2MASS). The first one is a 40 cm diameter telescope in Earth orbit. The second project is related to the U.S. Fred Lawrence Whipple Observatory on Mount Hopkins, Arizona, and at the Cerro Tololo Inter-American Observatory in Chile. 

The optical data was taken from the Sloan Digital Sky Survey (SDSS) 2.5-m wide-angle optical telescope located at Apache Point Observatory in New Mexico, United States. Fluxes in B (4361 \AA)\footnote{1 \AA = 0.1 nm = $10^{-10}m$} and R (6407 \AA) colors were performed by the Dupont 2.5 m telescope at Las Campanas Observatory (LCO) in the southern Atacama Desert of Chile. The ultraviolet range was taken from the Galaxy Evolution Explorer (GALEX), an orbiting ultraviolet space telescope (NUV -- 2267 \AA, FUV -- 1516 \AA).
Photometric points of WLQs at visible wavelengths are collected based mainly
on the SDSS optical catalog Data Release 7 \citep{SDSSDR7}. It contains the u (3551 \AA), g (4686 \AA) , r (6166 \AA), i (7480 \AA), and z (8932 \AA) photometry. In the case of PHL 1811, the measurements of fluxes have been based on B and R
colors \citep{Prochaska11}. The flux at U (3465 \AA) band was observed by the UVOT telescope on-board the {\it Swift} satellite
\citep{UVOT}. Near-infrared photometry in the W1-W4 bands have been taken from the WISE Preliminary Data Release \citep{wise,Wise_2} \footnote{Band {\it W1} – \SI{3.4}{\micro\metre}, Band {\it W2} – \SI{4.6}{\micro\metre}, Band {\it W3} – \SI{12}{\micro\metre}, Band {\it W4} – \SI{22}{\micro\metre}}. Those data were supplied by photometry in the J (\SI{1.25}{\micro\metre}), H (\SI{1.65}{\micro\metre}), and \Ks\ (\SI{2.17}{\micro\metre}) colors obtained from the 2MASS \citep{2mass}. Crucial points for the project are those detected in near- and far-ultraviolet (NUV, FUV, respectively) wavelengths. They are provided by the Galex Catalog Data Release 6 \citep{galex}. 


I have used the photometric points to cover wide SED. However, the photometric point, which lays on an emission line, may represent the flux of this line not of the continuum. Therefore I have used the spectra when possible. I have used the spectra observed by SDSS, to check photometric data positions in regard to the spectrum. Eight out of ten WLQs spectra have been taken from the SDSS catalog. In the case of PHL 1811 and PG 1407+265 the spectra have been taken from \citet{Leighly07a} and \citet{Mcdowell95}, respectively. 
To check photometric points, I used continuum fitting windows (see Tab. \ref{tab:windows_forster} adapted from \citealt{Forster_spectral_windows2001}). I have taken the spectrum of each WLQ, have set the range of each window, and have binned the spectral data to get the point representing the mean flux in the window. This is my new photometric point. Then I checked whether these points are comparable with photometric points taken from the mentioned catalogs and added them to the photometric set.

\begin{table}[ht]
    \centering
    \begin{tabular}{c}
        Rest frame wavelength range (\AA)  \\
        \hline
        Continuum \\
        \hline
        1140 - 1150 \\
        1275 - 1280\\
        1320 - 1330\\
        1455 - 1470\\
        1690 - 1700\\
        2160 - 2180\\
        2225 - 2250\\
        3010 - 3040\\
        3240 - 3270\\
        3790 - 3810\\
    \end{tabular}
    \caption[Continuum fitting windows]{Continuum fitting windows \citep{Forster_spectral_windows2001}}
    \label{tab:windows_forster}
\end{table}

The spectra and photometric points are in observed frame. Eventually, I have shifted them to a source rest frame based on wavelength relation:
\begin{equation}
1 + z_{spec} = \frac{\lambda_{obs}}{\lambda_{emit}}    
\end{equation}
where $z_{spec}$ is mainly emission redshift of the source; $\lambda_{obs}$ is the observed wavelength of a source by telescopes; $\lambda_{emit}$ is emitted by the source.

Basic observational properties of the sample of WLQs and the sources of their photometry points are listed in Tab.~\ref{tab:coordinates}.


\begin{sidewaystable}
\fontsize{10}{12}\selectfont
\caption{Sample of Weak Emission-Line Quasars and the sources of their photometry points
   \label{tab:coordinates}}

\vspace{2ex}
\noindent
\begin{center}
\begin{tabular}{|c|c|c|c|c|c|}
\hline 
\hline 
 Name &  RA & Dec & {$z_{spec}$} & {$A_{V}$} & Telescope/satellite \\
 & (degrees)& (degrees) &  & & (electromagnetic bands) \\
  & (J2000.0)  &  (J2000.0) &  & & \\
(1) & (2) & (3) & (4) & (5) & (6) \\
\hline 
 SDSS J083650.86+142539.0 & 129.211935 & +14.427527 & 1.749 & 0.129& WISE (W1,W2,W3), SDSS (u,g,r,i,z), Galex (NUV) \\
SDSS J094533.98+100950.1 & 146.391610 & +10.163912 & 1.683 & 0.062& 2MASS (J,H,\Ks), SDSS (u,g,r,i,z), Galex (NUV,FUV) \\
SDSS J114153.34+021924.3 & 175.472251 & +02.323508 & 3.550 & 0.065 & WISE (W1,W2,W3,W4), SDSS (u,g,r,i,z) \\
SDSS J123743.08+630144.9 & 189.429435 & +63.029141 & 3.490 & 0.032 & WISE (W1,W2,W3,W4), SDSS (u,g,r,i,z) \\
SDSS J141141.96+140233.9 & 212.924908 & +14.042742 & 1.754 & 0.064 & WISE (W1,W2,W3,W4), SDSS (u,g,r,i,z), Galex (NUV,FUV) \\
SDSS J141730.92+073320.7 & 214.378855 & +07.555744 & 1.716 & 0.084 &  WISE (W1,W2,W3,W4), SDSS (u,g,r,i,z), Galex (NUV,FUV) \\
SDSS J144741.76-020339.1 &  221.924048 & -02.060986 & 1.430 & 0.163 & WISE (W1,W2), SDSS (u,g,r,i,z), Galex (NUV,FUV)) \\
SDSS J152156.48+520238.5 & 230.485324 & +52.044062 & 2.238 & 0.052 &  WISE (W1,W2,W3,W4), 2MASS (J,H,\Ks), SDSS (u,g,r,i,z) \\
PHL 1811 & 328.756274 & -09.373407 & 0.192 & 0.133 & WISE (W1,W2,W3,W4), 2MASS (J,H,\Ks), LCO (B,R), Swift (U), \\
 & & & & & Galex (NUV,FUV) \\
PG1407+265 & 212.349634 & +26.305865 & 0.940 & 0.043 &WISE (W1,W2,W3), 2MASS (J,H,\Ks), SDSS (u,g,r,i,z) \\
\hline
\end{tabular}
\begin{quote}
  The coordinates (Col. 2 and 3), spectral redshift (Col. 4), and foreground Galactic extinction measured at the V color (Col. 5) are taken from the NASA/IPAC Extragalactic Database (NED). The column (6) contains references to the names of the relevant catalogs and photometric points.
\end{quote}
\end{center}
\end{sidewaystable}
\section{Data reduction}
Due to contamination either by internal or external effects such as the dust in our Galaxy, the influence of the intergalactic medium (\acrshort{IGM}), starlight from the host galaxy, or the dusty torus in the AGN, the observational data has to be corrected.
Let's imagine a beam of photons that are moving toward an observer. It has traveled a long way through many obstacles, mediums, and materials. Firstly, let's consider the beginning of that travel: the AD radiate and this light I want to observe. One of the biggest impacts in the vicinity of the black hole is different absorbers like a molecular torus, winds, or a warm absorber. During calculation, I have to take into account the radiation of them. Relatively close to the system is the host-galaxy region, which I have to take into account as well. Next, the light has to come through the intergalactic stellar medium.


Last, but not least, is our mother galaxy -- the Milky Way and the absorption in our atmosphere as well, which is mostly taken into consideration within observations and calibrated before each observation \citep{Cardelli89}.

\subsection{Dereddening}\label{dereddening}


Dust scatters, absorbs light, and contains heavy elements produced by the nuclear burning of the stars. It is important to take this into account to get data as close as possible to the truth. The first correction of the Spectral Energy Distribution (SED) of all objects was the Galactic reddening with an extinction law. Interstellar extinction influences the whole range of spectra, especially in the ultraviolet. The influence of our Galaxy is well understood. Interstellar medium were studied by \citet{Cardelli89, Schlegel98, Fitzpatrick99, Hutchings01, Czerny2004_extinction}. This interstellar reddening can be written as \citep{Zombeck}:
\begin{equation}
    E(B-V) := (m_{B}-m_{V})_{observed}-(m_{B}-m_{V})_{intrinsic}
    \label{zombec_eq}
\end{equation}
Where $B \equiv m_{B}$ and $V \equiv m_{V}$ are the apparent brightness measured in blue and visible color (e.g. at 4000 \AA\ and 5500 \AA\ respectively in the Johnson-Morgan system, in magnitude). Equation (\ref{zombec_eq}) is equal to
\begin{equation}
    E(B-V) = A_{B} - A_{V}
    \label{Pogson}
\end{equation}
where $A_{\lambda} = m_{\lambda}^{observed} - m_{\lambda}^{intrinsic}$ and $\lambda$ is for B and V.

Based on the Pogson relation we know that each $A_{\lambda}$ is equal to:
\begin{equation}
     A_{\lambda} = 2.5 \log \frac{F_{\lambda}^{intrinsic}}{F_{\lambda}^{observed}}
\label{Pogson_rew}
\end{equation}

In order to calculate proper $F_{\lambda}^{intrinsic}$ for an object (e.g. WLQ), we must know $A_{\lambda}$ of it.
\citet{Cardelli89} express Eq. (\ref{Pogson}) in the form :


\begin{equation}
    A_{\lambda} = E(B-V) (a(\lambda) \times R_{V} + b(\lambda)),
\end{equation}
where $a(\lambda)$ and $b(\lambda)$ are functions calculated for each wavelength $\lambda$.

$R_{V}$ is defined as $A_{V}/E(B-V)$. $a(\lambda)$ and $b(\lambda)$ are given by the authors. $E(B-V)$ and $A_{V}$ values for a specific direction are taken from NED \footnote{The NASA/IPAC Extragalactic Database (NED): {\tt\string ned.ipac.caltech.edu}} database, based on the dust map of the Milky Way created by \citet{Schlegel98}. 

Generally for a normal region $R_{V} = 3.2 \pm 0.2$; however, it varies from 2.6 to 5.5 in the measurements of the diffuse ISM \citep{Fitzpatrick99}.
 

In \citet{Cardelli89} the extinction curve has a cutoff at 1250 \AA\ and for a few photometric points of WLQs I have used $\lambda < 1200$\AA. Nevertheless, the extinction law examined in the range of 900-1200 \AA\ seems to follow the Cardelli et al. law \citep{Hutchings01}. In this way, I extrapolate the curve down to 900 \AA\ for my FUV photometric points by using the same formula. $A_{V}$ values in the direction of the examined WLQ are listed in Tab. \ref{tab:coordinates}.

\subsection{Photoelectric absorption in the IGM}
In the case of high-z quasars, the UV fluxes are very sensitive to
photoelectric absorption in the IGM. 
We do not know the attenuation of the flux by the IGM along each line of sight. Therefore, following \citet{Castignani13}, I have used the effective optical depth, $\tau_{\rm eff}(\nu, z)$, averaged over all possible
directions. The correction for the observed intensity is: $I_{\nu, intrinsic} = I_{\nu, observed} \exp(\tau_{\rm eff}(\nu, z))$, based on the values of $\tau_{\rm eff}$ collected in Castignani et al. (their Table 1). Following this formula, I recalculate the fluxes in the Galex FUV and NUV, the Swift U band, and the SDSS u, g filters. The effective optical depth in the three other SDSS filters (i.e. r, i, and z) vanishes for the redshift range considered there. For weak active galactic nuclei such as WLQ, correction in the UV range is crucial for the calculation of the contribution from the AGN.

\subsection{Host-galaxy and its starlight}
\label{contamination}

Contamination by starlight plays an important role in the low-luminosity AGNs \citep[e.g.][]{Greene_Ho_contamination04, Netzer14_contamination}. However, the contribution to the SED from stars in the QSO host galaxy is likely
to be negligible \citep{Shen11_starlight,Collinson15}. This is a common assumption for quasars at $z>0.5$ \citep{Shen11_starlight}. According to \cite{Shen11_starlight} the host contamination for the luminosity of a quasar is substantial at $log$L(5100 \AA) < 44.5, and becomes negligible toward higher luminosities. The WLQs which I have used all had luminosities higher than $10^{44.5}$ \ergs at 5100 \AA.

Nevertheless, to check its contribution, I have determined the level of starlight for each of the objects individually. Following \citeauthor{Collinson15}, a 5 Gyr-old elliptical galaxy template\footnote{SWIRE Template Library: \citet{Polletta07}; {\tt\string www.iasf-milano.inaf.it/$\sim$polletta/templates/}} which is not AGN, have been used as the stars' contamination to the fluxes. I have estimated the level of starlight in each WLQ using the \mbh--$L_{\rm bulge}$ relation \citep{Degraf15}, where $L_{\rm bulge}$ is the bulge luminosity in the V-band, and \mbh\ means the mass of SMBH in WLQ. The \mbh\ of WLQs was taken from previous literature studies (see Tab.\ref{tab:bestfit}). I assumed that they are correct to estimate the level of starlight in WLQs, even if their masses are biased due to the inappropriate value of the FWHM. For this approach only the order of a mass was important.

\begin{figure}[!htb]
\centering
    \includegraphics[width=1.0\textwidth]{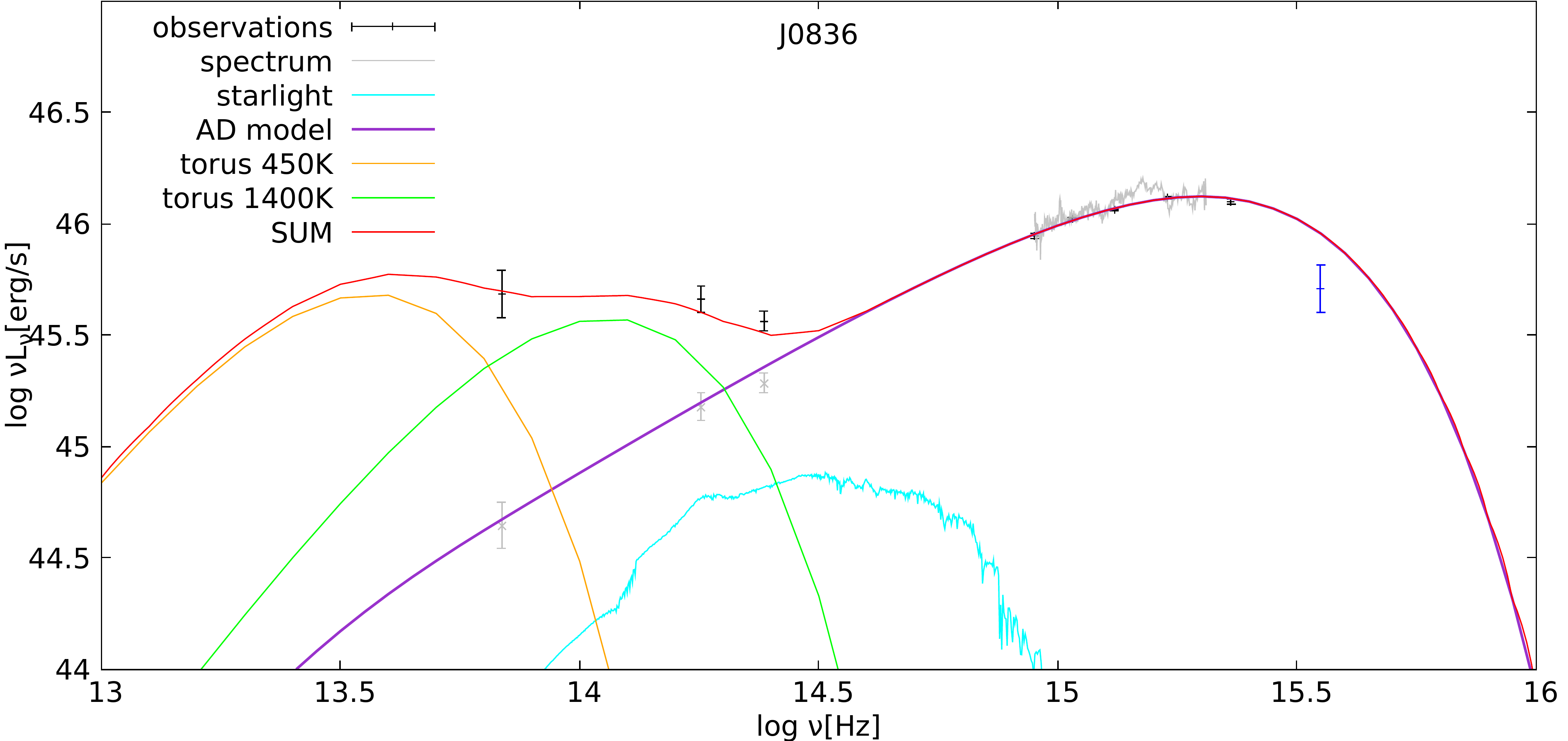}
    \caption[The best fit and components of SED fit for WLQ SDSS J083650.86+142539.0.]{The best fit of WLQ SDSS J083650.86+142539.0. Summary of all components. Black points with errors show photometry data and the grey line represents spectrum. The violet line shows the contribution of the AD model. Tori with temperatures 450K and 1400K, respectively, are displayed by orange and green lines. Cyan line points to the level of starlight. The sum of the components, and thus the best fit, is shown by the red line.}
    \label{fig:SUM_J0836}
\end{figure}

I have taken the \mbh--$L_{\rm bulge}$ relation given by \citet{Degraf15}:
\begin{equation}
    \frac{\mmbh}{\mMsun} = 10^{\alpha} \left(\frac{x}{x_{0}}\right)^{\beta}
\end{equation}
and parameters $\alpha$ and $\beta$ from their Table 2 and Table 5, x = $L_{\rm bulge, V}$ and $x_{0} = 10^{10.5} \mLsun$. For each of the objects, I found the best host galaxy starlight contribution to the SED. The level of starlight for WLQ SDSS J0836 is shown by cyan line in Fig. \ref{fig:SUM_J0836}. I have concluded that the host galaxy contribution to the total SED continuum is small in all the objects.

Starlight has a bigger contribution to infrared WISE points than to the optical/UV data and adding the contamination of the starlight to that data helps refine the fitting procedure.

\subsection{Torus contamination}
Torus is the dense, dusty molecular gas present in the tens of parsec away from a black hole. The contribution of the torus can be prominent for type 2 AGNs (obscured, Seyfert 2). To describe the torus contamination, I have used one or two single-temperature black-bodies (BB). It describes the thermal emission of tori visible at IR data, as a hot and cold component (see Fig.~\ref{fig:SUM_J0836}, orange and green lines). I have used the Planck function:
\begin{equation}
    B_{\nu}(T) = \frac{2h\nu^3}{c^2} \frac{1}{e^{\frac{h\nu}{kT}-1}}
\end{equation}
where  $\nu$ is the frequency, \textit{T} - absolute temperature, \textit{k} is the Boltzmann constant, \textit{h} - the Planck constant and \textit{c} is the speed of light.

This approach describes the contribution of torus well. The fitted effective temperatures are collected in 
Tab.~\ref{tab:Temperature}, the mean values of them for the two BB components are 1110 K and 460 K, respectively. Those values
are close to those referred to 'hot' and 'warm' BB components in AGNs (1100-2200K vs. 300-700K); \cite{Collinson17}. The temperatures of the 'hot' component are also similar to those seen in WLQs ($\mathrm{870\,K < T < 1240\,K}$); \citet{DS09}.
Finally, the photometric points and spectra of all WLQs have been corrected (see Fig. \ref{fig:SUM_J0836}) and prepared for the next step of study based on the fitting method.
\begin{table}
   \tablenum{2}
    \centering
    \caption {Fitted temperature of tori in the sample of WLQs}
    \begin{tabular}{cc}
    \hline   
    Name & Temperature [K]\\
    \hline
    J0836 & 450, 1400 \\
    J0945 & - \\
    J1141 & 450, 970 \\
    J1237 & 430, 970  \\
    J1411 & 410, 970  \\
    J1417 & 430, 950  \\
    J1447 & 410, 1250  \\
    J1521 & 750 \\
    PHL 1811 & 450, 1100  \\
    PG 1407 & 650, 1240  \\
    \hline
    \label{tab:Temperature}
    \end{tabular}
 \end{table}

\clearpage

\section{Sample selection of type 1 QSO and calibration of method}\label{sec:LBQS_method}
To check if my disk fitting method works concerning WLQs correctly, I am running a fitting method on a sample of normal type 1 quasars. For this purpose, I have selected a sample of objects taken from the Large Bright Quasar Survey (LBQS) \citep{Hewett95, Hewett01}. It is one of the largest published spectroscopic surveys of optically selected quasars at bright apparent magnitudes. It contains data including the positions and spectra of 1067 quasars. Additionally, \citet{Vestergaard09} give black hole masses and Eddington accretion rate estimates of 978 LBQS (see their Table 2). The disk fitting method gives results that we can trust as long as the spectrum in the ultraviolet is visible. My numerical program needs to see the UV pivot point in the Big Blue Bump regime of the SED to calculate \mbh\ and the accretion rate with errors. For this reason, I have chosen 27 quasars with the presence of a well visible big blue bump. The sample of the quasars is observed at redshifts between 0.254 and 3.36. Their supermassive black hole masses are in the range 8.09--10.18 in $\log M_{\rm BH}$ (M$_{\odot}$) and luminosities, $\log L_{\rm bol}$ (\ergs) =  45.25--47.89. The photometric points of the selected quasars come from the same catalogs mentioned earlier. I have performed the data correction process in the same way as for WLQs.


    

%
%


\section{Hypothesis testing approach}







\subsection{$\chi^2$ procedure}
In my fitting technique, I have used a $\chi^2$ procedure to find the best-fit model and evaluate the quality of the fit. In order to obtain the best method for hypothesis testing approach I followed methods that has been applied in statistical tests. It is based on directly matching the observed photometric points to the AD model (E, which I will generate). 
In this approach, I have calculated $ \chi^2 = \sum_{i=1}^n (O_i - E_i)^2/\sigma_i$ for SED of each quasar, $O_i$ is the $i$th photometric point of the corrected flux, $E_i$ is the continuum level taken from model. Modeled monochromatic flux $F_i$ which correspond to the $i$th photometric point, $\sigma_i$ is the observed error, and $n$ is the total number of observed data for the quasar \citep{chi_square}. The $\chi^2$ test determines if data sample matches with the model. If the $\chi^2$ statistic value is high (e.g. > 10) it means that a data does not fit with the model, whereas a small $\chi^2$ value (< 5) tells us how observed data fits with a model which I represent. It valid how well observations one can define by a model description. 

\subsection{Bayesian inference}
An alternative inference technique is to treat unknown parameters as if they were random variables. We start with an \textit{a priori} distribution of the parameters and then use the collected data and Bayes' theorem to obtain an updated \textit{a posteriori} distribution of the parameters. Instead of determining the probability associated with the test, we determine the probability of the parameters themselves. For this reason to use the information on yet determined data (e.g. black hole masses, accretion rates in quasar sample) one can carry out an analysis using the Bayesian method.

According to Bayes’ theorem the posterior probability can be assigned as \citep{sivia_rawlings_2009} :
\begin{equation}
P(H|D,I) = P(D|H,I) \frac{P(H,I)}{P(D,I)},
\label{bayes_chapter2}
\end{equation}

\noindent where $H$ is a hypothesis, which could be affected by the data $D$. In our case the hypothesis is a
particulate model $m$. $I$ is any prior information about $H$. The information $I$ corresponds to the
observed values e.g. for the black hole masses, accretion rates, black hole spins that influence the model. $P(H,I)$ is the prior probability of an uncertain quantity that express one’s beliefs before calculation of the posterior probability. The $P(D|H,I)$ is the probability of observing the measured data $D$ if the hypothesis $H$ was true. It measures the goodness of a fit of the model to data. In other word it refers to the likelihood of model $m$. Note that this likelihood function is an evidence function, whereas the posterior probability is a hypothesis function. $P(D,I)$ is the probability of $D$ occurring. 

To sum up, prior probability, in Bayesian statistical inference, is the best assessment of the probability of an outcome based on the current knowledge before a new observation/experiment is performed and collection of new data. The posterior probability is the revised probability of an event occurring after taking into consideration new information. The posterior probability is calculated by updating the prior probability by using Bayes’ theorem. In statistical terms, the posterior probability is the probability of the event $H$ occurring given that event $D$ has occurred \citep{gelmanbda04_bayes}.

\include{ch3}
\include{ch4}
\chapter{Results}

\section{Method - Novikov-Thorne model of accretion disk}

\begin{figure}[ht]
  \centering
    \includegraphics[width=1.0\textwidth]{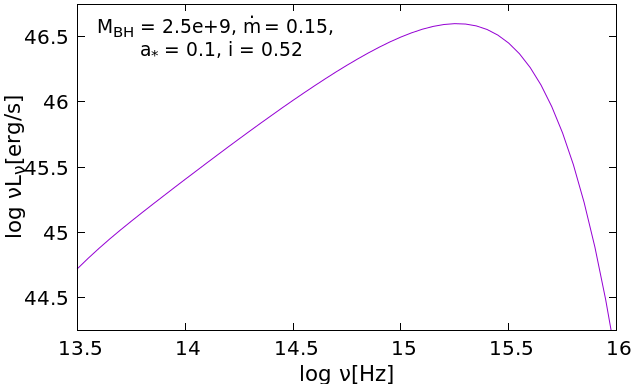}
 \caption[Example of the Novikov-Thorne model]{Example of the Novikov-Thorne model: $\mmbh = 2.5\times 10^9, \mdotm = 0.15, a_* = 0.1, i = 0.52.$}
\label{fig:NT_example_model}
\end{figure}

The primary goal of this work is to fit the SED of quasars by the geometrically thin and optically thick accretion disk (AD) model described by the \citeauthor{NT73} (\acrshort{NT}) equations. In the simplest approach, the AD continuum can be illustrated by the \citeauthor{SS73} model, nevertheless this attitude does not include a non-zero spin. The solution to this problem has resulted in the NT equations that I use in my numerical code. The code was written in the FORTRAN language. Equations (\ref{eq:NT_flux} - \ref{eq:three_zero_of_eq}) were included and the model output of the continuum of the AD were produced. An example of the model is shown on Fig. (\ref{fig:NT_example_model}). As the spin of the black hole increases, the innermost stable circular orbit (ISCO) decreases and the disk produces more high-energy radiation. The output continuum of the NT model is fully specified by four parameters, which I determine. These 4 parameters are: the black hole mass -- \mbh, the mass accretion rate -- $\dot M$, the dimensionless spin\footnote{$\mas = \frac{cJ}{GM^2}$} -- \as, and the inclination -- \incl\ at which an observer looks at the AD. The mass of the black hole is expressed in units of mass of the Sun (\Msun)\footnote{1 \Msun = 1.99 $\times10^{30}$ kg}, and the accretion rate in the form of the dimensionless Eddington rate, i.e. $\dot m$\footnote{$\dot m := \frac{\dot M}{\dot M_{Edd}}$, where $\dot M_{Edd} = \frac{4 \pi G \mmbh \mu m_p}{\sigma_{T} c \eta}$, where: G - gravitational constant, c - speed of light, $m_p$ - mass of proton, $\mu$ - mean molecular weight, $\sigma_{T}$ - Thomson cross-section $\eta$ - radiative efficiency}.

I construct a grid of 366000 models of AD, for evenly spaced values of \mbh, \dotm, \as, and \incl.  The $\log \mmbh$ range is from 6.0 to 12.0, the Eddington accretion rate covers the band 0--1, and the dimensionless spin $0 \leq \mas \leq 0.9$ with the step 0.1.  The inclination is fixed for 6 values that cover a range from 0$^{\circ}$ to 75$^{\circ}$ in steps of 15$^{\circ}$ (see Tab. \ref{tab:grid}). 
\begin{table}
\centering
\caption {Parameter values for the grid of the AD models}
\begin{tabular}{lcc}
\hline
Parameter & $\Delta$ & min-max values\\
\hline
 $\log$ \mbh  & 0.1 & 6--12\\ 
 \dotm & 0.01 & 0--1 \\
 \as & 0.1 & 0--0.9 \\
 \incl & 15$^{\circ}$ & 0$^{\circ}$--75$^{\circ}$\\
\hline
\label{tab:grid}
\end{tabular}
\end{table}
It is important to determine the radiative efficiency, $\eta$, in the SED fitting method. There are many approaches to estimate this. The $\eta \simeq 0.057$ is analytical calculation (i.e. taken from theory) for non-rotating BH. \citet{Shankar09} suggest $\eta = 0.05-0.1$ in relation to AGNs based on observations of the power emitted by them. Observational constraints on growth of BHs made by {\citet{Yu}} give us reasonable argument that $\eta$ should be $\gtrsim 0.1$. Even more, \citet{Cao08} proposed $\eta = 0.18$ for QSOs with BH masses above $10^9$~\Msun. 

My performed analysis (i.e. calibration on LBQS sources) allows us to conclude that $\eta$ should be in the range of $0.15-0.20$. Those values are required by me to obtain the conformity of SMBH masses in LBQS if I use my SED fitting and single-epoch virial methods. I adopt the value of $\eta=0.18$ in relation to both types of quasars -- LBQS and WLQs in my PhD thesis.

\section{Description of results}

\begin{figure}[ht]
  \centering
    \includegraphics[width=1.0\textwidth]{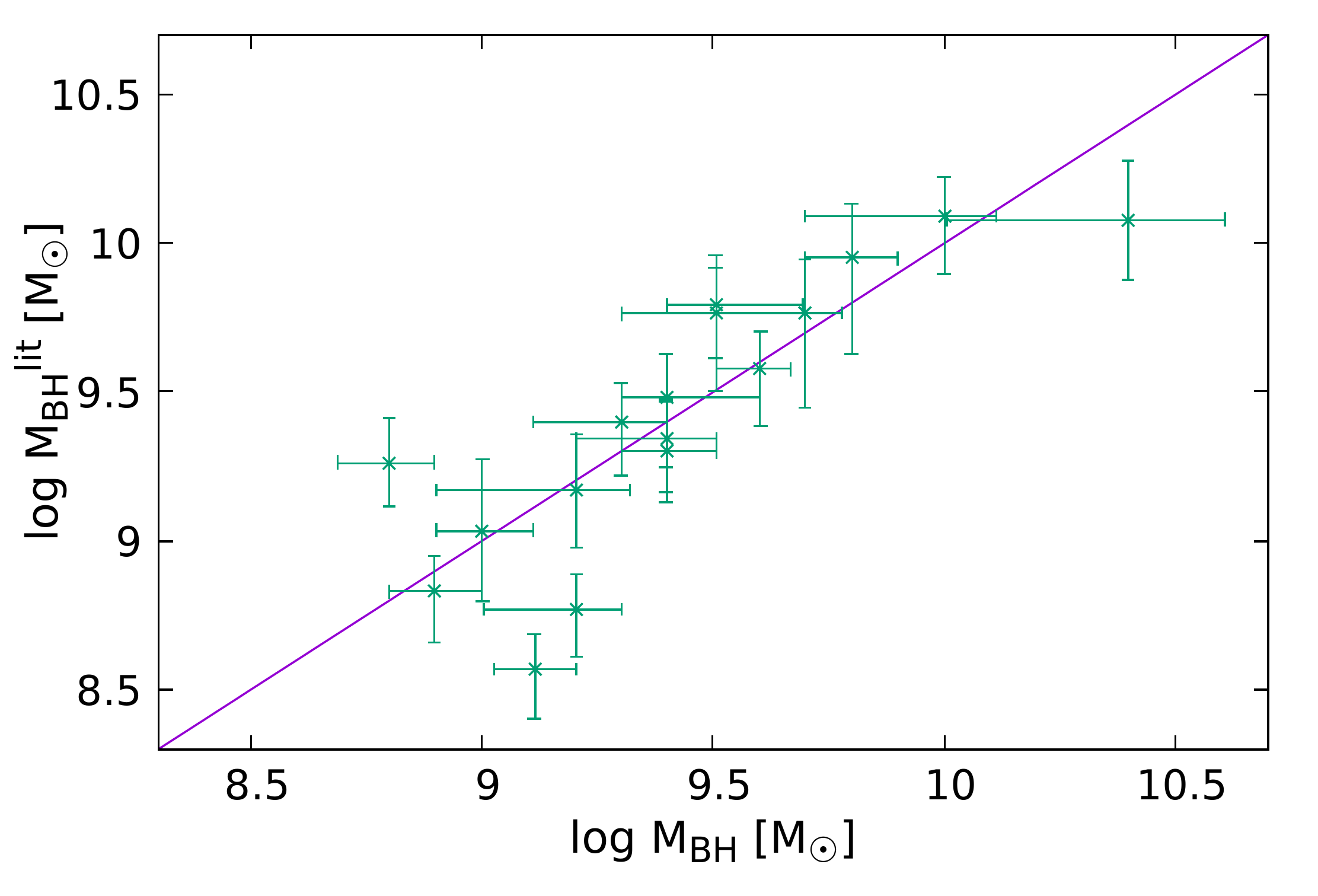}
 \caption[Comparison of LBQS masses with results from the continuum-fitting method]{Comparison of LBQS masses (on the y-axis) with \mbh\ from my models based on the continuum fitting method (on the x-axis). Violet solid line is identity 1:1 line.}
\label{fig:LBQS_plot}
\end{figure}

For the initial analysis, I have used 27 quasars from the LBQS survey (see Sec. \ref{sec:LBQS_method}). I have fitted their photometric points to the generated models of the ADs. I would like to note that I use the same grid of 366,000 models. Fig. \ref{fig:LBQS_plot} shows us a comparison of the supermassive black hole masses determined by \citet{Vestergaard09} -- \mbhlit\ (on the y-axis), to those obtained by us -- \mbh\ (on the x-axis). Both masses are given in mass units of the Sun. The solid violet line is a 1:1 identity line. \citet{Vestergaard09} used the black hole mass determination based on their formula (1), which is proportional to \FWHM (line) and luminosity $\nu L_{\nu}$. The compliance of masses and relatively small distribution of errors means that the continuum fitting method applies to quasars. 

\begin{figure}[!htb]
\centering
    \includegraphics[width=0.85\textwidth]{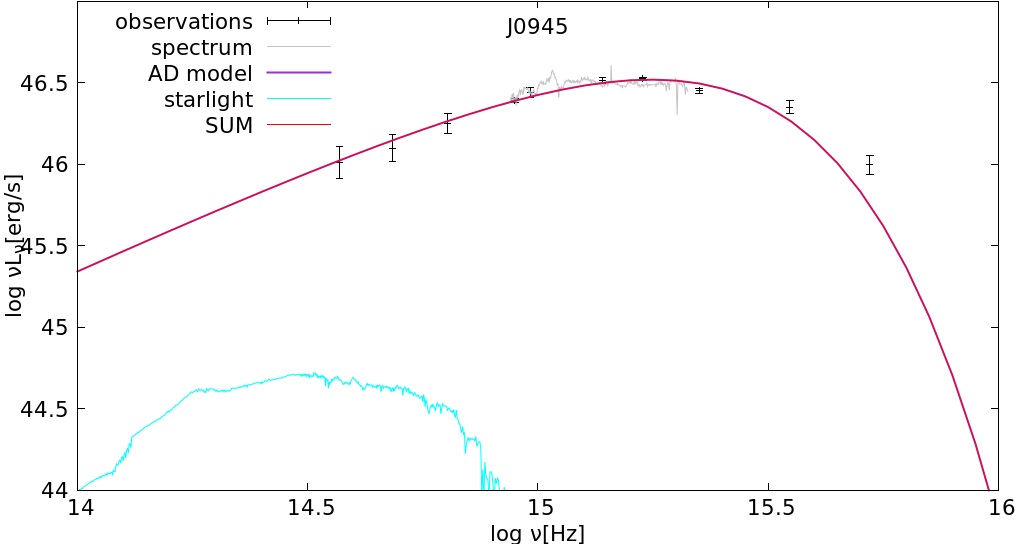}
    \caption[The best fit and components of SED fit for WLQ SDSS J094533.98+100950.1]{The best fit of WLQ SDSS J094533.98+100950.1. Summary of all components. Black points with errors show photometry data and the grey line represents spectrum. The violet line shows the contribution of the AD model. Cyan line points to the level of starlight. The sum of the components, and thus the best fit, is shown by the red line. Only in this case I did not need to model tori. A violet and a red lines are very close to each other.}
    \label{fig:SUM_J0945}
    \includegraphics[width=0.85\textwidth]{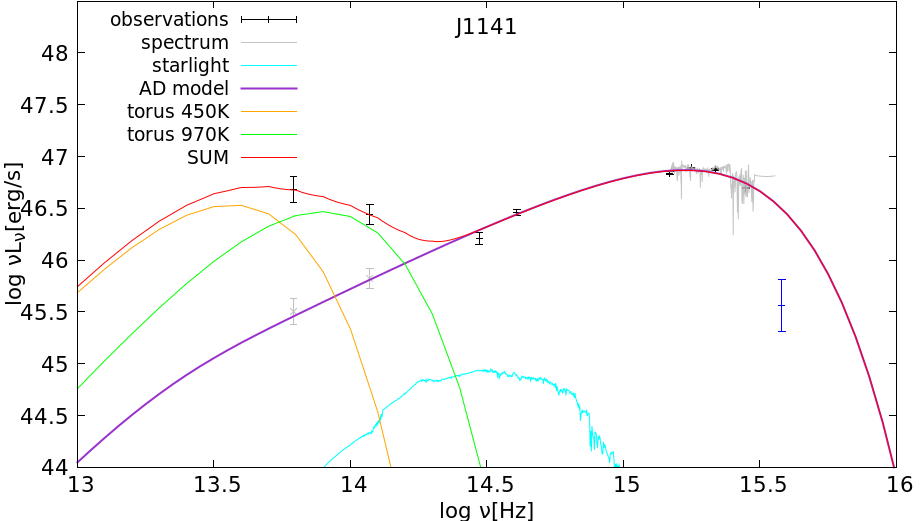}
    \caption[The best fit and components of SED fit for WLQ J114153.34+021924.3 ]{The best fit of WLQ J114153.34+021924.3. Summary of all components. Black points with errors show photometry data and the grey line represents spectrum. The violet line shows the contribution of the AD model. Tori with temperatures 450K and 970K, respectively, are displayed by orange and green lines. Cyan line points to the level of starlight. The sum of the components, and thus the best fit, is shown by the red line.}
    \label{fig:SUM_J1141}
\end{figure}

\begin{figure}[!htb]
\centering
    \includegraphics[width=0.85\textwidth]{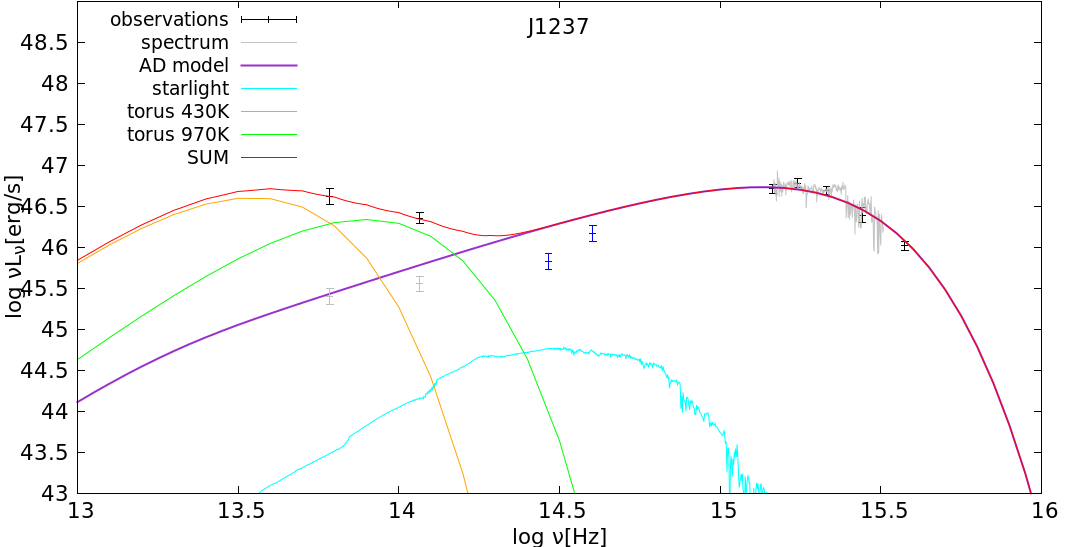}
    \caption[The best fit and components of SED fit for WLQ SDSS J123743.08+630144.9]{The best fit of WLQ SDSS J123743.08+630144.9. Summary of all components. Black points with errors show photometry data and the grey line represents spectrum. The violet line shows the contribution of the AD model. Tori with temperatures 430K and 970K, respectively, are displayed by orange and green lines. Cyan line points to the level of starlight. The sum of the components, and thus the best fit, is shown by the red line.}
    \label{fig:SUM_J1237}
    \includegraphics[width=0.85\textwidth]{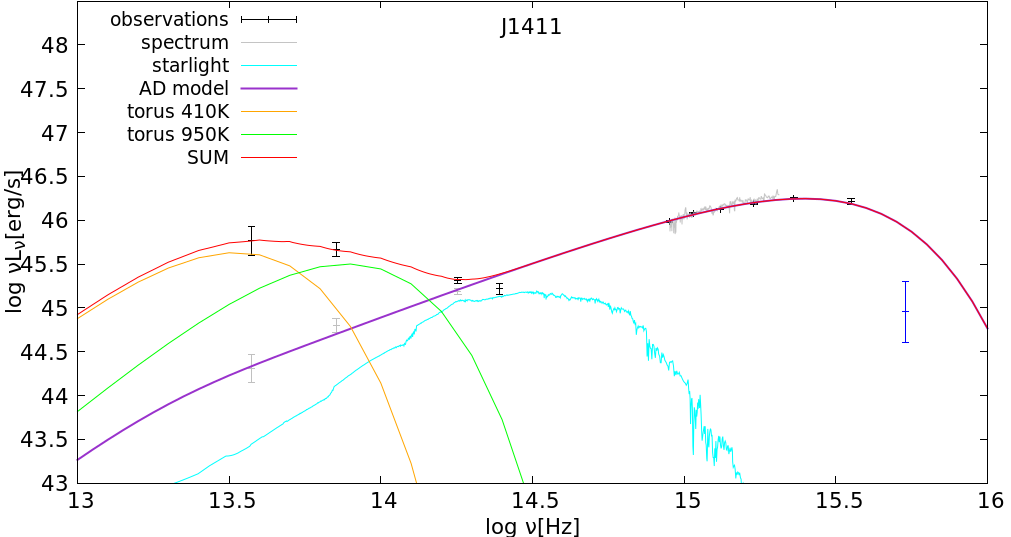}
    \caption[The best fit and components of SED fit for WLQ SDSS J141141.96+140233.9.]{The best fit of WLQ SDSS J141141.96+140233.9. Summary of all components. Black points with errors show photometry data and the grey line represents spectrum. The violet line shows the contribution of the AD model. Tori with temperatures 410K and 950K, respectively, are displayed by orange and green lines. Cyan line points to the level of starlight. The sum of the components, and thus the best fit, is shown by the red line.}
    \label{fig:SUM_J1411}
\end{figure}

\begin{figure}[!htb]
\centering
    \includegraphics[width=0.85\textwidth]{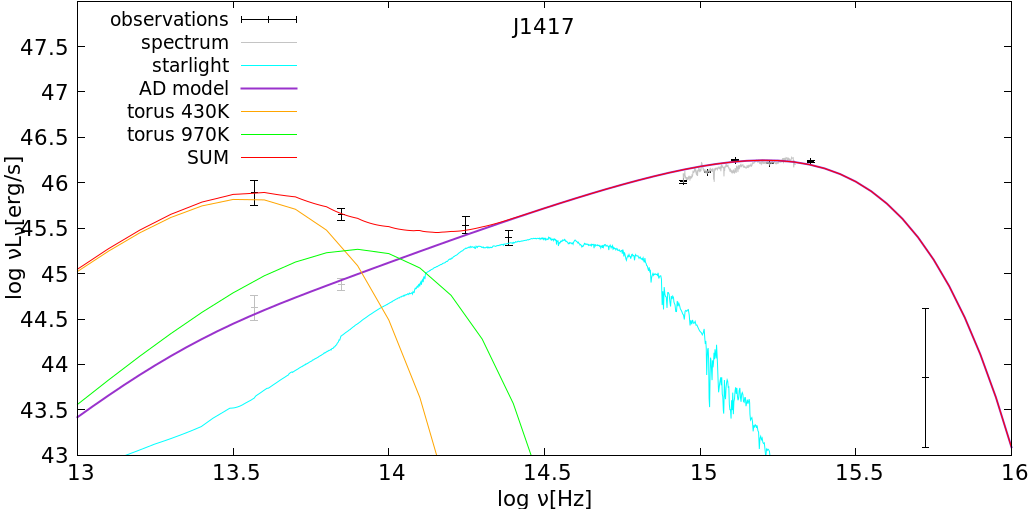}
    \caption[The best fit and components of SED fit for WLQ SDSS J141730.92+073320.7]{The best fit of WLQ SDSS J141730.92+073320.7. Summary of all components. Black points with errors show photometry data and the grey line represents spectrum. The violet line shows the contribution of the AD model. Tori with temperatures 430K and 970K, respectively, are displayed by orange and green lines. Cyan line points to the level of starlight. The sum of the components, and thus the best fit, is shown by the red line.}
    \label{fig:SUM_J1417}
    \includegraphics[width=0.85\textwidth]{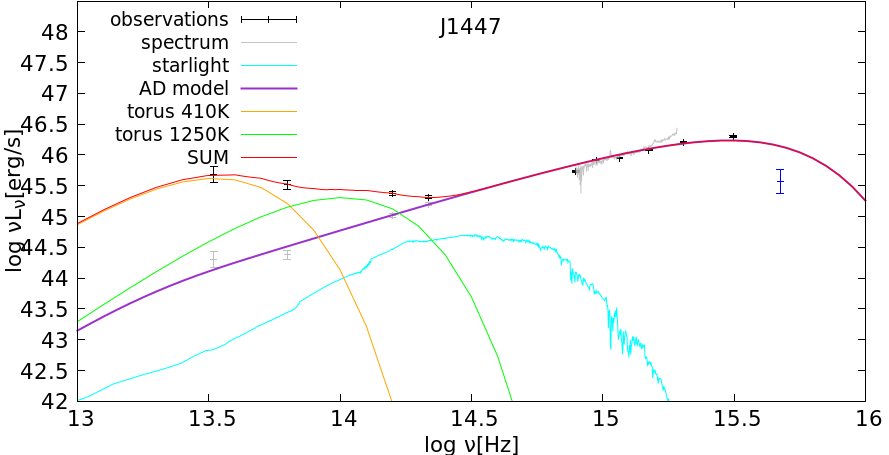}
    \caption[The best fit and components of SED fit for WLQ SDSS J144741.76-020339.1.]{The best fit of WLQ SDSS J144741.76-020339.1. Summary of all components. Black points with errors show photometry data and the grey line represents spectrum. The violet line shows the contribution of the AD model. Tori with temperatures 410K and 1250K, respectively, are displayed by orange and green lines. Cyan line points to the level of starlight. The sum of the components, and thus the best fit, is shown by the red line.}
    \label{fig:SUM_J1447}
\end{figure}

\begin{figure}[!htb]
\centering
    \includegraphics[width=0.85\textwidth]{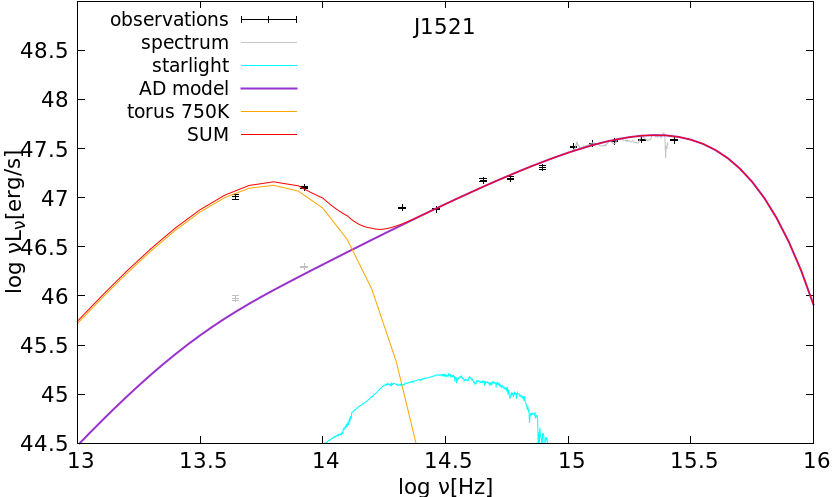}
    \caption[The best fit and components of SED fit for WLQ SDSS J152156.48+520238.5]{The best fit of WLQ SDSS J152156.48+520238.5. Summary of all components. Black points with errors show photometry data and the grey line represents spectrum. The violet line shows the contribution of the AD model. Tori with temperatures 750K displayed by orange line. Cyan line points to the level of starlight. The sum of the components, and thus the best fit, is shown by the red line.}
    \label{fig:SUM_J1521}
   \includegraphics[width=0.85\textwidth]{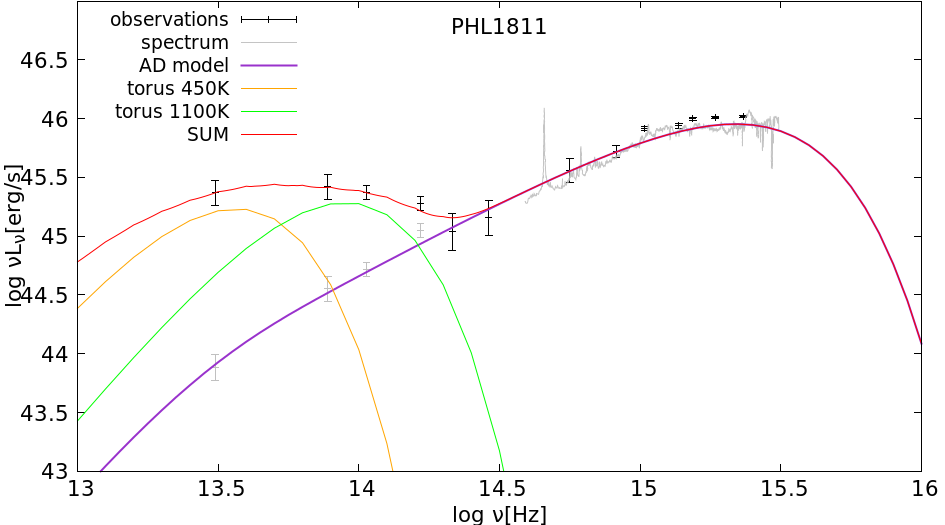}
    \caption[The best fit and components of SED fit for WLQ PHL1811]{The best fit of WLQ PHL1811. Summary of all components. Black points with errors show photometry data and the grey line represents spectrum. The violet line shows the contribution of the AD model. Tori with temperatures 450K and 1100K, respectively, are displayed by orange and green lines. Cyan line points to the level of starlight. The sum of the components, and thus the best fit, is shown by the red line.}
    \label{fig:SUM_PHL1811}
\end{figure}

\begin{figure}[!htb]
\centering
    \includegraphics[width=0.85\textwidth]{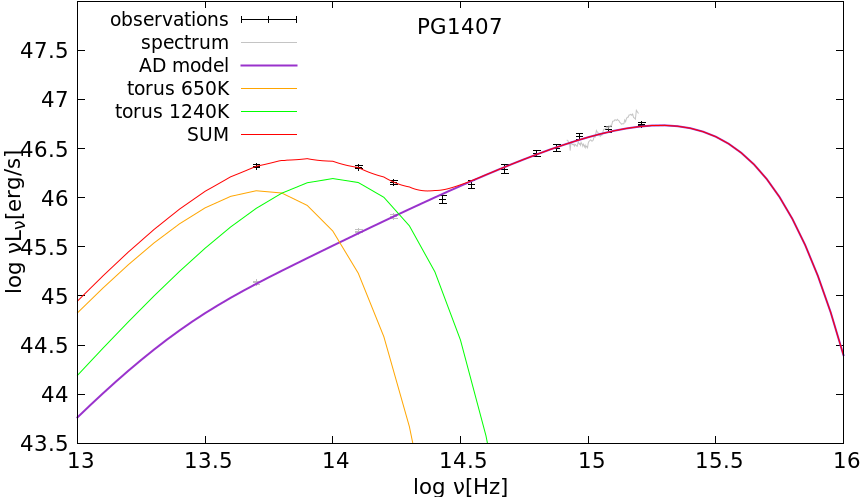}
    \caption[The best fit and components of SED fit for WLQ PG1407]{The best fit of WLQ PG1407. Summary of all components. Black points with errors show photometry data and the grey line represents spectrum. The violet line shows the contribution of the AD model. Tori with temperatures 650K and 1240K, respectively, are displayed by orange and green lines. Cyan line points to the level of starlight. The sum of the components, and thus the best fit, is shown by the red line.}
    \label{fig:SUM_PG1407}
\end{figure}

Next, I switch to the main project task -- determining the global parameters of 10 WLQs. The sample of Weak emission-line Quasars contains 10 objects. Fig. \ref{fig:SUM_J0836} and \ref{fig:SUM_J0945} - \ref{fig:SUM_PG1407} show in detail how the fitting procedure works. The different lines show the individual components: accretion disk, tori, starlight level (the solid red line shows the sum of those components). Fig. \ref{fig:SED_1}  presents the best fits of disk continua that match the quasars SED. On the x-axis of the figures is the logarithmic value of frequency in Hertz, while on the y-axis is the logarithmic value of $\nu L_{\nu}$ in \ergs. The accretion disk continuum is marked with a solid purple line, the photometric data are shown by black crosses and blue points with errors. The blue points observed in UV bandpass of 5 WLQs (J0836, J1141, J1411, J1417, and J1447 (Fig. \ref{fig:SED_1}) and in optical range (J0836) could suggest an absorption seen in some quasars (e.g. KVRQ 1500-0031 \citealt{Heintz18}, SDSS J080248.18+551328.9 \citealt{Ji15,Liu15}).
The absorption is caused by intrinsic gas in the host galaxy and/or the greater influence of the assumed UV photoelectric absorption. The blue points are outliers and for this reason, I model the best fits in both cases: taking into consideration all points with and without outliers ($\chi^2$ values in parenthesis of Tab.~\ref{tab:bestfit}).

In the next step, I would like to compare the obtained SMBH and \dotm\ with values from the literature.
Satisfactory fits are defined as those showing reduced $\chi^2 \lesssim 5.5$. The values of BH masses and the accretion rates (\mbhlit\ and \dotmlit, respectively) of 9 WLQs were collected by different authors (see Tab. \ref{tab:bestfit}, Col.~9 for references). 





Those values for PG 1407 were determined below because there is a lack of them. I estimate its SMBH mass based on the equation (7.27) from \citet{Netzer13}, which states:

\begin{equation}\label{eq:hb_eq}
    \mmbh (\mHbeta) = 1.05 \times 10^8 \left[\frac{L_{5100}}{10^{46}\mergs}\right]^{0.65} \times \left[\frac{\mFWHM(\mHbeta)}{10^3\mkms}\right]^2 \mMsun,
\end{equation}
The spectrum and the level of continuum at 5100 \AA\ is given by \cite{Mcdowell95}. This level is $\nu L_{\nu} = 3.16 \times 10^{46}$ \ergs. Unfortunately, the \Hb\ emission-line is almost undetectably weak \citep{Mcdowell95}. For this reason, I estimate the \FWHM\ of the \MgII\ line as follows. Using the \FeII\ template taken from \citet{Vestergaard01}, I subtract contribution of the \FeII\ pseudo-continuum from the \MgII\ presented in the spectrum and fit a Gauss function to the emission line. The \FWHM(\MgII) calculated in this way is equal to $4300^{+1400}_{-530}$ \kms. Next, I have converted the width of the magnesium to appropriate the hydrogen based on the equation (6) by \citet{Wang09}:

\begin{equation}\label{eq:wang_mg}
   \log \left[\frac{\mFWHM(\mMgII)}{1000\mkms}\right] = (0.81 \pm 0.02) \times  \log \left[\frac{\mFWHM(\mHbeta)}{1000\mkms}\right] + (0.05 \pm 0.01),
\end{equation}
\\
Thus, \FWHM(\Hb)$= 5400^{+2240}_{-810}$ \kms. 
Finally, the calculated BH mass is $M_{\mathrm{BH}}^{\mHbeta} = (2.62^{+2.61}_{-0.73}) \times 10^9$~\Msun. I take this value as \mbhlit\ (Tab.~\ref{tab:bestfit}). Additionally, I have calculated BH mass based on the \MgII\ line following the similar procedure. I have used equation (7.28) from \citet{Netzer13}:

\begin{equation}\label{eq:mg_eq}
    \mmbh (\mMgII) = 8.9 \times 10^7 \left[\frac{L_{3000}}{10^{46}\mergs}\right]^{0.58} \times \left[\frac{\mFWHM(\mMgII)}{10^3 \mkms}\right]^2 \mMsun,
\end{equation}

As a result of Eq. (\ref{eq:mg_eq}) I have obtained $M_{\mathrm{BH}}^{\mathrm{Mg(II)}} = (3.69^{+2.79}_{-0.85})\times\ 10^9$ \Msun.
The accretion rate in PG 1407 is calculated using the relationship \citep[see equation 2 in][]{Plotkin15}:
\begin{equation}\label{eq:plotkin15}
      \frac{L_{bol}}{L_{Edd}} = 0.13 f(L) \times \left[\frac{L_{5100}}{10^{44}\mergs}\right]^{0.5}
\times \left[\frac{\mFWHM(\mHbeta)}{10^3\mkms}\right]^{-2},
\end{equation}
where $f(L)$ is the luminosity-dependent bolometric correction; it equals to 5.7 \citep{Shemmer10}. Eventually, $\mdotmlit = 0.45$ based on calculated \FWHM(\Hb) and luminosity at 5100 \AA\ in PG~1407. 
The reason for calculating the \mbh\ based on \Hb\ in all WLQs is due to the fact that, in the WLQ spectrum, neither \MgII\ nor \CIV\ lines are well built. Still, weaker \Hb\ than in normal quasars is a better choice. Due to the weak emission-lines, the measurement of the proper \FWHM value of the line might be biased. It seems that the \FWHM value of \Hb\ in the WLQ is biased but with lower contamination; however, \MgII\ and especially \CIV\ are very different compared to the normal, well-build lines in the quasar. The continuum fit method is an unbiased way to describe the global parameters. 

\begin{sidewaystable}
\caption{The best fit parameters
  \label{tab:bestfit}}
\vspace{2ex}
\noindent
\begin{center}
\begin{tabular}{|lccccl|ccc|}
\hline
\tablehead{
\colhead{Name} & \colhead{\mbh} & \colhead{\dotm} & \colhead{\as} & \colhead{\incl} & \colhead{$\chi^2$/d.o.f} &
\colhead{\mbhlit} & \colhead{$\dot{m}_{\rm lit.}$} & \colhead{ref.} }

{Name} & {\mbh} & {\dotm} & {\as} & {\incl} & {$\chi^2$/d.o.f} &
{\mbhlit} & {$\dot{m}_{\rm lit.}$} & {ref.}\\

(1) & (2) & (3) & (4) & (5) & (6) & (7) & (8) & (9)\\
\hline
J0836 & $(1.30 ^{+5.00} _{-0.30}) \times 10^{9}$ & ${0.38 ^{+0.28}_{-0.25}}$ & ${0.00 ^{+0.30}_{-0.00}}$ & ${0.26 ^{+0.52}_{-0.26}}$ & 2.48 (1.91) & $(3.89^{+13.11}_{-1.75})\times 10^8$ & ${0.87 ^{+1.36}_{-0.65}}$ & 1/1 \\
J0945 & $(2.00 ^{+4.30} _{-0.50})\times 10^9$ & $0.47^{+0.19}_{-0.16}$ & ${0.00 ^{+0.30}_{-0.00}}$ & $0.52 \pm 0.26 $ & 3.73 & $(1.12 ^{+0.43}_{-0.19})\times 10^9$ & $0.51 \pm 0.15$ & 1/1 \\ 
J0945 &  &  &  &  & & $(3.08 \pm 0.27) \times 10^9 \dagger$ &  & 2 \\ 
J1141 & $(6.30 ^{+13.70} _{-1.30}) \times 10^9$ & $0.56^{+0.28}_{-0.36}$ & ${0.10 ^{+0.40}_{-0.10}}$ & $0.26 \pm 0.26$ & 4.58 (2.67) & $(3.16^{+1.41}_{-0.97}) \times 10^9$ & $0.38^{+0.20}_{-0.10}$ * & 3/3 \\  
J1237 & $(5.00 ^{+6.00} _{-1.80}) \times 10^9$ & $0.30^{+0.29}_{-0.21}$ & ${0.10}^{+0.60}_{-0.10}$ & ${0.26 ^{+0.52}_{-0.26}}$ & 6.64 (3.93) & $(2.00^{+1.39}_{-0.65})\times 10^9$ & $0.40^{+0.21}_{-0.16}$ * & 3/3 \\
J1411 & $(2.50 ^{+0.70} _{-1.20})\times 10^9$ & $0.31^{+0.44}_{-0.08}$ & ${0.80 ^{+0.10}_{-0.40}}$ & $0.26 ^{ +0.52}_{-0.26}$ & 5.58 (1.25) & $(5.25^{+3.87}_{-2.23})\times 10^8$ & $0.34^{+0.42}_{-0.17}$ & 1/1 \\
J1417  & $(3.20 ^{+0.70} _{-1.90})\times 10^9$ & $0.54^{+0.19}_{-0.41}$ & $0.20^{+0.30}_{-0.10}$ & $0.00 ^{ +0.52}_{-0.00}$ & 2.88 & $(3.55^{+3.53}_{-2.15})\times 10^8$ & $0.92 \pm 0.50$ & 1/1 \\
J1447 & $(1.30 ^{+1.20}_{-6.30}) \times 10^9$ & $0.35^{+0.41}_{-0.07}$ & $0.80 \pm 0.10 $ & $0.26 \pm 0.26$ & 5.37 (4.41) & $(1.12^{+2.51}_{-1.25})\times 10^8$ & $1.30 ^{+0.17}_{-0.78}$ & 1/1 \\
J1521 & $(2.00^{+0.50}_{-5.70})\times 10^{10}$ & $0.48^{+0.19}_{-0.35}$ & ${0.80 ^{+0.10}_{-0.20}}$ & $0.52^{+0.26}_{-0.52}$ & 3.68 & $(6.20^{+1.73}_{-1.51}) \times 10^9$ & $0.81^{+0.27}_{-0.17}$ * & 4/4 \\
PHL 1811 & $(7.90^{+2.10}_{-3.90})\times 10^8$ & $0.34^{+0.50}_{-0.03}$ & $0.00^{+0.10}_{-0.00}$ & $ 0.00^{+0.26}_{-0.00}$  & 1.87 & $(1.14^{+2.57}_{-2.57})\times 10^8$ & $1.30^{+0.03}_{-0.02}$ * & 5/5 \\
PG 1407 & $(7.90 ^{+5.10}_{-2.90})\times 10^9$ & $0.26^{+0.28}_{-0.09}$ & $0.90^{+0.00}_{-0.10}$ & $0.78 \pm 0.26 $ & 1.42 & $(2.62^{+2.61}_{-0.73})\times 10^9$ & $0.45^{+0.17}_{-0.23}$ * & a/a \\
PG 1407 &  &  &  &  & & $(3.69^{+2.79}_{-0.85})\times 10^9 \dagger$ &  & a \\ 
\hline
\end{tabular}
\begin{quote}
Black hole masses, Eddington accretion rates, spins, and fitted inclinations are in Col. (2)-(5), respectively. Col. (6) contains the normalized $\chi^2$ values in two cases: numbers without parenthesis -- all photometric points are taken into account, numbers in parenthesis -- data with removed outliers (only black points in Fig.~\ref{fig:SED_1}, \ref{fig:SED_2}, \ref{fig:SED_3}) are considered. The values in Col. (2)-(5) refer to the case where all points are fitted. The values of the parameters in the absence of outliers are the same as before within the errors. Black hole masses and accretion rates taken form literature are in Col.(7)-(8). They are based on \FWHM(\Hb) measurements (values without $\dagger$). $\dagger$ -- \mbh\ is based on \MgII\ line. \mbh\ and \mbh$_{\rm ,lit.}$ are in units of $M_{\odot}$. * -- errors of $\dot{m}_{\rm lit.}$ are estimated by us. Numbers refer to articles: 1) \cite{Plotkin15}, 2) \cite{Hryniewicz10}, 3) \cite{Shemmer10}, 4) \cite{Wu11}, 5) \cite{Leighly07b}, 6) \cite{Mcdowell95}, a) this work.
\end{quote}
\end{center}
\end{sidewaystable}

The results are collected in Tab. \ref{tab:bestfit}. The black hole masses, 
accretion rates, spins, inclinations of each WLQs, and $\chi^2$ values are in Columns (2)-(6). Degeneration of solutions due to the spin parameter takes place. Two groups of the best fit for zero and non-zero spin are difficult to distinct \citep{Sun_Malkan89}. Thus, I also perform an additional fit with the fixed \as\ equals 0  (Tab.~\ref{tab:Schwarzschild}).  In this approach, the photometric points without outliers are taken into account (only the black points in Fig. \ref{fig:SED_1}).

\begin{table}
    \centering
    \begin{threeparttable}
    \caption [The Schwarzschild black hole solutions]{The Schwarzschild black hole solutions (\as=0)}
    \begin{tabular}{ccccc}
    \hline   
    Name & \mbh& \dotm & \incl & $\chi^2$/d.o.f\\
    \hline
    J0836 & $(1.30 ^{+5.00} _{-0.30}) \times 10^{9}$ & ${0.38 ^{+0.28}_{-0.25}}$ & ${0.26 ^{+0.52}_{-0.26}}$ & (1.91)\\
    J0945 & $(2.00 ^{+4.30} _{-0.50})\times 10^9$ & $0.47^{+0.19}_{-0.16}$  & $0.52 \pm0.26$ & (3.73)\\
    J1141 & $(7.90 ^{+2.10} _{-2.90}) \times 10^{9}$ & ${0.84 ^{+0.03}_{-0.28}}$ & ${1.30 ^{+0.00}_{-0.52}}$ & (4.11)\\
    J1237 & $(4.00 ^{+2.30} _{-0.80}) \times 10^{9}$ & ${0.30 ^{+0.06}_{-0.29}}$ & ${0.00 ^{+0.26}_{-0.00}}$ & (4.54)\\
    J1411 & $(7.90 ^{+3.40} _{-1.60}) \times 10^{9}$ & ${0.50 ^{+0.03}_{-0.11}}$ & ${0.00 ^{+0.26}_{-0.00}}$ & (1.43)\\
    J1417 & $(2.50 ^{+0.70} _{-0.50}) \times 10^{9}$ & ${0.66 ^{+0.19}_{-0.22}}$ & ${1.30 ^{+0.00}_{-0.26}}$ & (1.98)\\
    J1447 & $(5.00 ^{+2.90} _{-1.00}) \times 10^{9}$ & ${0.49 ^{+0.20}_{-0.05}}$ & ${0.78 ^{+0.52}_{-0.26}}$ & (4.96)\\
    J1521 & $(1.30 ^{+0.70} _{-3.40}) \times 10^{10}$ & ${0.86 ^{+0.11}_{-0.63}}$ & ${1.30 ^{+0.00}_{-0.52}}$ & (3.99)\\
    PHL 1811 & $(7.90^{+2.10}_{-3.90})\times 10^8$ & $0.34^{+0.50}_{-0.03}$ &
    $ 0.00^{+0.26}_{-0.00}$  & (1.87)\\
    PG 1407 & $(2.50 ^{+0.70} _{-0.50}) \times 10^{9}$ & ${0.85 ^{+0.06}_{-0.19}}$ & $0.78\pm 0.26$  & (0.93)\\
    \hline
    \label{tab:Schwarzschild}
    \end{tabular}
    \end{threeparttable}
    \begin{minipage}{\linewidth}\small
    \textit{Note.} -- The normalized $\chi^2$ values in parenthesis means the data with removed outliers
    \end{minipage}
    \end{table}


    

\newpage

\newpage

\newpage

\begin{sidewaysfigure}
    \centering
    \subfloat[]{\includegraphics[width=0.45\textwidth]{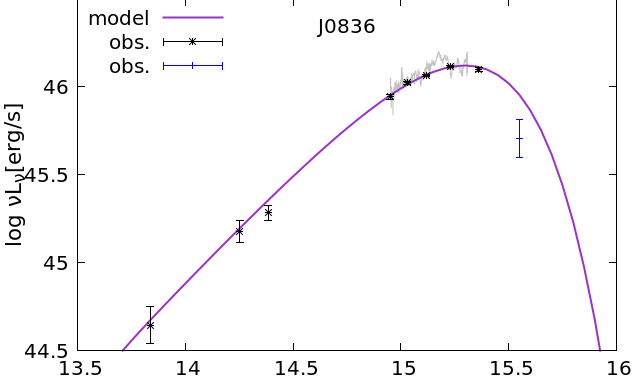}}
    \qquad
    \subfloat[]{\includegraphics[width=0.45\textwidth]{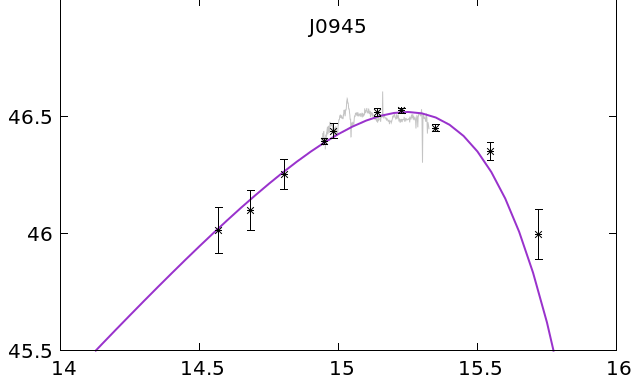}}
    \qquad
    \subfloat[]{\includegraphics[width=0.45\textwidth]{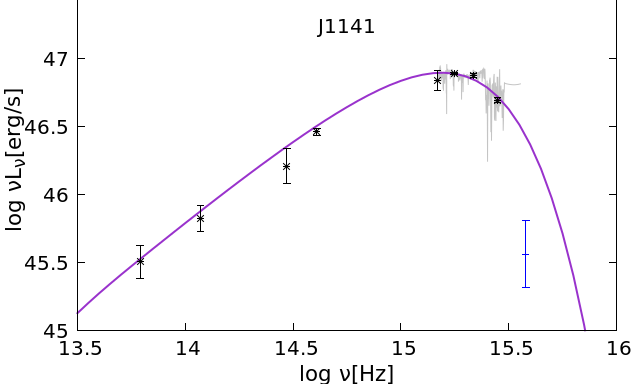}}
    \qquad
    \subfloat[]{\includegraphics[width=0.45\textwidth]{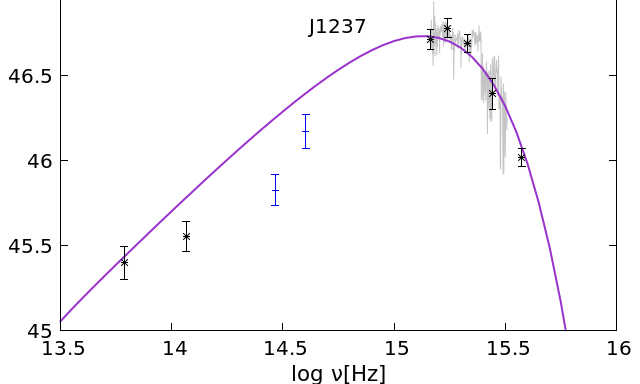}}
    \caption[{The best fit of SED to photometric points of 10 WLQs, part 1}]{The best fit of SED to photometric points of 10 WLQs. Black crosses and blue points with errors show corrected observational data, a grey line -- spectra (optic/UV). The solid violet line represents the theoretical curve of the AD continuum model.}%
    \label{fig:SED_1}
\end{sidewaysfigure}

\begin{sidewaysfigure}
\centering
\ContinuedFloat 
    \subfloat[]{\includegraphics[width=0.45\textwidth]{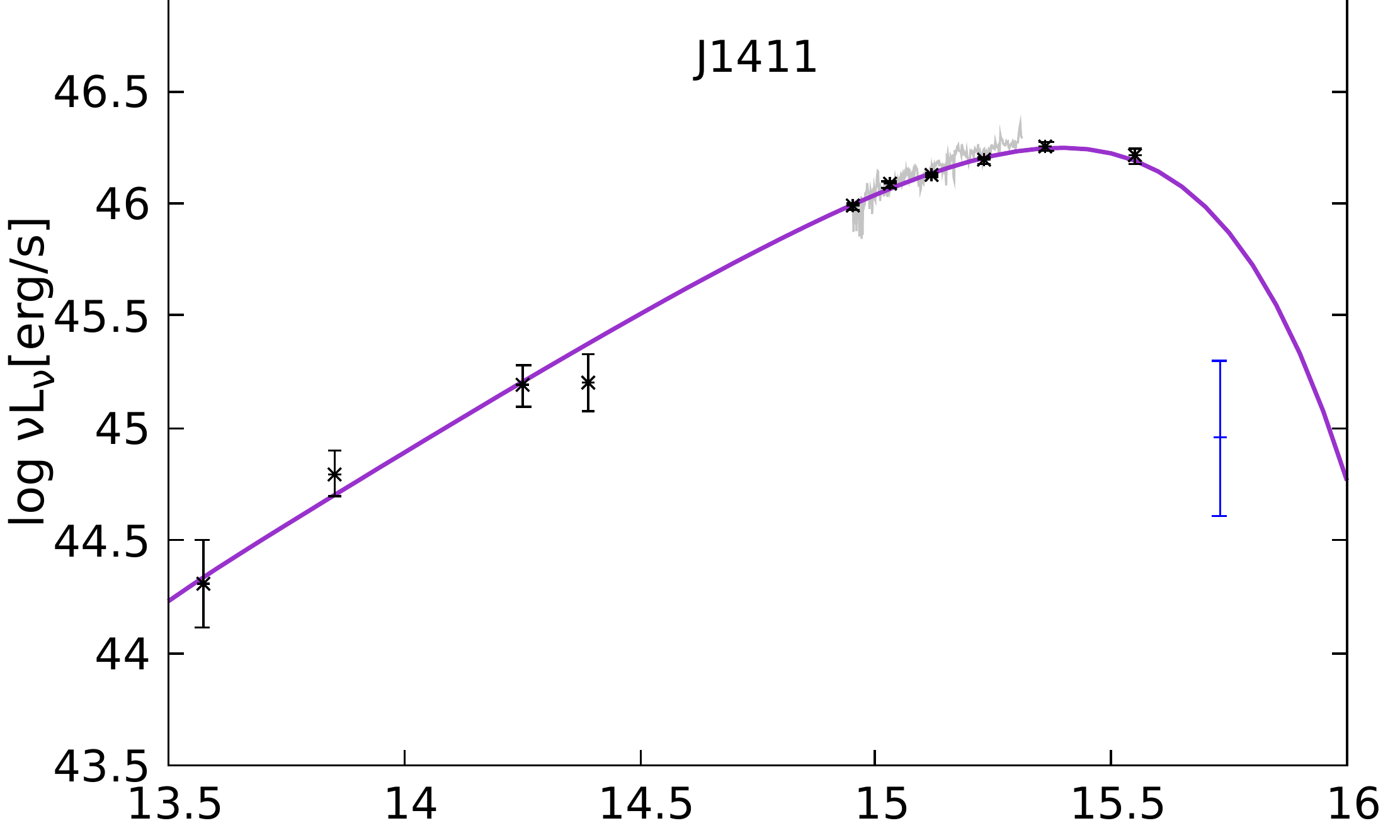}}
    \qquad
    \subfloat[]{\includegraphics[width=0.45\textwidth]{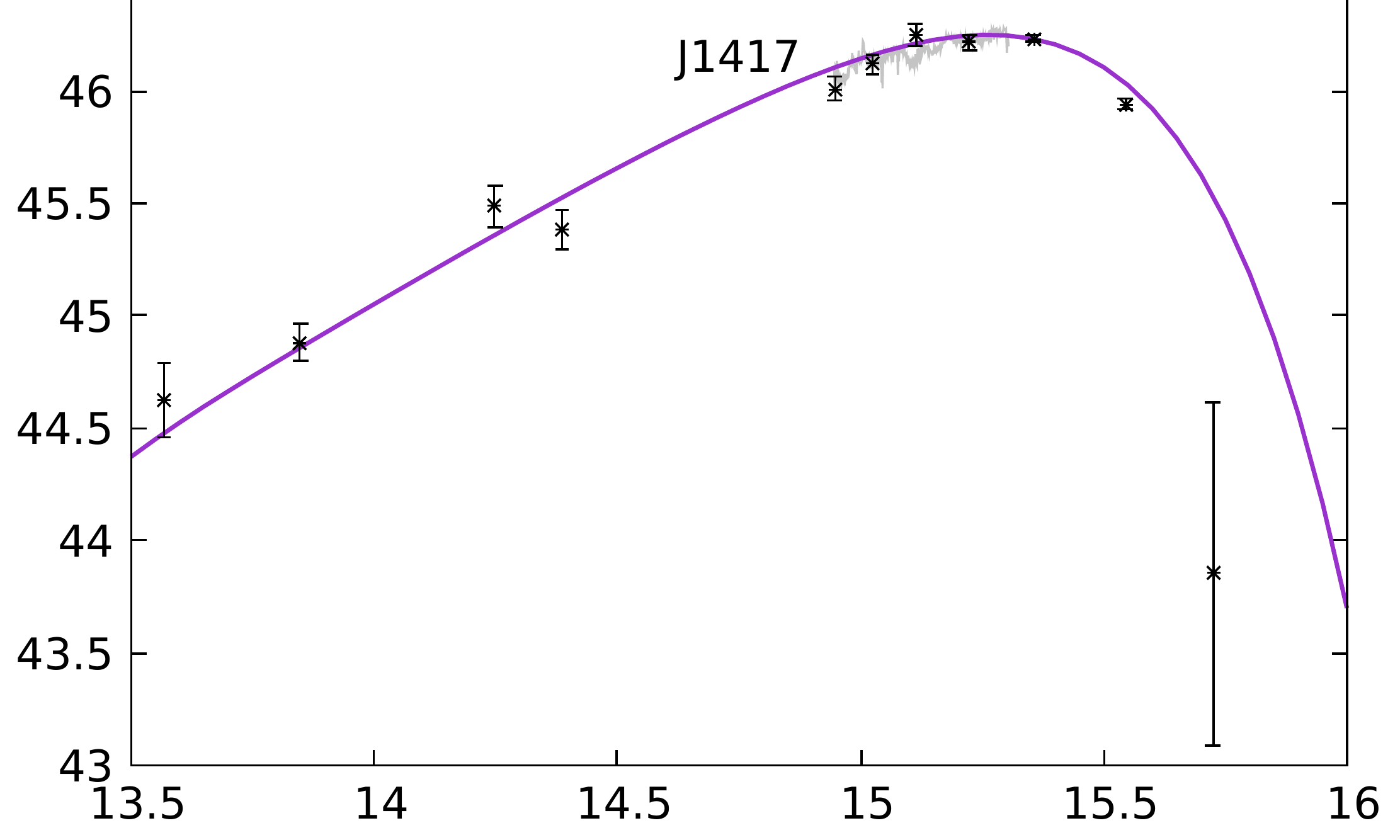}}
    \qquad
    \subfloat[]{\includegraphics[width=0.45\textwidth]{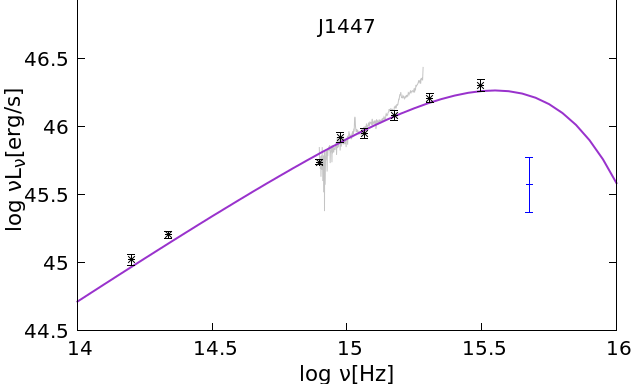}}
    \qquad
    \subfloat[]{\includegraphics[width=0.45\textwidth]{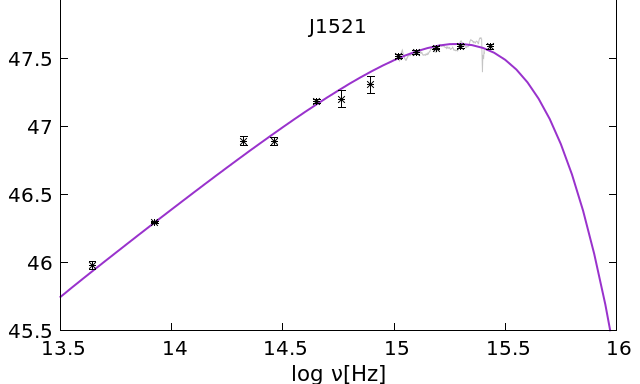}}
    \caption[{The best fit of SED to photometric points of 10 WLQs, part 2}]{The best fit of SED to photometric points of 10 WLQs. Black crosses and blue points with errors show corrected observational data, grey line -- spectra (optic/UV). The solid violet line represents the theoretical curve of the AD continuum model.}
  \label{fig:SED_2}
\end{sidewaysfigure}

\begin{sidewaysfigure}
\centering
\ContinuedFloat 
        \subfloat[]{\includegraphics[width=0.45\textwidth]{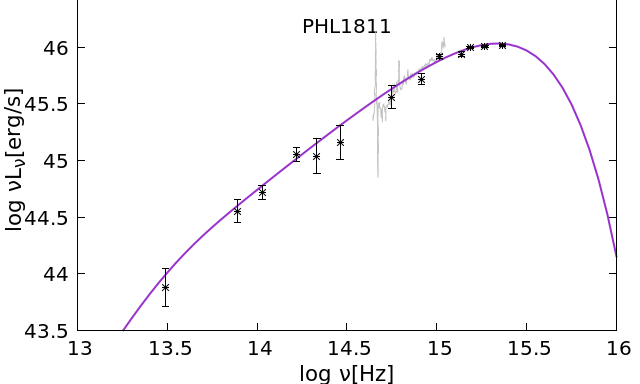}}
        \subfloat[]{\includegraphics[width=0.45\textwidth]{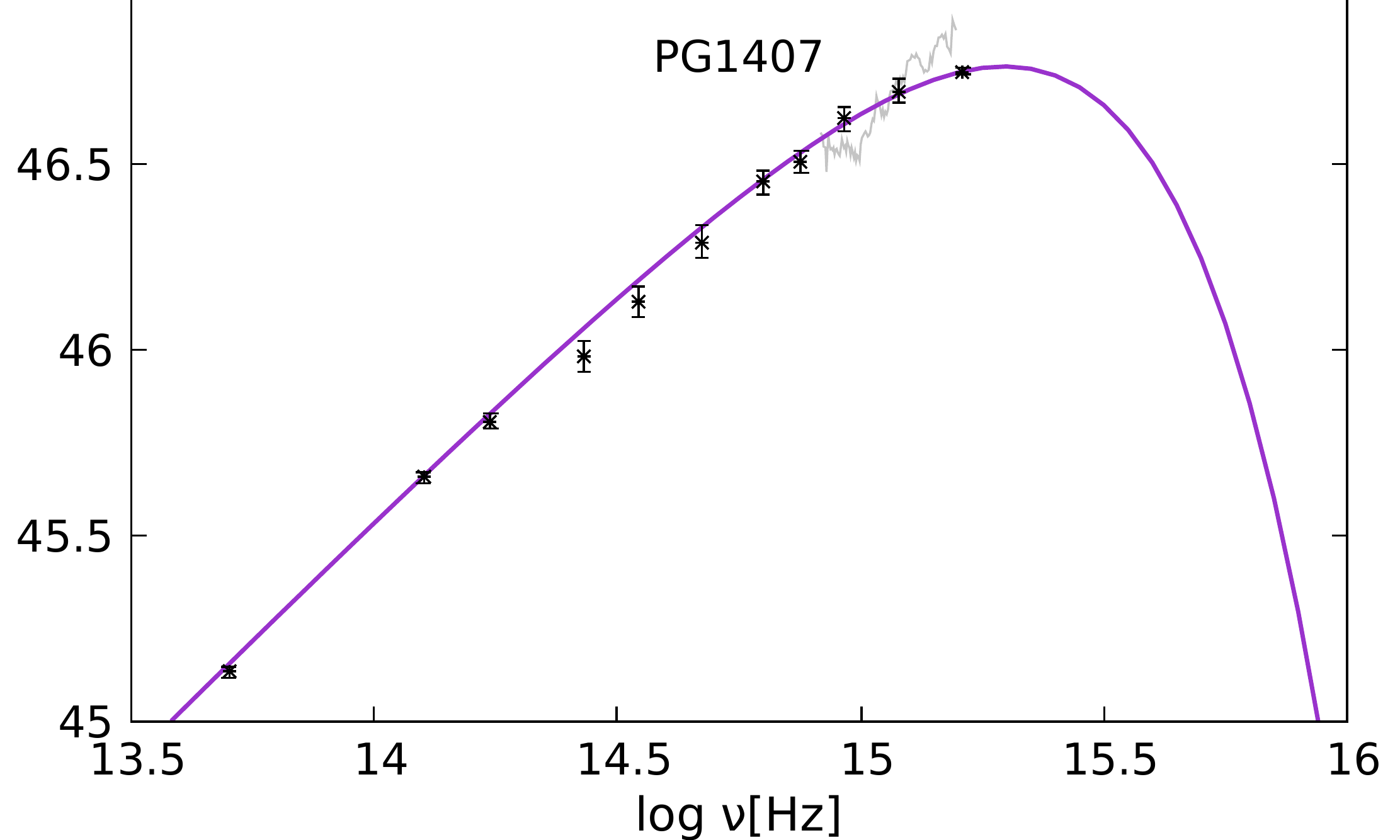}}
    \caption[{The best fit of SED to photometric points of 10 WLQs, part 3}]{The best fit of SED to photometric points of 10 WLQs. Black
      crosses and blue points with errors show corrected observational data, grey line 
      -- spectra (optic/UV). The solid violet line represents the theoretical curve of the AD continuum model.}
  \label{fig:SED_3}
\end{sidewaysfigure}

\clearpage
\section{Analysis of results}
\subsection{Bayes' theorem}
I determine the best fits based on the $\chi^2$ procedure. Additionally, I would like to estimate the parameters in more sophisticated way using Bayes' theorem. The posterior probability, $P(H|D,I)$ is calculated using Eq. (\ref{bayes_chapter2}). The hypothesis is a model $m=m$(\mbh, $\mdotm$, $a$, $i$). As mentioned earlier, the model is affected by the data –- the SED and spectra of quasars. For each model $m$, I derive its likelihood  $P(D|m,I) \propto exp(-\chi^2/2)$. The information, $I$, should be $M^{obs}_{BH}$, $\mdotm_{obs}$, $a_{obs}$, $i_{obs}$ which are observed. Nevertheless, I do not have those real parameters and therefore, $I$, is based on earlier estimated $M^{lit}_{BH}$, $\mdotm_{lit}$, $a_{lit}$, $i_{lit}$, and their errors (i. e. $\mdotm_{obs} = \mdotm_{lit}$, etc.). 


Assuming a Gaussian probability distribution for
\mbhlit\ with standard deviations equal to $\sigma_M$, the prior can
be written as $P(H|\mmbhlit) \propto \exp{(-(\mmbh - \mmbhlit)^2/2\sigma_M^2)}$. The prior probability related to \dotm\ takes a similar Gaussian form. We do not have prior knowledge of either BH spin or inclination. I have assumed a delta function probability distribution for both parameters.

Finally, the posterior probability is determined for each of the 366,000 models, as
the product of the likelihood ($\propto exp(-\chi^2/2))$ and the priors on \mbh\ and
\dotm\ \citep[for details see][Appendix A]{Capellupo15}. This is given
by:
\begin{equation}
  P\left( H|D,I \right)=N \times \exp{\left( \frac{-\chi^2}{2}\right)} \times
   \exp{\left( -\frac{(\mmbh-\mmbhlit)^2}{2\sigma_M^2} \right)} \times \exp{\left( -{\frac{(\mdotm-\mdotmlit)^2}{2\sigma^2_{\dot{m}}}} \right)}, 
   \end{equation}
where $N$ is the normalization constant.

The Bayesian analysis is helpful for saying which model has the highest probability of explaining the observed SED \citep{Capellupo15}.

For Bayesian analysis, it is needed to determine the errors of the accretion rates of those WLQs, which the literature does not provide. I have used the mentioned Eq. \ref{eq:plotkin15} \citep[see][]{Plotkin15}, the upper and lower limits of \FWHM(\Hb) and $L_{5100{\angstrom}}$.
Errors of \dotmlit\ are listed in Tab. \ref{tab:bestfit}.

The results for the model with the highest posterior probability are shown in Tab. \ref{tab:Bayes}. I have found that the values of BH masses with a satisfactory fit are the same using the $\chi^2$ or Bayes' theorem, within their errors. No significant differences in black hole masses calculated from both the $\chi^2$ and the Bayesian analysis are indicated. This suggests that the determined global parameters are correct and describe the overall SED shape of these objects. I take the \mbh\ calculated from the $\chi^2$ from Tab.~\ref{tab:bestfit} for further analysis.



\begin{table}
    \centering
    \begin{threeparttable}
    \captionsetup{width=1.0\textwidth}
    \caption [The BH masses from the Bayesian analysis]{The BH masses from the Bayesian analysis (\as $\neq$0)}
    \begin{tabular}{cccccc}
    \hline   
    Name & \mbhbay\ [\Msun] & $\mdotm_{Bay}$ \\
    \hline
    J0836 & $(2.00^{+4.30}_{-1.00})\times 10^9$ & $(0.32^{+0.31}_{-0.19})$  \\
    J0945 & $(3.20^{+3.10}_{-1.20})\times 10^9$ & $(0.25^{+0.22}_{-0.18})$ \\
    J1141 & $(7.90^{+4.10}_{-2.90})\times 10^9$ & $(0.39^{+0.45}_{-0.19})$ \\
    J1237 & $(6.30^{+3.70}_{-1.30})\times 10^9$ & $(0.27^{+0.33}_{-0.26})$\\
    J1411 & $(1.30^{+1.90}_{-1.30})\times 10^9$ & $(0.48^{+0.51}_{-0.28})$ \\
    J1417 & $(2.50^{+2.50}_{-1.20})\times 10^9$ & $(0.32^{+0.41}_{-0.31})$ \\
    J1447 & $(1.00^{+1.50}_{-6.00})\times 10^9$ & $(0.37^{+0.50}_{-0.23})$ \\
    J1521 & $(1.30^{+1.20}_{-5.00}) \times 10^{10}$ & $(1.27^{+0.22}_{-0.82})$ \\
    PHL 1811 & $(7.90^{+3.10}_{-3.90}) \times 10^8$ & $(0.40^{+0.14}_{-0.05})$ \\
    PG 1407 & $(7.90^{+3.40}_{-2.90}) \times 10^9$ & $(0.29^{+0.25}_{-0.13})$ \\
    \hline
    \label{tab:Bayes}
    \end{tabular}
    \end{threeparttable}
    \begin{minipage}{\linewidth}
    \textit{Note.} -- I have calculated spins and inclination but the errors were so large ($\geq$ 60 \%) that they were within the entire range of the mesh
    \end{minipage}
\end{table}


\subsection{The \mbh\ of the thesis vs literature - \mbh\ comparison}

\begin{figure}
    \centering
    \plotone{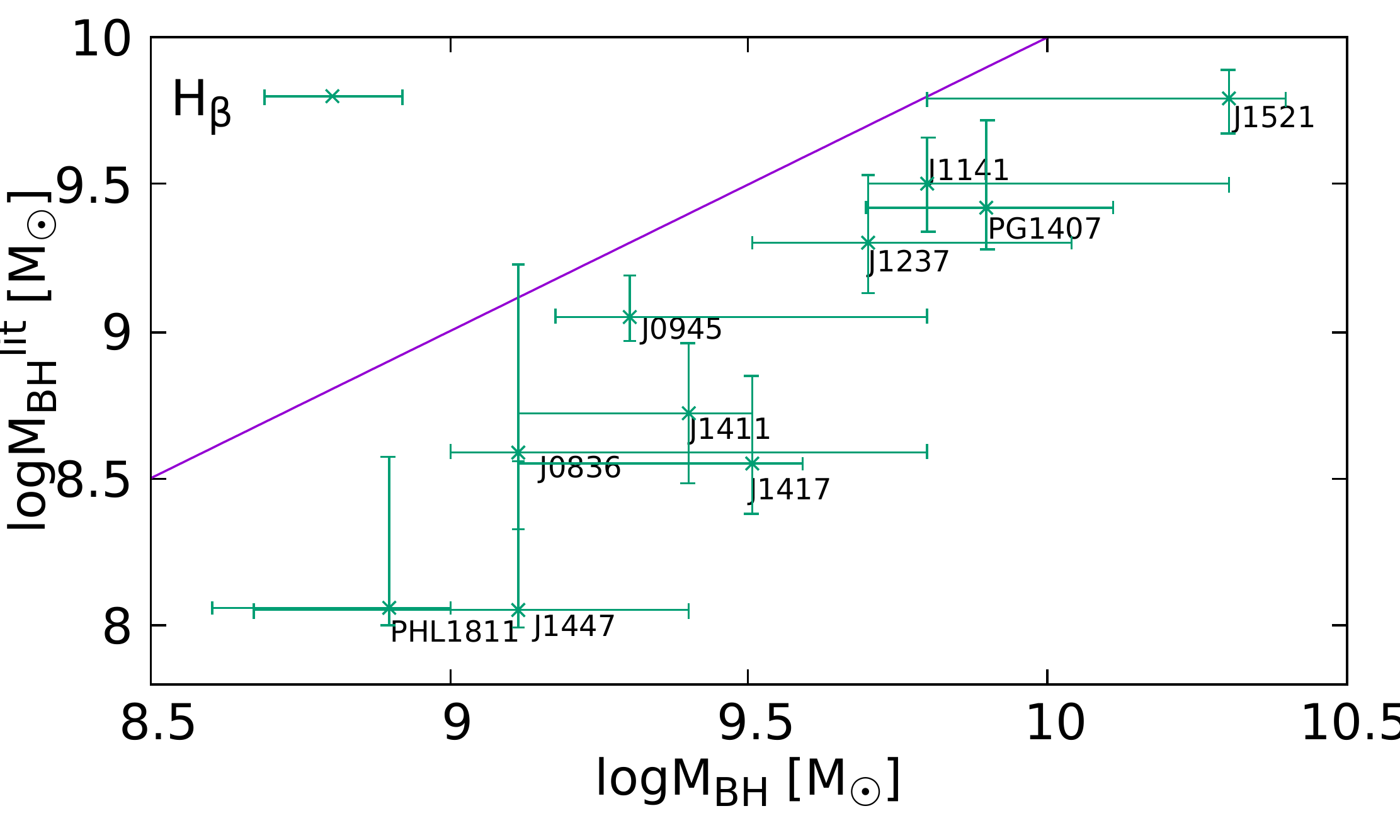}
    \caption[{Comparison of SMBH masses of WLQs.}]{Comparison of SMBH masses (\mbhlit) of WLQs based on
      \FWHM(\Hb) estimations with \mbh, which come from spectral fitting
      method. The solid violet line is identity 1:1 line.} 

    \label{fig:Comparison}
\end{figure}

\begin{figure}
    \centering
    \plotone{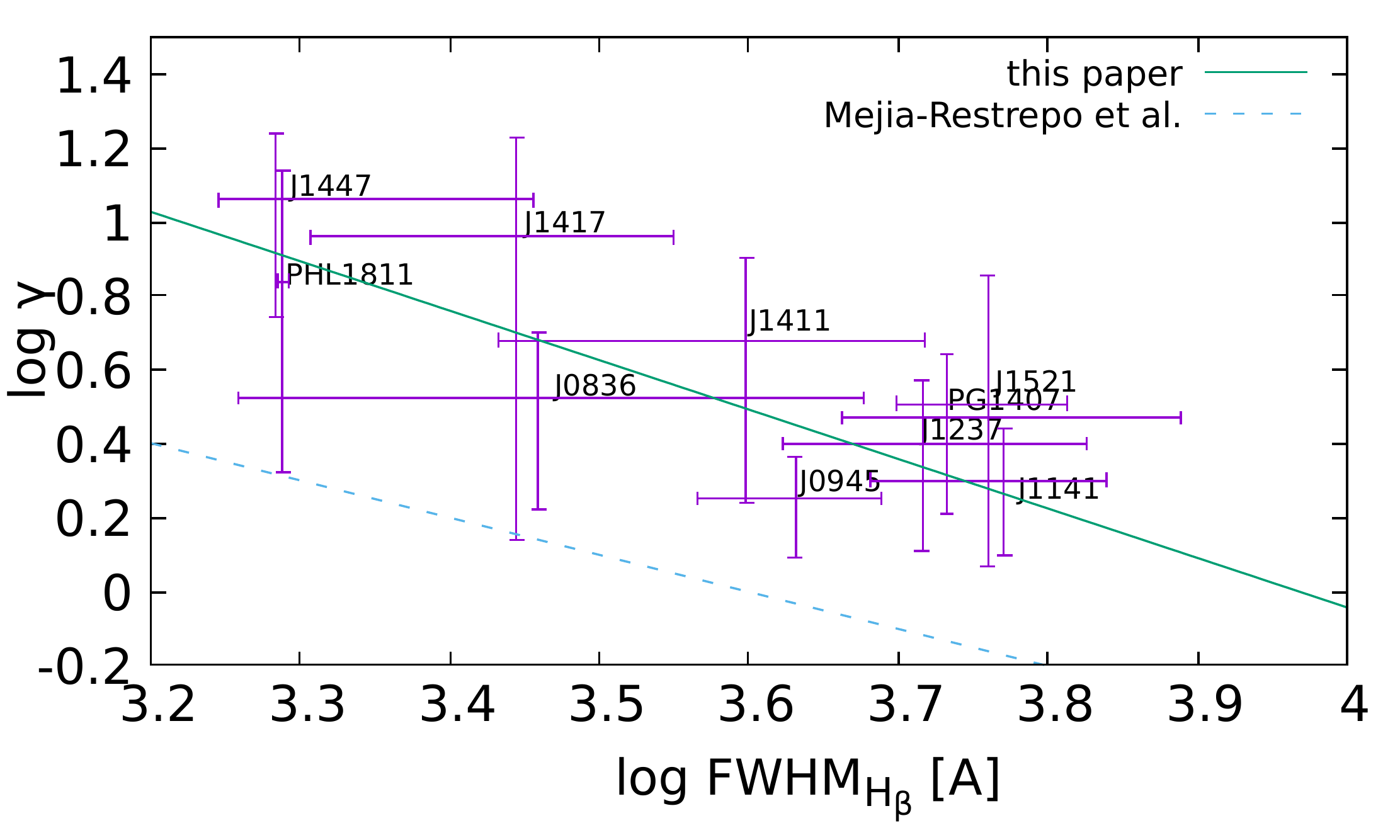}
    \caption[{The $\gamma$ factor ($= \mmbh/\mmbhlit$) versus
      \FWHM(\Hb).}]{The $\gamma$ factor ($= \mmbh/\mmbhlit$) versus
      \FWHM(\Hb). The best fit (i.e. $\gamma \propto \mFWHM^{-1.34}$)
      is shown by the solid green line. Points show data of 10 WLQs.  The
      blue dash line represents the fit of the virial factor (i.e. $f
      \propto \mFWHM^{-1.17}$) to 37 AGNs made by \citet{Mejia19}. In
      my case $\gamma = \mathrm{const} \times f$, where const.~= 0.75. The dash line is
      shifted down for better visibility.}
    \label{fig:factor+}
\end{figure}

In order to compare my \mbh\ and \dotm\ with those values taken from the literature, Fig.~\ref{fig:Comparison} shows the mass distribution of black holes. Identity 1:1 line is marked as a solid purple line. The presented mass comparison suggests that the literature determinations of black hole masses, \mbhlit, based on \FWHM(\Hb), are generally underestimated. I also determine the difference between \mbhlit\ and my \mbh\ values. The $\gamma$ factor is calculated ($\gamma := \mmbh/\mmbhlit$). In Fig.~\ref{fig:factor+}, I present the relationship between the logarithmic value of \FWHM(\Hb) in \kms\ (see Tab. \ref{tab:fwhm and gamma}) and the logarithmic value of the $\gamma$ factor. The green solid line shows the best fit between those variables. The fit is made using the nonlinear least-squares (NLLS) Marquardt-Levenberg algorithm, which takes into account errors in both the $x$ and $y$ directions. The relationship is:
\begin{equation}
            \log \gamma =\left(-1.338 \pm 0.366\right) \times 
            \log \left( \frac{\mFWHM (H_{\beta})}{10^3 \, \mkms} \right)+ 
        \left(1.294 \pm 0.234\right).
        \label{eq:gamma_eq}
\end{equation}

    \begin{table}
    \centering
    \begin{threeparttable}
    \captionsetup{width=1.0\textwidth}
    \caption {The Full Width at Half Maximum of \Hb\ emission-line and the $\gamma$ factor}
    \begin{tabular}{cccc}
    \hline   
    Name & \FWHM(\Hb) [\kms] & $\log \gamma$ & ref. \\
    \hline
    J0836 & $2880^{+1877}_{-1069}$ & $0.52^{+0.18}_{-0.31}$ & 1\\
    J0945 & $4278 \pm 598$ & $0.25^{+0.11}_{-0.16}$ & 1\\
    J1141 & $5900^{+1000}_{-1100}$ & $0.30^{+0.14}_{-0.20}$ & 2\\
    J1237 & $5200^{+1500}_{-1000}$ & $0.40^{+0.17}_{-0.29}$ & 2\\
    J1411 & $3966 \pm 1256$ & $0.68^{+0.22}_{-0.44}$ & 1\\
    J1417 & $2784 \pm 759$ &  $0.96^{+0.27}_{-0.82}$ & 1\\
    J1447 & $1923^{+933}_{-164}$ & $1.06^{+0.18}_{-0.32}$ & 1 \\
    J1521 & $5750 \pm 750$ * & $0.51^{+0.36}_{-0.44}$ & 3\\
    PHL 1811 & $1943\pm 19$ & $0.84^{+0.3}_{-0.52}$ & 4\\
    PG 1407 & $5400^{+2240}_{-810}$ $\sharp$ & $0.47^{+0.17}_{-0.26}$ & a\\
    \hline
    \label{tab:fwhm and gamma}
    \end{tabular}
    \end{threeparttable}
    \begin{minipage}{\linewidth}\small
    \textit{Note.} -- $\gamma$ = \mbh /\mbhlit. * -- errors are estimated by us. $\sharp$~\Hb\ is weak and almost not visible, its \FWHM\ is estimated based on \FWHM(\MgII). Numbers refer to \FWHM\ sources: 1) \cite{Plotkin15},  2) \cite{Shemmer10}, 3) \cite{Wu11}, 4) \cite{Leighly07b}, a) this work.
    \end{minipage}
    \end{table}

    
    

For a better assessment of my calculations, I have determined
the Spearman coefficient, which is $r_{s} = -0.806$ and the linear correlation coefficient, $r = -0.82$. It can be seen that three of the ten objects (J0945, J1141, and J1237) are close to the 1:1 line and the $\gamma$ factor is $\lesssim 2.5$. The masses of three other sources (PHL 1811, J1417, J1447)  should be multiplied by $\gamma > 7$. The mean $\gamma$ factor is 4.7, and the median is 3.3. The dashed blue line in Fig. \ref{fig:factor+} represents the best fit obtained by \citet{Mejia19}. They used a sample of 37 Type I AGNs, which lie in the range of redshifts $\sim1.5$. Eq. (\ref{eq:gamma_eq}) allows us to correct the black hole masses \mbhlit\ which has been determined so far and based on the \FWHM\ values of the \Hb\ line.

\begin{equation}
    \mmbh = \mmbhlit \times 10^{1.294} \times \left(\frac{\mFWHM(\mHbeta)}{10^3\mkms}\right)^{-1.1338}
\end{equation}



%
%

\chapter{Discussion}

\section{The virial factor and black hole masses in Weak emission-line Quasars}

The virial factor (\acrshort{f} in Eq.~\ref{eq:blr}) is often assumed to be
constant, with values of 0.6-1.8 \citep[e.g.][]{Peterson04RM, Onken04, Nikolajuk06}, where 3/4 corresponds to the spherical geometry of the BLR \citep{Netzer_AGNbook_Peterson_SAAS1990}.  Generally, $f$ dependents on non-virial velocity components such as winds, the relative thickness $(H/\mRblr)$ of the Keplerian BLR orbital plane, the line-of-sight inclination angle ($i$) of this plane, and the radiation pressure, and should be a function of those phenomenon \citep{Wills86, Gaskell09, Denney09, Denney10, Shen14, Runnoe14}. The analysis carried out by \citet{Mejia19} indicates the low influence of radiation pressure on the $f$ factor; however, this mechanism cannot be excluded. Whether I omit the influence of radiation pressure influence or not, the line-of-sight of a gas in a planar distribution of the BLR plays an important role in calculating black hole mass. The measured \FWHM\ of emission lines depends on the velocity of the radiated gas/source. Unfortunately, the nature of the velocity component responsible for the thickness of the BLR, and thus its geometry, is unclear (e.g. \citealt{Done96,Collin06,Czerny16}; for a recent review, see \citealt{Czerny19}).  

My results support those gotten by \citet{Mejia19}. Instead of quasars, they study 37 AGNs
at redshifts $\sim 1.5$. They based of different analysis that performed by me find similar behavior. They fitted the AD model to the spectra of AGNs. Their AD model was the standard, geometrically thin, optically thick with general relativistic and disc atmosphere corrections. They examined alternative scenarios, including radiation pressures and luminosity dependency \citep[for details see,][]{Mejia19}. Similar to my approach, \cite{Mejia19} method and modeling is independent of the BLR geometry and kinematics. The authors indicate the dependency of the virial factor on the observed \FWHM\ of the broad emission-line (such as $H\alpha$, $H\beta$, \MgII, \CIV) in the form of an anti-correlation,  

\begin{equation}
    \mmbh \propto f(\mFWHM) \times R_{BLR} \times \mFWHM^2,
\end{equation}
where $f(\mFWHM) \propto \mFWHM^{-a}$, $a$ is some positive index (for the \Hb\ line $a$ = 1.17).

This implies that the BH mass estimations based on the reverberation or the single-epoch virial BH mass
method are systematically overestimated for AGN systems with larger \FWHM\ (e.g. $\gtrsim 4000$
\kms\ for $H\alpha$) and underestimated for systems with small $\mFWHM(H\alpha)\lesssim 4000$\kms.  It is worth to note that the opposite rule applies to the Eddington accretion rates (because
$\mdotm \propto \mmbh^{-1}$). I~have found a similar underestimation of \mbhlit\ values (taken from the literature) in the sample of WLQs (see Fig.~\ref{fig:factor+}). The SMBH masses of WLQs which show \FWHM(\Hb)$\simeq 5000\ \mkms$ need to be multiplied by a small factor of 1.5, while the rest of them (i.e. \FWHM(\Hb) > 7000 \kms) requires a larger factor up to 12. This means that the masses of about 50-60\% of WLQs are underweight based on \FWHM(\Hb) values.
This fact is also seen in Fig.~\ref{fig:Comparison} where \textbf{correction of the SMBH masses based on \FWHM\ estimation is needed}. I modify the equation used to calculate \mbhlit\ (Equation 1 in \citealt{Plotkin15}):
\begin{equation}
    \frac{\mmbhlit}{10^6\mMsun} = 5.05 \left(\frac{\nu L_{\nu}(5100)}{10^{44} \mergs}\right)^{0.5}\left(\frac{\mFWHM(H_{\beta})}{10^3 \mkms}\right)^2
\end{equation}
using the definition of the $\gamma$ factor ($\mmbh=\gamma \times \mmbhlit$)
and Eq.~(\ref{eq:gamma_eq}).
The corrected formula for the SMBH masses in WLQs is:
\begin{equation}
    \frac{\mmbh}{10^7  \mMsun}=  (9.94^{+7.09}_{-4.13}) \times
    \left(\frac{\nu L_{\nu}(5100)}{10^{44} \mergs}\right)^{0.5} 
    \times \left(\frac{\mFWHM(H_{\beta})}{10^3 \mkms}\right)^{0.66\pm0.37}
\label{eq:mbhwlq}
\end{equation}
The weaker dependence on \FWHM(\Hb) (index 2 vs 0.66) can be realized when the BLR is elongated and parallel to the accretion disk. There is accumulated evidence in the literature favoring a disk-like geometry for the BLR \citep{Wills86, Laor06, Decarli08, Pancoast14, Shen14, Mejia18N,Wang19}. On the other hand, the BLR may be also dominated by outflows, which are perpendicular to the line-of-sight. This scenario will favor the idea of quasar reactivation. The outflow could rebuild the \Hb\ region \citep{Hryniewicz10, Li2020, Andika2020}. The systematic underestimation of \FWHM\ (and \mbhlit) may also be caused by the strong influence of the \FeII\ pseudo-continuum in optics. In this case observed \FWHM\ of line is smaller. Such phenomena are noticed by \citet{Plotkin15} for their sample of WLQs, which have larger $R_{\rm opt,FeII}$ and narrower \Hb\ than most reverberation mapped quasars. 

\cite{Mejia19} dependence of \mbh\ on the observed \FWHM\ of the Balmer lines for AGNs is close to linear (\mbh $\propto$ \FWHM(\Hb)$^{0.82 \pm 0.11}$, when $f$ is a function of the Full Width) rather than quadratic (\mbh $\propto$ \FWHM$^2$ when $f=$ const). In my case, this relationship for WLQs is a bit weaker (\mbh $\propto$ \FWHM(\Hb)$^{0.66 \pm 0.37}$, when $f$ depends on \FWHM), but still compatible with the \citeauthor{Mejia19} result within the 1$\sigma$ error. The similar or even same behavior of normal AGNs and WLQs suggests that \textbf{both kinds of sources have the same dim nature of the velocity component and similar geometry of the BLR.}
 
The relationship between the widths of \MgII\ and  \Hb\ emission-lines can be noticed in the AGNs \citep[see][Eq.\ref{eq:wang_mg}]{Shen08,Wang09}. The estimation of BH masses based on those lines is consistent with each other. However, a bias between the estimation of \CIV\ and \MgII\ mass suggests that the \CIV\ estimator is severely affected by an outflow \citep{Baskin05, Trakhtenbrot12, Kratzer15}. The authors argue that using \Hb\ or \MgII\ is better for BH mass estimation than \CIV. A similar note is made by \citeauthor{Shen08}, whose results are based on 58,643 quasars from the SDSS catalog and who claim that the bias may be too large for individual objects when using a CIV estimator, but it is still consistent with \MgII\ and \Hb\ on average. On the other hand, \citet{Wang09} have found that \FWHM(\MgII) is systematically smaller than \FWHM(\Hb), and the BH masses based on the \MgII\ estimator show subtle deviations from those commonly used. Referring to WLQs, \citet{Plotkin15} suggest that using the \MgII\ line could cause bias to the mass measurements due to the large contribution of the \FeII\ pseudo-continuum. Thus, in this work, I base the estimations of BH masses on FWHM(\Hb) by trying to avoid \CIV\ and \MgII.

In this thesis, I have fixed the value of the radiative efficiency $\eta=0.18$. However, I would like to check its influence on the results. I again perform simulations and compare values of the best fits of BH masses and accretion rates in two cases, when $\eta=0.18$ and $\eta=0.36$. The accretion rates are 40-70\% higher and the BH masses are on average 20-30 \% less massive for higher~$\eta$ (for the first order approximation I have $\mmbh^2 \mdotm/\eta \simeq$ const). Please note that such a decline in BH masses cannot explain the underestimation of \mbhlit\ and the choice of solutions with $\eta \nsim 0.18$ increases the $\chi^2$.



    
\section{Inclination dependency}

\begin{figure}
    \centering
    \includegraphics[width=0.9\textwidth]{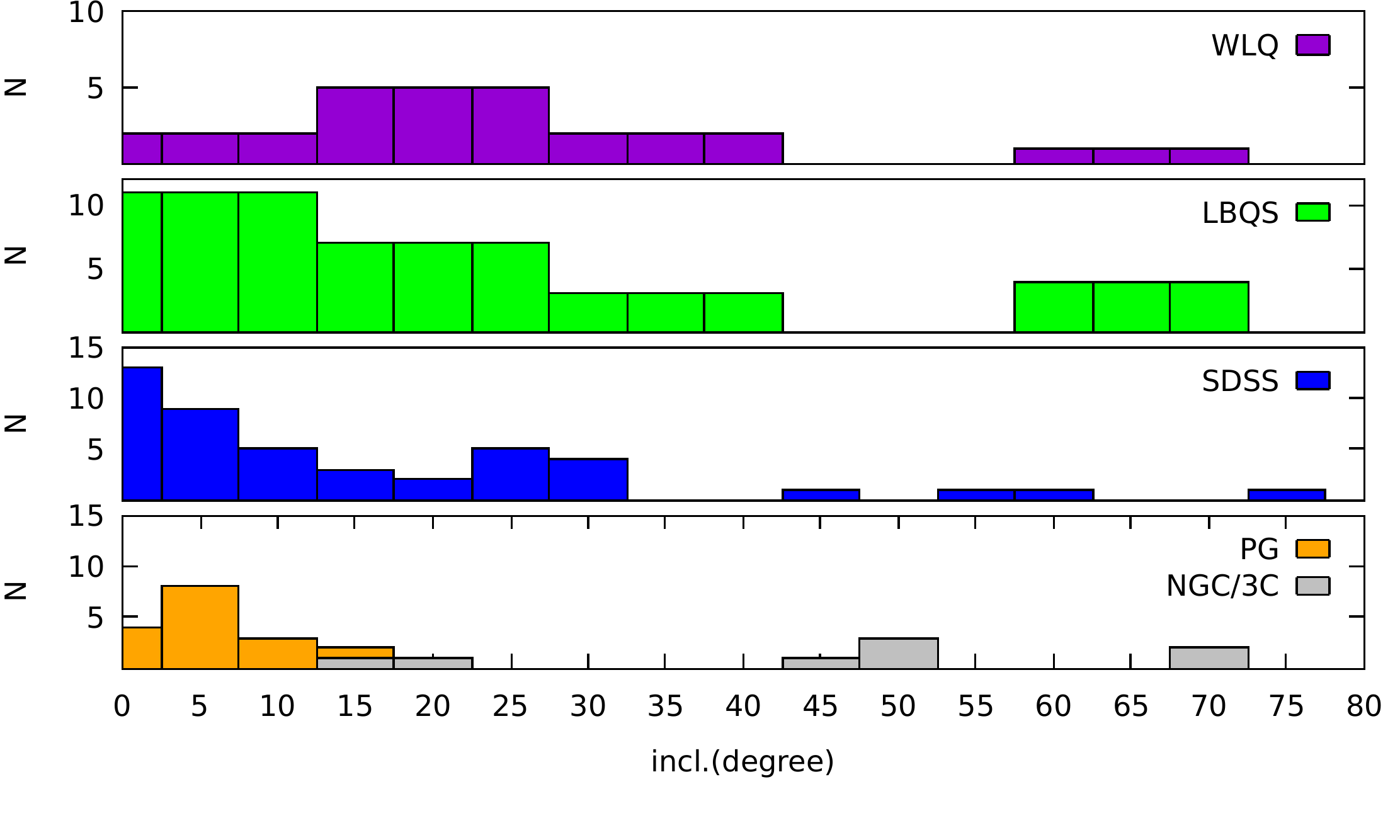}
    \caption[The distribution of inclinations in AGNs and WLQs.]{The distribution of inclinations in AGNs. My sample of 10 WLQs and 25 LBQS quasars (the top panel and below).
    Blue histogram indicates SDSS quasars \citep{Wildy18}. The lowest panel shows PG quasars (orange histogram), 3C and Seyfert 1 galaxies (grey color). 
    The data of them come from \citet{Bian2002}.}
    \label{fig:incl_hist}
\end{figure}

The agreement for LBQS and discrepancy for WLQ quasars between their virial and SED masses may suggest that \FWHM\ is a biased indicator of the virial velocity, due to the inclination of the emission region of the \Hb\ line. The BLR in WLQs could be less face-on (i.e. seen more in parallel with the accretion disk) than those in LBQS and other quasars. I compare the inclinations of 10 WLQs and LBQS quasars in my sample with those derived from 44 SDSS quasars \citep{Wildy18}, 17 PG quasars, and 7 Seyfert 1 galaxies \citep{Bian2002}. Generally, the main inclinations in LBQS are located in the range 0-15$^{\circ}$, whereas WLQs are shifted toward higher values with the main peak at 15-30$^{\circ}$ (Fig. \ref{fig:incl_hist}). I point out the same average inclination values in SDSS and LBQS quasars. It is worth mentioning that fitted errors are significant. Thus, inclinations may be the same in all quasars. According to \citet{Collin04} (see their equation 11 ($M_{real}/M_{obs} = 1/[(H/R_{BLR})^2 + sin (i)^2]$) and figure 8), the bigger inclination in WLQs (e.g. $i_{LBQS} = 5^\circ \rightarrow i_{WLQ}=20^\circ$) together with the assumption of at flat BLR geometry ($H/\mRblr \lesssim 0.3$, where H is the height of the BLR, and $R_{BLR}$ - the distance from BH) causes the underestimation of \FWHM\ by a factor $\gtrsim 2.5$. If we assume that the widths with values in the order of 2000 \kms are produced in the disk-like geometry (see \ref{fig:factor+}), then BH masses should be 6-10 times heavier. This is comparable with my analysis. Higher widths, like those in LBQS quasars, could be formed in the BLR with spherical geometry. In this case, the calculated BH masses do not require correction. It seems that the flatness of the BLR in weak emission-line quasars \textbf{plays a more important role than the inclination.}

\cite{Collin04} use equations with factors suitable for normal quasars which show strong lines and broad \FWHM. However, this is not true for many WLQs. For this reason, both values \mbhlit\ and \dotmlit\ could have been wrongly calculated. 

\section{Shielding gas as an explanation of the nature of WLQs}
Shielding of the BLR by a geometrically thick disk is thus perhaps generally applicable to quasars with lower Eddington ratios (which supports my result, see Tab. \ref{tab:bestfit}). \cite{Luo15} analysis indicate that my results of the accretion rate values i.e. 0.3-0.6 could be explained as a shielding gas phenomena of the weak emission-lines. In my opinion, the shielding gas is a plausible explanation for the WLQs. Apart from the explanation that WLQs are QSOs in a reactivation phase, along with the development of the BLR, the shielding gas seems like a good explanation of the nature of WLQs.

Based on our geometrically thick disk scenario for PHL 1811 analogs and WLQs, these extreme and rare quasars are likely not a distinct population but are instead extreme members of the continuous population of quasars. The shielding effect from a puffed-up inner disk likely exists beyond these extreme objects, and it might be applicable at a milder level to quasars with lower Eddington ratios. In either the slim-disk model or the \cite{Jiang14} simulations, the disk is unlikely to be as thin and flat as what a standard disk model describes as long as $L_{Bol}/ L_{Edd} \leq 0.3$. When the Eddington ratio is smaller than those of PHL 1811 analogs and WLQs, the radius and scale height of the puffed-up disk decreases, and its covering factor to the BLR also decreases, leading to a larger C IV EW than those of the PHL 1811 analogs and WLQs.

It is likely the blue outliers in the UV range might be related to the shielding gas (see Fig. \ref{fig:SED_1}). The absorption is caused by intrinsic gas in the host galaxy and/or the bigger influence of the assumed UV photoelectric absorption. 

\chapter{The corona investigation in the high mass SMBH.}
I have scoured the \cite{Shen2011qsocatalog} catalog for the most massive SMBH. The catalog represents properties of the 105,783 quasars in the SDSS Data Release 7 quasar catalog \citep{SDSSDR7}. It contains the continuum and emission line measurement of the $H\alpha$, $H\beta$, \MgII\, and \CIV.

I have found a few unusual and promising candidates. In the following chapter, I will describe one of them, SDSSJ110511.15+530806.5 quasar with the abnormal, broad absorption. 
\section{Introducing the SDSS J110511.15+530806.5 quasar.}
\begin{figure}[ht!]
  \centering
   \includegraphics[width=1.0\textwidth]{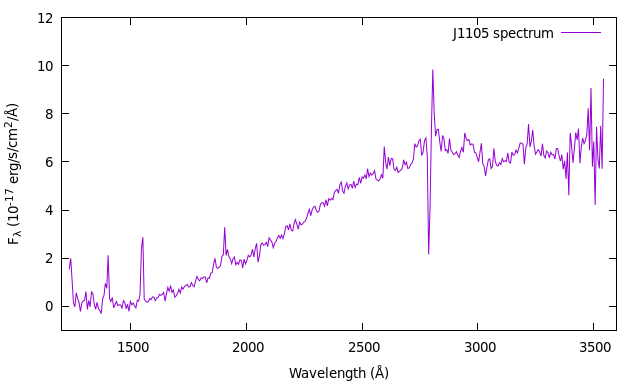}
 \caption{SDSS J110511.15+530806.5 spectrum.}
\label{fig:sdss_J1105}
\end{figure}

SDSS J110511.15+530806.5 (thereafter J1105) lays in the Ursa Major constellation at a distance of $8574.8 \pm 600.3$ Mpc from us. Its redshift ($z_{spec}$) equals to $1.9386 \pm 0.0015$. Galactic Extinction toward J1105 is $A_V$ = 0.027 mag and $A_K$ = 0.003 mag\footnote{NASA/IPAC Extragalactic Database (NED)}. The quasar's parameters are listed in Tab. \ref{tab:J1105_para}. The spectrum of J1105 is shown in Fig. \ref{fig:sdss_J1105}.

\begin{table}[ht!]
\caption{SDSS J110511.15+530806.5 parameters.}
\vspace{2ex}
\noindent
\begin{center}
\begin{tabular}{ c|c|c|c|c|c }
\hline 
RA & Dec  & {$z_{spec}$} & {$A_{V}$} & log{\mbh} & log \dotm  \\
(J2000.0) & (J2000.0) &  & [mag]  & [log \Msun] \\
(degrees) & (degrees) &  &  & \\

\hline 
166.296486 & 53.135186  & $1.9386\pm0.0015$ & 0.027 & $11.205 \pm 0.216$ & -4.713  \\
\hline
\end{tabular}
\label{tab:J1105_para}
\begin{quote}
In \cite{Shen11} quasar catalog is postulated that the virial BH mass is NOT true mass. They strongly recommend caution and pointed out that it may be biased.
\end{quote}
\end{center}
\end{table}



\begin{figure}[ht!]
  \centering
\includegraphics[width=1.0\textwidth]{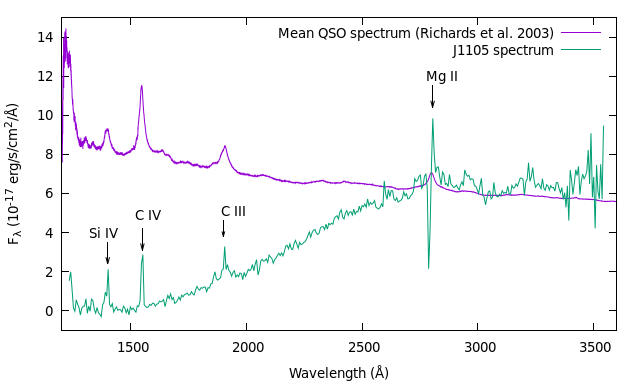}
 \caption[{J1105 spectrum with mean QSO spectrum}]{J1105 spectrum with mean QSO spectrum by \cite{Richards_2003}. Both spectra were correlated regarding the \MgII.}
 \label{fig:sdss_J1105_plus_richards}
\end{figure}



 

\section{The Global Absorption}

From Fig. \ref{fig:sdss_J1105} and Fig. \ref{fig:sdss_J1105_plus_richards} one can see the significant contribution of an absorption. Therefore, my first attempt concerned the extinction in the Milky Way as an explanation of the unusual, broad absorption. The first and the simplest explanation is the \cite{Cardelli89} extinction law. Unfortunately, it does not work in such deep absorption. Further, the mean extinction curve was taken from the \cite{Czerny_extinction_2004}. Using Czerny's code, I was able to include a more sophisticated extinction law. They assumed different carbon grain and silicate dust temperatures. Their extinction mainly takes the amorphous carbon grains into account as an explanation of the red quasars from SDSS. This extinction implicates that the grain radius may explain various extinction curves for AGNs. Progressing further, I have rebuilt the code to implement Wickramasinghe \& Hoyle's approach (\citeyear{Wickramasinghe1998}), which postulated that microdiamonds may imply on the interstellar medium, thus the excess of the interstellar extinction at the UV can be observed. Unfortunately, none of the laws of extinction work for the absorption explanation. 

This suggests the existence of intervening systems (e.g. a galaxy or IGM that lies on the line of sight to the quasar) or intrinsic/associated systems (e.g. large-scale outflows, accretion wind)\citep{Stone_abs_2019}. In other words, the second system is related to absorption from a matter associated with the central engine and/or host galaxy. Additionally, if we focus on absorption lines alone, we can divide quasars into the broad absorption lines (BALs) sources associated with outflowing matter, and the narrow absorption lines (NALs) assigned with the host galaxy or the intervening systems.

The NALs are associated with the EW value of just a few hundred \kms \citep[e.g.][]{Wild_abs2008}; BAL QSOs have those values larger \citep{Stone_abs_2019}. Around 60 \% of QSOs spectra prevail the NALs \citep[e.g.][]{Vestergaard2003_CIV_EW}, whereas the BAL are seen in only 20 \% \citep[e.g.][]{Knigge2008}. Moreover, \cite{Stone_abs_2019} suggest the division of NALs with regard to velocity of the absorbed matter. They distinguish between intervening - regarding the number density per unit redshift ($dN/dz$) - or intrinsic - regarding the number density per unit velocity ($dN/d\beta$, $\beta = \frac{V_{abs}}{c}$). The idea that a NAL is represented by intrinsic or intervening material is still under debate. 

\begin{figure}[H]
  \centering
    \includegraphics[width=1.0\textwidth]{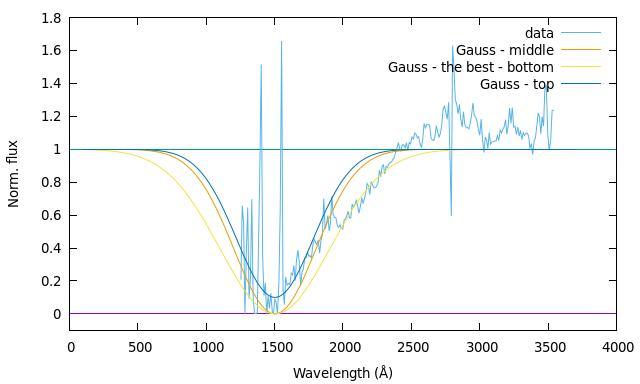}
 \caption[{Comparison of Gaussian functions}]{Comparison of three Gaussian's, the most bottom is the best. Its EW is 1002~\AA.}
\label{fig:3_gaussians}
\end{figure}

Let's come back to our quasar. I model the global absorption in J1105 by the Gaussian function. Fig. \ref{fig:3_gaussians} shows the normalized spectrum with three fitted Gaussian curves. EW of the top is 561 \AA, the middle - 751 \AA\, and the bottom - 1002 \AA. $\lambda_{min}$  for all Gaussian functions $\sim$ 1500 \AA. The sigma values are: 280 \AA, 300 \AA, 400 \AA, respectively. The bottom line is \textbf{the most plausible fit} to the data based on one Gauss function. However, we see that the multiple Gaussian functions could fit the data in a better way.

\section{The Influence of the Corona}

\begin{figure}[H]
  \centering
    \includegraphics[width=0.8\textwidth]{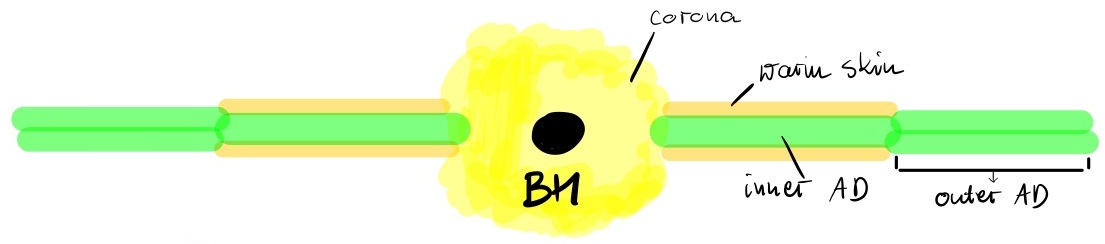}
 \caption[{The vicinity of the BH with AD + corona system.}]{The vicinity of the BH with AD + corona system. The corona (yellow), AD (green), and warm skin (dark yellow)}
 \label{fig:AD_corona_vicinity_general}
\end{figure}

The above considerations about extinction as an explanation of broad absorption were not satisfied. Moreover, the BAL and NAL do not explain such global absorption. Therefore I decided to make a more complex exploration including an influence of the corona and warm skin on the AD continuum (Fig. \ref{fig:AD_corona_vicinity_general}). 

The corona (hot medium) has $kT_{e,hot}$ $\sim$ 40 - 100 keV with Thomson optical depth in the range of 1-2, whereas the warm skin is an optically thick area with $kT_{e,warm}$ $\sim$ 0.1 - 1 keV and $\tau$ $\sim$ 2-25 \citep{Rozanska_2015, KubotaDone_2018}. The corona/warm skin has interesting properties. One is that it is a notably ionized medium. It is mainly seen in the observer's line of sight towards the source. The observations indicate the presence of narrow absorption lines and absorption continua in the soft X-ray \citep{Czerny2003corona_highacc}.

There are two mechanisms concerning warm skin: a back-scattering of the light that emerges from a disk and the Comptonization inside it (a change in the spectrum of light due to scattering from electrons) \citep{MN_BCZ_skin2004}. The concerns of the \cite{Czerny2003corona_highacc} work might be implied for the case of J1105, namely if the disk is dominated by radiation pressure, the warm skin has stabilized it.

\section{The Fitting Procedure for J1105}
For the next step, I included the influence of the corona and warm skin on the AD spectrum. Furthermore, I add clouds as a global absorption explanation. One of the most plausible models to do so is the work of \cite{KubotaDone_2018}. The model is based on the outer standard accretion disk, the inner warm Comptonising region which produces an excess of soft X-rays, and a hot corona. This assumption was taken for further examination. 

To model the J1105 spectrum I use XSPEC tools, a command-driven, interactive, X-ray spectral-fitting program, designed to be completely detector-independent so that it can be used for any spectrometer \citep{XSPEC1996}. The XSPEC package offers a few dozen models.

\newpage
\noindent In this project I have used my combination of XSPEC models: 
\begin{itemize}
    \item \textbf{redden*QSOSED*gabs*gabs}
    \item \textbf{redden*AGNSED*gabs*gabs}
\end{itemize}
The input is the observed spectrum.

\begin{figure}[H]
  \centering
    \includegraphics[width=0.85\textwidth]{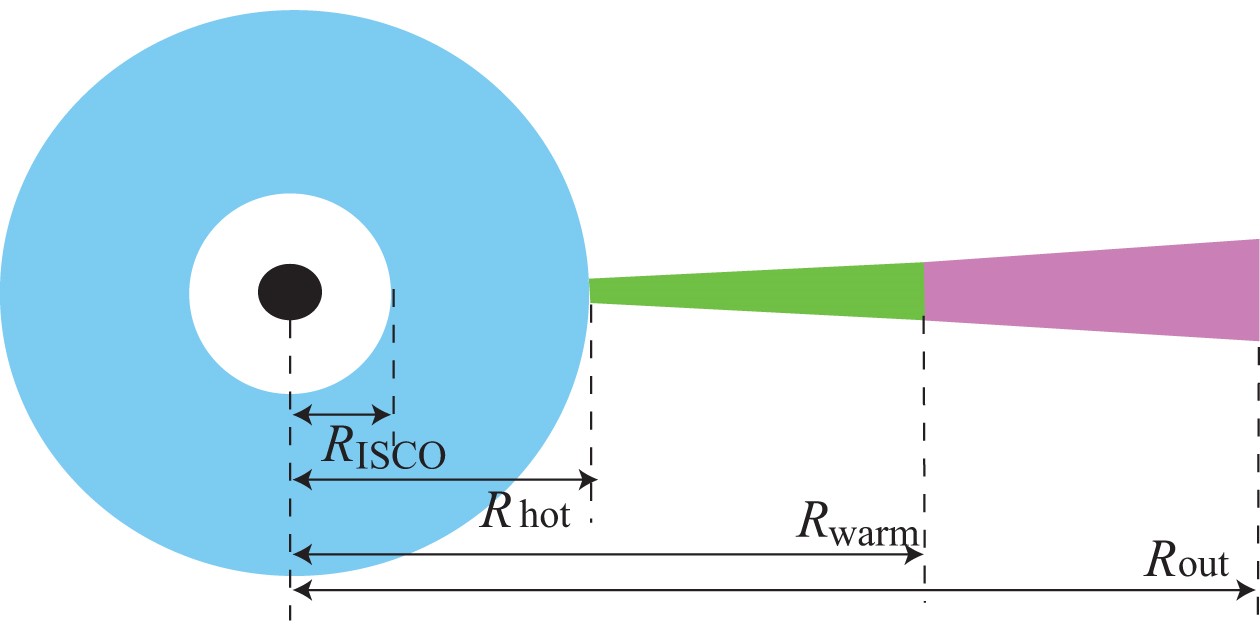}
 \caption[{The geometry of the \cite{KubotaDone_2018} model.}]{The geometry of the \cite{KubotaDone_2018} model. It represents geometry for hot inner flow (blue), warm Compton emission (green), and outer AD (magenta).}
 \label{fig:KD18_fig2}
\end{figure}

The \textbf{redden} model is the reddening in the Milky Way described by \cite{Cardelli89}. The E(B-V) for J1105 quasar is 0.008. The second component is the \textbf{QSOSED/AGNSED} \citep{KubotaDone_2018}. It models the intrinsic spectrum of the QSO. The SED model has three characteristic regions: the outer standard disc region (AD), the warm Comptonising region (warm skin), and the inner hot Comptonising region (corona)(see Fig. \ref{fig:KD18_fig2}). The AD is optically thin and geometrically thick described by NT equations. The corona is a hot medium, the skin of the AD is a warm gas. The QSOSED model assumptions are listed below:

\begin{table}[ht!]
    \centering
    \caption{QSOSED model with fixed parameters.}
    \begin{tabular}{|c|c|c|c|c|c|c|c|}
    \hline
    Parameter & Value & & Parameter & Value & & Parameter & Value \\
    \hline
    \hline
    {\mbh}(\Msun) & free & & cos {\incl}  & free & & $R_{hot}$ ($r_g$) & ** \\
    \hline
    D (Mpc) & free & & $kT_{e, hot}$ ( keV) & 100 & & $R_{warm}$ ($r_g$) & $=2R_{hot}$  \\
    \hline
    {\dotm} & free & &  $kT_{e, warm}$ ( keV) & 0.2 & & $R_{out}$ ($r_g$) & 100000 \\
    \hline
    {\as} & free & & $\Gamma_{hot}$ & * & & $H_{tmax}$ ($r_g$) & 100 \\
    \hline
    E(B-V) & free  & & $\Gamma_{warm}$ & 2.5 & & z & free \\
    \hline

    \end{tabular}
    \label{tab:qsosed_parameters}
    \begin{quote}
    * - $\Gamma_{hot}$ is calculated via  \cite{KubotaDone_2018} Eq.(6) 
    \newline ** - $R_{hot}$ is calculated to satisfy $L_{diss, hot}$ = 0.02 $L_{Edd}$
    \end{quote}
\end{table}


\noindent where: $kT_{e, hot}$ is electron temperature for the corona; 
$kT_{e, warm}$ - electron temperature for the warm skin;
$\Gamma_{warm}$ - the spectral index of the warm skin component;
$\Gamma_{hot}$ - the spectral index of the corona component;
$R_{hot}$ - outer radius of the corona component in $r_g$; \footnote{$r_g = \frac{G\mmbh}{c^2}$}
$R_{warm}$ -  outer radius of the warm skin component in $r_g$;
$R_{out}$ - the outer radius of the disc in units of $r_g$;
$H_{tmax}$  - the upper limit of the scaleheight for the hot Comptonisation component in $r_g$ unit;

For the warm Comptonising region, this model adopts the passive disc scenario tested by \cite{Petrucci2018comptonization}. Here, the flow is assumed to be completely radially stratified, emitting as a standard disc blackbody from $R_{out}$ to $R_{warm}$, as warm Comptonisation from $R_{warm}$ to $R_{hot}$, and then makes a transition to the hard X-ray emitting hot Comptonisation component from $R_{hot}$ to $R_{ISCO}$. The warm Comptonisation component is optically thick, so it is associated with the material in the disc. 

At a radius below $R_{hot}$, the energy is emitted in the hot Comptonisation component. This has a much lower optical depth, so it is not the disc itself. In the model, the albedo is fixed at a = 0.3, and the seed photon temperature for the hot Comptonisation component is calculated internally. This model does not take the color temperature correction into account \citep{XSPEC1996,KubotaDone_2018}. The geometry of the model is shown in Fig. \ref{fig:KD18_fig2}.

The third component \textbf{gabs*gabs} contains two absorptions Gaussian's and represents deep absorption in J1105. Each Gaussian function is represented by three parameters: the line energy in keV, line width (sigma) in keV, and line depth. 
\begin{equation}
    M(E) = exp \left(-\left(\frac{strength}{\sqrt{2\pi} \sigma}\right)exp\left(-0.5\left(\frac{(E-E_{line})}{\sigma}\right)^2\right)\right)
    \label{eq:gaussian_xspec}
\end{equation}
\section{Results}
The fitting of the spectrum of J1105 by \textbf{redden*QSOSED*gabs*gabs} model and the parameters are presented below.

\begin{figure}[H]
  \centering
    \includegraphics[width=1.0\textwidth]{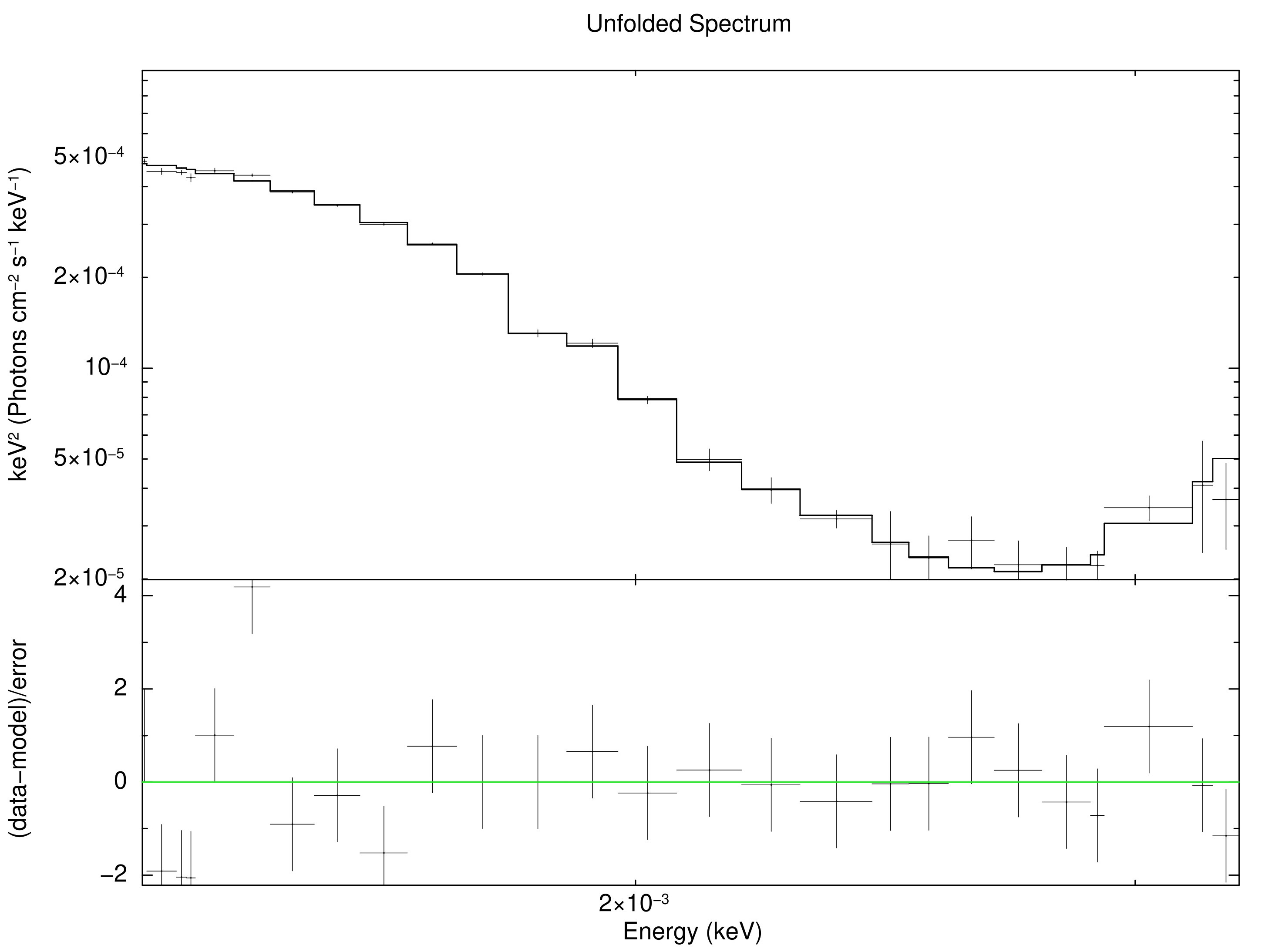}
 \caption[{The fit of the SDSSJ110511.15+530806.5. - redden*QSOSED*gabs*gabs}]{The fit of the SDSSJ110511.15+530806.5. Top panel: the best fit of J1105 using the \textbf{redden*QSOSED*gabs*gabs} model. Bottom panel: residuum of $\chi^2$. The $\chi^2$/ndf = \textbf{2.80}.}
\label{fig:best_fit_J1105_qsosed}
\end{figure}

The free parameters in the QSOSED model are black hole mass, accretion rate, the spin of the black hole, and the inclination angle. Additionally, the Gaussian positions were free throughout the modeling. The Tab. \ref{tab:agnsed_fitting_parameters} shows the best fit parameters for modeling QSOSED with two Gaussian functions. Fig. \ref{fig:best_fit_J1105_qsosed} represents the best fit of the J1105 using QSOSED by the XSPEC. The reduced $\chi^2$ for the fit is: \textbf{2.80} with 15 degrees of freedom. 




\begin{table}[H]
    \centering
    \caption{The fitting model \textbf{redden*QSOSED*gabs*gabs} parameters for J1105.}
    \begin{tabular}{|c|c|c|}
    \hline
    Model &  Parameter & Value \\
    \hline
    redden & E(B-V) & 0.08 (fixed) \\
    \hline
    \hline
    \hline
    \multirow{7}{*}{QSOSED}  & \mbh (\Msun) & $(3.524 \pm 0.012)\times 10^{9}$\\
    \cline{2-3}
    &   \dotm & $0.276 \pm 0.002$\\
    \cline{2-3}
    & cos \incl & $0.836 \pm 0.008$ \\
    \cline{2-3}
    & $a_*$ & $0.35 \pm 0.02$ \\
    \cline{2-3}
    & distance (Mpc)  & 5075 (fixed) \\
    \cline{2-3}
    &  redshift & 1.929 (fixed)\\
    \cline{2-3}
    & normalization & $ 0.4299 \pm 0.0014$ \\
    
    \hline
    \hline
    \multirow{3}{*}{First Gaussian} & Line E(keV) & $(2.791 \pm 0.012) \times 10^{-3}$\\
    \cline{2-3}
    &$\sigma$ (keV) & $(5.840\pm 0.065) \times 10^{-4}$\\
    \cline{2-3}
    &strength & $ (4.697 \pm 0.096) \times 10^{-3}$\\
    
    \hline
    \multirow{3}{*}{Second Gaussian} & Line E(keV) & $(2.125 \pm 0.022) \times 10^{-3}$\\
    \cline{2-3}
    & $\sigma$ (keV) & $(3.875 \pm 0.120)\times 10^{-4}$\\
    \cline{2-3}
    & strength & $(7.023 \pm 0.365) \times 10^{-4}$ \\
    \hline

    \end{tabular}
    \label{tab:qsosed_fitting_parameters}
    \end{table}

%
    
It is worth noting that I have tried a similar approach by using only one Gaussian function to describe the deep absorption. However, the reduced $\chi^2$ was not promising and the values were very high (a few hundred).

In the next step, I have improved the fit. The QSOSED model is limited. I have used the AGNSED code to improve the fit and to have free parameters describing the corona, the warm skin, and their respective arrangements (Tab. \ref{tab:agnsed_fitting_parameters}). Regarding the free parameters of the corona, warm skin, and disk I was able to constrain the nature of the deep absorption well. The reduced $\chi^2$ for the fit is: \textbf{2.056} for 10 degrees of freedom.

\begin{table}[H]
    \centering
    \caption{The fitting model \textbf{redden*AGNSED*gabs*gabs} parameters for J1105.}
    \begin{tabular}{|c|c|c|}
    \hline
    Model &  Parameter & Value \\
    \hline
    redden & E(B-V) & 0.08 (fixed) \\
    \hline
    \hline
    \hline
    \multirow{14}{*}{AGNSED}  & \mbh (\Msun) & $(3.522 \pm 0.010)\times 10^{9}$\\
    \cline{2-3}
    &  \dotm & $0.274 \pm 0.001$\\
    \cline{2-3}
    & cos \incl & $0.841 \pm 0.003$ \\
    \cline{2-3}
    & $a_*$ & $0.36 \pm 0.02$ \\
    \cline{2-3}
    & $kT_{e, hot}$ ( keV) & 100 (fixed)* \\
    \cline{2-3}
    & $kT_{e, warm}$ ( keV) & $0.356^{+4.876}_{-0.350}$ \\
    \cline{2-3}
    & $\Gamma_{hot}$ & $1.300^{+1.545}_{-1.300}$ \\
    \cline{2-3}
    & $\Gamma_{warm}$ & $2.722 \pm 0.153$ \\
    \cline{2-3}
    & $R_{hot}$ ($r_g$) &  $58.54 \pm  3.36$ \\
    \cline{2-3}
    & $R_{warm}$ ($r_g$) & $430.0^{+50.0}_{-11.0}$ \\
    \cline{2-3}
    & $r_{out}$ ($r_g$) & 100000 (fixed) \\
    \cline{2-3}
    & $H_{tmax}$ ($r_g$) &= $R_{hot}$ \\
    \cline{2-3}
    & distance (Mpc) & 5075 (fixed) \\
    \cline{2-3}
    & redshift & 1.0 (fixed)\\
    \cline{2-3}
    & normalization & $0.5063\pm0.0017$ \\
    \hline
    \hline
    \multirow{3}{*}{First Gaussian} & Line E(keV) & $(2.792 \pm 0.015) \times 10^{-3}$\\
    \cline{2-3}
    &$\sigma$ (keV) & $(5.842\pm 0.069) \times 10^{-4}$\\
    \cline{2-3}
    &strength & $ (4.686 \pm 0.098) \times 10^{-3}$\\
    \hline
    \multirow{3}{*}{Second Gaussian} & Line E(keV) & $(2.125 \pm 0.022) \times 10^{-3}$\\
    \cline{2-3}
    & $\sigma$ (keV) & $(3.874 \pm 0.117)\times 10^{-4}$\\
    \cline{2-3}
    & strength & $(7.025 \pm 0.367) \times 10^{-4}$ \\
    \hline
    \end{tabular}
    \label{tab:agnsed_fitting_parameters}
    \begin{quote}
    * - at the beginning this parameter was free. However, it had large errors (>200 \%). For this reason, I followed \cite{KubotaDone_2018} and fixed this parameter according to the author's assumptions
    \newline  NOTE: reprocess is ON
    \end{quote}
    \end{table}
   
\begin{figure}[H]
  \centering
    \includegraphics[width=1.0\textwidth]{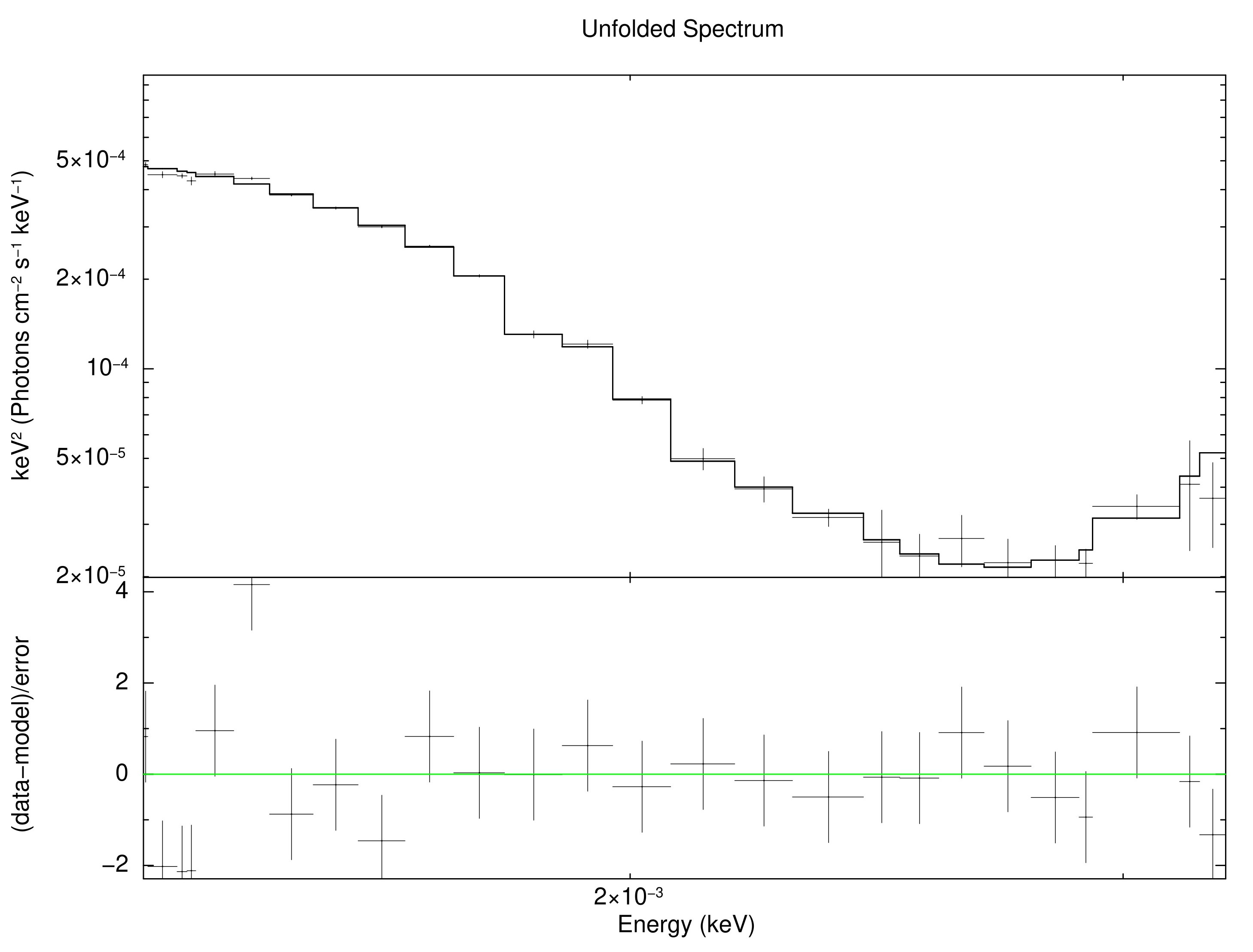}
 \caption[{The fit of the SDSSJ110511.15+530806.5. - redden*AGNSED*gabs*gabs}]{The fit of the SDSSJ110511.15+530806.5. Top panel: the best fit of J1105 using the \textbf{redden*AGNSED*gabs*gabs} model. Bottom panel: residuum of $\chi^2$. The $\chi^2$/ndf = \textbf{2.056}.}
 \label{fig:best_fit_J1105_agnsed}
\end{figure}    

For both approaches QSOSED and AGNSED (as a core of the modeling) I have used the same distance to the quasar and E(B-V) parameters. I was able to make a better fitting procedure in the second case and obtained the reduced $\chi^2$ = \textbf{2.056}. Not much improvement regarding the global parameters of the objects was observed.

\chapter{Discussion}
\section{Calculations for the Potential Absorption Source}\label{sec:calc_abs}
\begin{figure}[H]
  \centering
    \includegraphics[width=1.0\textwidth]{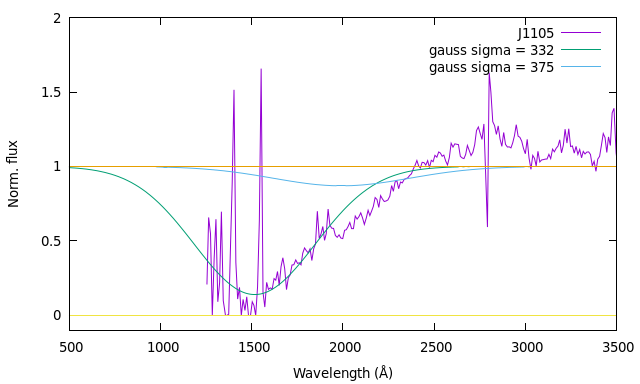}
 \caption[{The illustration of the two Gaussian's}]{The illustration of the two Gaussian's, with $\sigma = 332$ \AA, EW = 714 \AA; $\sigma = 375$ \AA, EW = 124 \AA.}
 \label{fig:4_gaussians}
\end{figure}

The origin of such abnormal deep absorption in the J1105 is interesting. In this paragraph, I have focused on two possible scenarios for the origin of the absorber. I have calculated two of the most plausible explanations.
I have presented the two Gaussian function descriptions (Fig. \ref{fig:4_gaussians}). The values of each of them were converted from the XSPEC fitting to wavelength in rest frame interpretation. The Gaussian function observed in the rest frame at 1516 \AA\ is the one that dominates the fit. However, all two of them are needed for the goodness of fit. 

\begin{table}[ht]
    \centering
    \caption[{Gaussian function in the observed frame and the rest frame.}]{Comparison of the observed frame energies and sigmas (keV) establish from the fitting of Gaussian functions with converted wavelength to the rest frame in \AA.}
    \begin{tabular}{c|c|c|c}
     & & observed frame (keV) & rest frame (\AA) \\
    \hline
    \multirow{2}{*}{First Gaussian} & Line E & $(2.792 \pm 0.015) \times 10^{-3}$&  $1516\pm10$\\
    \cline{2-4}
    &$\sigma$ & $(5.842\pm 0.069) \times 10^{-4}$ & $332\pm69$ \\
    \cline{2-4}
    
    \hline
    \multirow{2}{*}{Second Gaussian} & Line E & $(2.125\pm0.022)\times 10^{-3}$ & $1992\pm20$ \\
    \cline{2-4}
    & $\sigma$ & $(3.874 \pm 0.117)\times 10^{-4}$ &  $375\pm 64$ \\
    \cline{2-4}
    \hline
    \end{tabular}
    
    \label{tab:observed_energy_rest_wave}
\end{table}

\section{How far from the galaxy is the absorption cloud? }

The minima of the Gauss functions (Fig. \ref{fig:4_gaussians}) for J1105 in the rest frame are at $1516\pm10$ \AA, $1992\pm20$ \AA\ with $\sigma$ $332\pm69$ and $375\pm 64$, receptively.
If we assume that the global absorption in J1105 is caused by an extended cloud that absorbs \MgII\ (2796 \AA, 2803 \AA), then from equation $z = \frac{\lambda_{obs}}{\lambda_{em}} - 1 = \frac{\lambda_{abs}}{2800 \small\angstrom} - 1$ we are able to establish its $z_{1, \mathrm{MgII}} = - 0.4585$, $z_{2, \mathrm{MgII}} = - 0.2885$. This is the distance of this cloud/absorber from the host galaxy. However, if we assume that the cloud absorbs \CIV\ (1548 \AA, 1551 \AA), then its $z_{1, \mathrm{CIV}} = - 0.0213$, $z_{2, \mathrm{CIV}} = + 0.2859$.


\begin{figure}[H]
  \centering
    \includegraphics[width=0.8\textwidth]{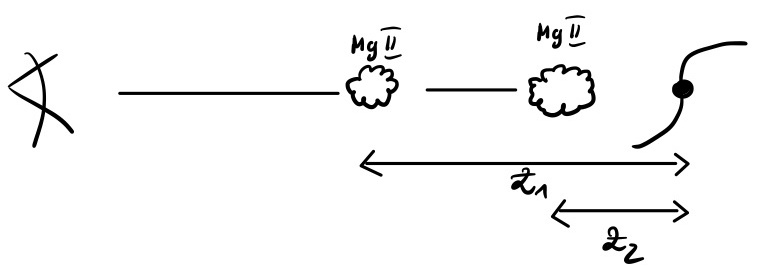}
 \caption{A scheme of potential absorption \MgII\ clouds in front of the quasar}
  \label{fig:MgII_clouds}
\end{figure}

\begin{figure}[H]
  \centering
    \includegraphics[width=0.8\textwidth]{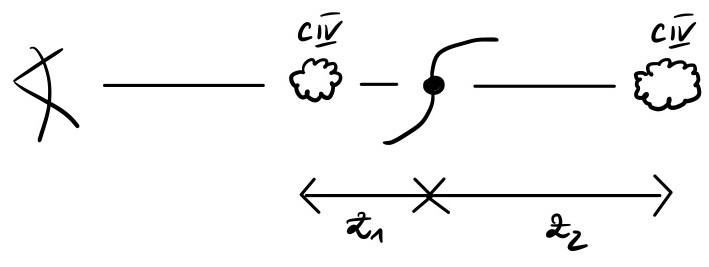}
 \caption{A scheme of potential absorption \CIV\ clouds}
 \label{fig:CIV_clouds}
\end{figure}

\subsection{If We Assume the Wind is the Absorber}

An absorption line in the spectrum of a NAL or BAL quasar is redshifted or blueshifted according to its lab position. Assuming the shift is caused by the Doppler effect and the absorber is a large-scale outflow, we can estimate the velocity of the absorbing material.



\begin{equation}
    \nu_{obs} = \nu_{em} \sqrt{\frac{1-\beta}{1+\beta}},
\end{equation}
\noindent  where $\beta = \frac{V}{c}$ and $V$ is velocity of a wind,
\noindent  then after rearrangement:
\begin{equation}
    \lambda_{em} = \lambda_{abs} \sqrt{\frac{1-\beta}{1+\beta}} = > \frac{1-\beta}{1+\beta} = \left(\frac{\lambda_{em}}{\lambda_{abs}}\right)^2
\end{equation}
\noindent From definition of redshift we know that $\lambda_{obs} = \lambda_{em} (1+z)$\\
\noindent therefore:
\begin{equation}
    \frac{1-\beta}{1+\beta} = \frac{1}{(1+z)^2} => (1-\beta)(1+z)^2 = (1+\beta)
\end{equation}
\begin{equation}
    (1+z)^2 - 1 = \beta (1+(1+z)^2)
\end{equation}
\noindent and final equation for $\beta$ looks like:
\begin{equation}
    \beta = \frac{(1+z)^2-1}{(1+z)^2+1}
\end{equation}

\noindent The final parameters for this section look like:
\begin{table}[ht]
    \centering
    \begin{tabular}{c|c|c|c}
        & redshift & $\beta$ & V (km/s) \\
        \hline
        \multirow{2}{*}{Mg II}
          & $-0.4585\pm 0.0032$ & $-0.3605^{+0.0018}_{-0.0021}$  & $-108135^{+555}_{-615}$  \\
          & $-0.2885\pm 0.0071$ & $-0.2482^{+0.0051}_{-0.0053}$ & $-74447^{+1543}_{-1577}$ \\
          \hline
        \multirow{2}{*}{CIV}
        & $-0.0213\pm0.0058$ & $-0.0211^{+0.0055}_{-0.0057}$  & $-6330^{+1698}_{-1710}$  \\
          & $+0.2859\pm 0.0013$ & $0.2463^{+0.0094}_{-0.0096}$ & $73887^{+2827}_{-2871}$ \\
    \end{tabular}
    \caption{The parameters of winds of \MgII\ and \CIV}
    \label{tab:redshift_beta_velocityl}
\end{table}

\noindent The negative velocity indicates that the wind is blowing in our direction. 

\cite{Stone_abs_2019} classified the Quasar absorption lines systems in view of their velocity distribution. Structures within $ V \leq$ 3000 \kms\ are called associated absorption lines. Systems with velocity separation 3000 - 12 000 \kms\ are dominated by intrinsicality or outflows from the QSO. Systems with velocity $\geq$ 12 000 \kms\ are disconnected from the QSO (see reference therein).

According to \cite{Stone_abs_2019}, our first case (i.e. \MgII\ absorption clouds) points out the intervening system. In the second case (i.e \CIV), the outflow is the first Gaussian with $V = -6330^{+1698}_{-1710}$ \kms.


\section{The column density of J1105}\label{sec:NH}
In this section, I have estimated the ion column densities, $N$, of the absorbers. Recent results of \cite{Saez2021_NH} about wind properties in PG 2112+059 was taken into account. Equation (9), (10), and (12) in \cite{Saez2021_NH} indicated the relation of EW and $N_j$ for the single Gaussian. However, in the case of J1105, I have to use two doublets of absorption, namely the \CIV\ and \MgII\ lines.
Using their Equation (12) for line doublet I was able to calculate $N_j$ of each Gaussian function in the optically thin regime for a fixed value of the ionization parameter and metallicity. 

\begin{equation}
    N_j \approx \frac{m_e c^2}{\pi e^2 (f_b \lambda^2_b+f_r \lambda^2_r)} EW,
\end{equation}
\noindent where b and r stands for parameters of the blue and red doublet respectively. $f$ is an oscillator strengths taken from \cite{Griesmann2000Carbon, Morton2003Magnesium}. The results are presented in the Tab. \ref{tab:NH_J1105}.

\begin{table}[ht]
    \centering
    \begin{tabular}{c|c|c|c}
    & & log$(N_j)$ & log$(N_j)$\\
    & EW(\AA)$^*$ & \CIV ($cm^{-2}$) & \MgII ($cm^{-2}$) \\
    \hline
    First Gaussian & $714^{+145}_{-147}$  & $17.027^{+0.080}_{-0.010}$ & $16.003^{+0.080}_{-0.010}$ \\
    Second Gaussian & $124^{+20}_{-21}$ & $16.266^{+0.065}_{-0.080}$ & $15.242^{+0.065}_{-0.080}$ \\
    \end{tabular}
    \caption{The column density of J1105}
    \begin{quote}
        * EW in the rest frame
    \end{quote}
    \label{tab:NH_J1105}
\end{table}

\cite{Saez2021_NH} indicate that the logarithmic values of column densities of CIV in BAL QSO PG 2112+059 vary during 13 years from 15.38 to 15.81. If they assumed the covering factor $\sim$ 60 \% then log $N_{CIV}$ = 16.12.

\cite{Dobrzycki2007} have determined the ion column densities of the individual components in HS1603+3820 quasar. For a mini-BAL system in this quasar, the mean \CIV\ column density is $15.265\pm0.135$. For system B, the mean \MgII\ column density is $15.106\pm0.155$. The \CIV\ column densities vary through systems of the quasar from $13.200\pm0.035$ to  $15.471\pm0.085$.

\section{Final remarks}

\begin{figure}[H]
  \centering
    \includegraphics[width=0.8\textwidth]{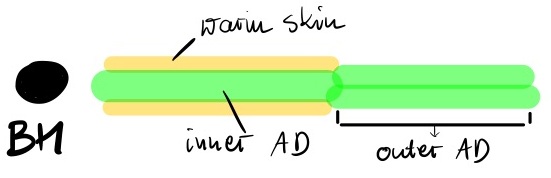}
 \caption[{The vicinity of the J1105}]{My view how the vicinity of the J1105 might look like}
 \label{fig:AD_corona_vicinity}
\end{figure}

In Fig. \ref{fig:AD_corona_vicinity} I present the potential idea of an accretion disk and warm skin concept, which might be seen in J1105. \cite{Czerny2003corona_highacc} represent the concept of the bright AGN with a warm skin above the AD. A similar concept of warm skin, which wraps around the inner part of AD is presented here.

The AGNSED is a more sophisticated code with the same approach as QSOSED, but with all the free parameters presented in Tab. \ref{tab:agnsed_fitting_parameters}. However, as a first order approach, the warm skins from \cite{Czerny2003corona_highacc, MN_BCZ_skin2004} and \cite{KubotaDone_2018} model, look good enough to examine the nature of J1105.

Primary I have fitted the QSOSED model with two Gaussian functions. The first Gaussian is the most dominant. The second seems to be crucial. The QSOSED model assumes the fixed parameters of the corona, the warm skin, and the disk size. I have chosen the AGNSED model with free parameters to get to know J1105 nature better. Moreover, I tried to find the best $kT_{e,hot}$ parameters. In this case, I have also tried many variants. It turned out that the best value is fixed to 100 keV, same as \cite{KubotaDone_2018} approach.
I was able to make an even better fit procedure and obtained the reduced $\chi^2$ from \textbf{2.800} to: \textbf{2.056}. Not much improvement regarding the global parameters of the objects was observed.
 SMBH from AGNSED is \mbh $\simeq 3.5 \times 10^9 \mMsun$, wheres \mbhlit $\simeq 1.6\times 10^{11} \mMsun$ \citep{Shen2011qsocatalog}.

In the first order of approximation of the modeling procedure, I have fitted only one Gaussian function, the reduced $\chi^2$ was 7.0912. It seems that this object in particular needed more functions that described the absorption. Another Gaussian's were needed. However, when I have tried to narrow the Gaussian's down I obtained a not well fitted model. The reduced $\chi^2$ of the worse fit was > 10. 

Besides the improved value of the reduced $\chi^2$, one of the most crucial take away is that \textbf{the AGNSED provide a better understanding of the physics and boundaries of the corona, the warm skin, and the disk size}. The mentioned parameters are within the assumptions of \cite{Czerny2003corona_highacc} and \cite{KubotaDone_2018} approaches (for more details please see their Table 1 and  Table 2 respectively).

In this section, I would like to compare parameters from the literature. I would like to remind the \dotm\ of J1105 is equal to $0.274 \pm 0.001$. The corona/warm skin concept of Mrk 359 by \cite{Czerny2003corona_highacc} indicate the \dotm = 0.3. Additionally, \cite{KubotaDone_2018} studied three close by objects ($z<0.16$). One of them PG 1115+407 has a similar accretion rate $\sim$ 0.4.

Moreover, the $\Gamma_{warm}$ = $2.722\pm0.153$ of J1105 is similar to one obtained for PG 1115+407; $\Gamma_{warm}$ = 3.06 and Mrk 359 with $\Gamma_{warm}$ = 3.32. 



Comparing study with QSOSED vs AGNSED as a main component of modeling, I have found that the $kT_{e, warm}$ = $0.356^{+4.876}_{-0.350}$ (keV). In Czerny's the value is fixed at 0.2 keV and \cite{KubotaDone_2018} give $kT_{e, warm}$ 0.50 keV (fixed). Regarding the corona/warm skin size, J1105 has $R_{hot}$ = $58.54\pm3.36$ $r_g$  and $R_{warm}$  = $430.0^{+50.0}_{-11.0}$ $r_g$, comparing to $R_{warm}$ of Mrk 359 = $300 R_{Schw}$ ($R_{Schw} \sim 2 r_g$). 

The difference may be caused due to the divergence between \mbh. \cite{Czerny2003corona_highacc} and \cite{KubotaDone_2018} objects do not contain such massive SMBH as J1105. In their studies, the \mbh\ are $\sim$ $10^8\mMsun$. However, the physics and corona/warm skin/disk description should not vary very much. Nevertheless, the description of the corona/warm skin for J1105 is well established and comparable with literature studies.

Contrary to the massive black holes, \cite{Zhang2000warmskin} modeled stellar BH: GRS 1915+105 and GRO J1655-40. Their study revealed a three-layered atmospheric structure in the inner region of AD. They found out a warm skin with T = 1.0-1.5 keV and an optical depth $\sim$ 10. They sum up, the temperatures of the cold disk of $\sim 10^6K$, the warm skin $\sim 10^7K$, and corona $\sim 10^9K$ indicated the complicated phenomena around such objects (i.e. corona/warm skin explanation). However, they focused on the AD of the stellar mass black holes. GRO J1655-40 has a black hole $\sim 7$ \Msun, GRS 1915+105 $\sim 12 $ \Msun. 


\cite{Meusinger2016} study of J1105 indicated that it has reddening steeper than in the Small Magellanic Cloud, perhaps even steeper than for IRAS 14026+4341 galaxy. They could not either find a good reddening/extinction solution for the case of J1105. They classify it as a strongly reddened narrow-line quasar. \cite{Jiang_2013} argued that such an unusual reddening law is based on a speculative assumption of an exotic dust grain size distribution that lacks large grains.

My result indicated that the deep absorption in J1105 could be explained by two Gaussian functions which may describe two different clouds. The \CIV\ as an absorbing material is more plausible due to the fact that $\lambda_{min}$ of the main Gaussian is 1516 \AA\, whereas the laboratory line of \CIV\ lies at 1548\AA\ and 1551 \AA. The clouds or even a wind contains mostly \CIV\ which absorbs radiation from an AD/corona of the J1105. The most dominant Gaussian with $\sigma = 375$ \AA\ has a strength of nearly 90\% of the whole recess. Regarding the redshifts of the potential clouds, one can see that for \MgII\ case two clouds along the line of sight of observer occurred. Surprisingly in the \CIV\ case (Fig. \ref{fig:CIV_clouds}) one of the clouds is 'behind' the J1105! It seems that it is the wind component or Failed Radiatively Accelerated Dusty Outflow (FRADO) observed by Czerny's group. This wind falls to the AD. 

Using \cite{Weymann1979} approach, \cite{Stone_abs_2019} distinguish three components of the velocity distribution of NALs: the uniform distribution of very high velocity components which is unassociated with the QSO; the influence of the virialized material clustered around the QSO itself, or associated with material ejected from the QSO. Regarding the values from Sec. \ref{sec:calc_abs}, the plausible explanation of the origin of the absorption in J1105 is \textbf{the strong outflow represented by the first Gaussian function} and \textbf{the intervening material} e.g. dwarf galaxy \textbf{along the line of sight} seen as the second Gaussian curve in the spectrum. Nevertheless, the intervening material may explain both Gaussian functions. Regarding my results, I cannot exclude either NAL or BAL phenomena as an explanation of the nature of J1105.
However, such wide absorption has not been observed in the sources of BALs (\CIV\ column density $\sim$ 17 cm$^-2$). Further investigation was needed. The explanation of the abnormal properties of J1105 might lay at the root of the warm skin concept.

\begin{figure}[H]
  \centering
    \includegraphics[width=0.65\textwidth]{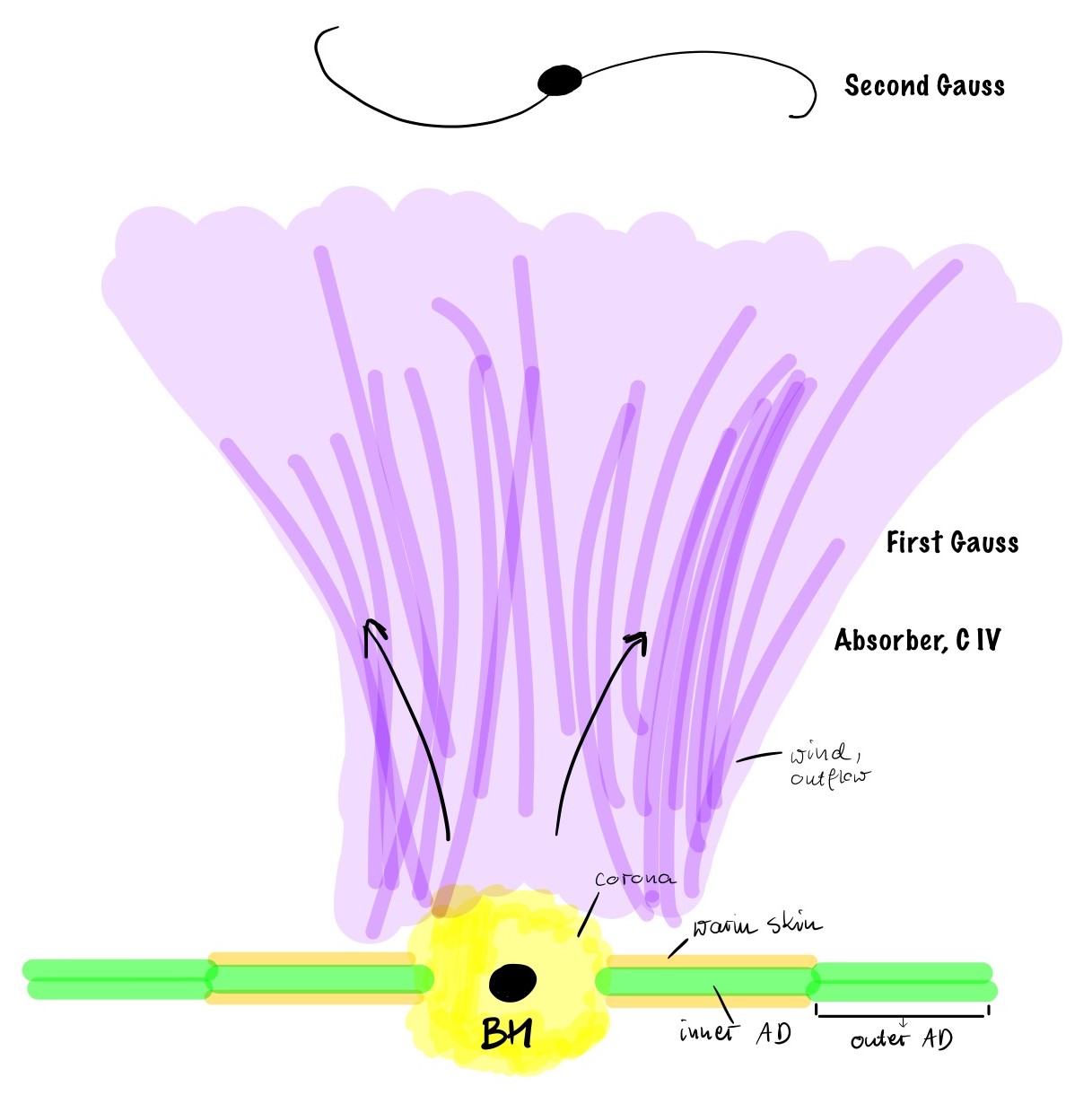}
 \caption{An explanation of the nature of J1105 - winds.}
 \label{fig:J1105_wind}
\end{figure}

\begin{figure}[H]
  \centering
    \includegraphics[width=0.65\textwidth]{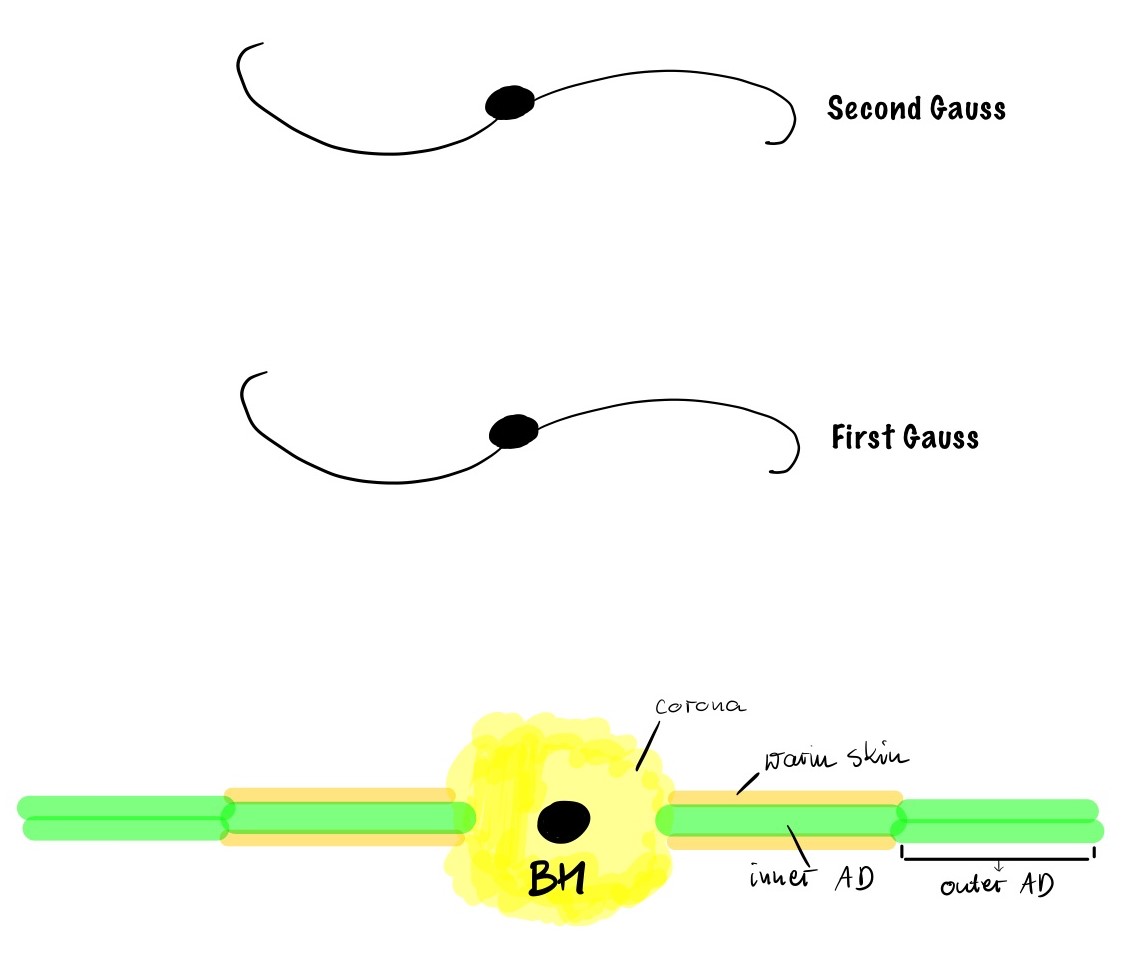}
 \caption{An explanation of the nature of J1105 - two dwarf galaxies.}
 \label{fig:J1105_2galaxies}
\end{figure}

In summary of the nature of the SDSSJ110511.15+530806.5 quasar I have presented two visual explanation of this phenomena (Fig. \ref{fig:J1105_wind}, Fig. \ref{fig:J1105_2galaxies}).

This chapter will be concluded as a paper soon. I have found more objects with very high \mbh\ and abnormal features with their spectra. This work is in progress and I will examine it further.


\chapter{Conclusion}

I have studied the accretion disk continua of 10 WLQs. The SMBH masses of those objects were estimated
previously based on the single-epoch virial \mbh\ method (\mbhlit ). Generally, the RM and the single-epoch virial BH mass method are inadequate for BH mass estimation in WLQs, due to the weakness of emission-lines in these objects. I have created a grid of 366,000 accretion disk models using the Novikov--Thorne formulas. The phenomena of the WLQs were expressed by four parameters (\mbh, $\mdotm$, spin of BH, and the line-to-sight inclination) to describe the observed SED.

My main findings are as follows:
\begin{enumerate}
    \item Using the pure Novikov–Thorne model, we can describe very well the SED of WLQs,
    \item I have compared obtained BH masses with those obtained from the literature. The SMBH masses of WLQs, which are estimated based on \FWHM(\Hb), are underestimated. On average, the masses are undervalued by 4–5 times. The median of this correction factor is 3.3,
    \item I have proposed the new formula to estimate \mbh\ in WLQs based on their observed \FWHM(\Hb) and luminosities at 5100 \AA\ (Equation \ref{eq:mbhwlq}). This equation helps to calculate proper weight of BH masses,
    \item My results suggest that selected WLQs have accretion rates in the range of $\sim$ 0.3–0.6 (see Tab. \ref{tab:bestfit}),
    \item  My results support \cite{Mejia18N} results and confirm that the virial factor, $f$, depends on \FWHM. In this paper, $f \propto \mFWHM(\mHbeta)^{-1.34 \pm 0.37}$. The BLR is a non-spherical region. The BLR region has the disk-like geometry.
    
\end{enumerate}
I suggest that some WLQs are normal quasars in a reactivation stage. However, the shielding gas is another plausible explanation for the WLQs. Apart from the previous mentioned finding that WLQs are QSOs in a reactivation phase, along with the development of the BLR, the shielding gas seems like a good explanation of the nature of WLQs.

The abnormal absorption of SDSS J110511.15+530806.5 quasar is explained in the two Gaussian functions description. The fitting procedure of the model \textbf{redden*AGNSED*gabs*gabs} concluded with the global parameters:
\begin{itemize}
    \item \mbh = $(3.522 \pm 0.010)\times 10^{9}$,
    \item \dotm = $0.274 \pm 0.001$,
    \item cos \incl = $0.841 \pm 0.003$, 
    \item $a_*$ = $0.36 \pm 0.02$
\end{itemize}

The corona and the warm skin investigation stipulated in:
\begin{itemize}
    \item $kT_{e, hot}$ = 100 keV, $\Gamma_{hot}$ = $1.300^{+1.545}_{-1.300}$, $R_{hot}$ = $58.54 \pm  3.36$ $r_g$,
    \item $kT_{e, warm}$ = $0.356^{+4.876}_{-0.350}$ keV, $\Gamma_{warm}$ = $2.722 \pm 0.153$, $R_{warm}$ = $430.0^{+50.0}_{-11.0}$ $r_g$. 
\end{itemize}
 
\newpage
In summary:

\begin{enumerate}
   \item The investigation of the Weak Emission-line Quasars.
   \begin{itemize}
     \item The continuum fit method is an unbiased way to describe the global parameters,
\item The flatness of the BLR in weak emission-line quasars plays a more important role than the inclination,
\item The correction of the SMBH masses based on \FWHM\ estimation is needed,
\item A similar or even the same behavior of normal AGNs and WLQs suggests that both kinds of sources have the same dim nature of the velocity component and similar geometry of the BLR.
   \end{itemize}
   \item The spectrum of SDSS J110511.15+530806.5 with extensive absorption.
    \begin{itemize}
 \item The nature of the deep absorption of the J1105 found an explanation in the two Gaussian function with $\sigma = 332$ \AA, EW = 714 \AA and $\sigma = 375$ \AA, EW = 124 \AA,
\item Column density of the first Gaussian with log$(N_j)$ of \CIV\ =  $17.027^{+0.080}_{-0.010}$ $cm^{-2}$ or log$(N_j)$ of \MgII\ $16.003^{+0.080}_{-0.010}$ $cm^{-2}$,
\item Column density of the  second Gaussian with log$(N_j)$ of \CIV\ =  $16.266^{+0.065}_{-0.080}$ cm$^{-2}$ or log$(N_j)$ of \MgII\ $15.242^{+0.065}_{-0.080}$ cm$^{-2}$
\item The explanation of the origin of the absorption in J1105 by strong outflow/wind and/or intervening material along the line of sight (Fig. \ref{fig:J1105_wind} and Fig. \ref{fig:J1105_2galaxies})
\end{itemize}
\end{enumerate}

\newpage
\section{Future work/prospect of WLQs research}

I will focus on describing the AD continua models for the broader sample to constrain \mbh\ and \dotm\ for WLQs. It will include a recently developed sample of X-ray WLQs \citep[e.g.][]{Luo15, Ni18, Ni20, Tim20}. Using more sophisticated codes (e.g. hardening factor, the influence of the corona) -- such as: QSOSED, KerrBB from XSPEC package -- I will develop a grid of models with a range of \mbh, \dotm, and spin parameters. Exploiting available data I will scour databases to create a SED of each WLQ. The potential new observations for WLQs are crucial for the disk-fitting method to have a spectrum with UV points. It allows us to observed the bend point in the SED which is important for the model constraints. The Swift's Ultraviolet/Optical Telescope (UVOT) and/or Hubble Space Telescope (HST) time will be significant.


\begin{figure}[ht]
\begin{center}
    \includegraphics[width=0.7\textwidth]{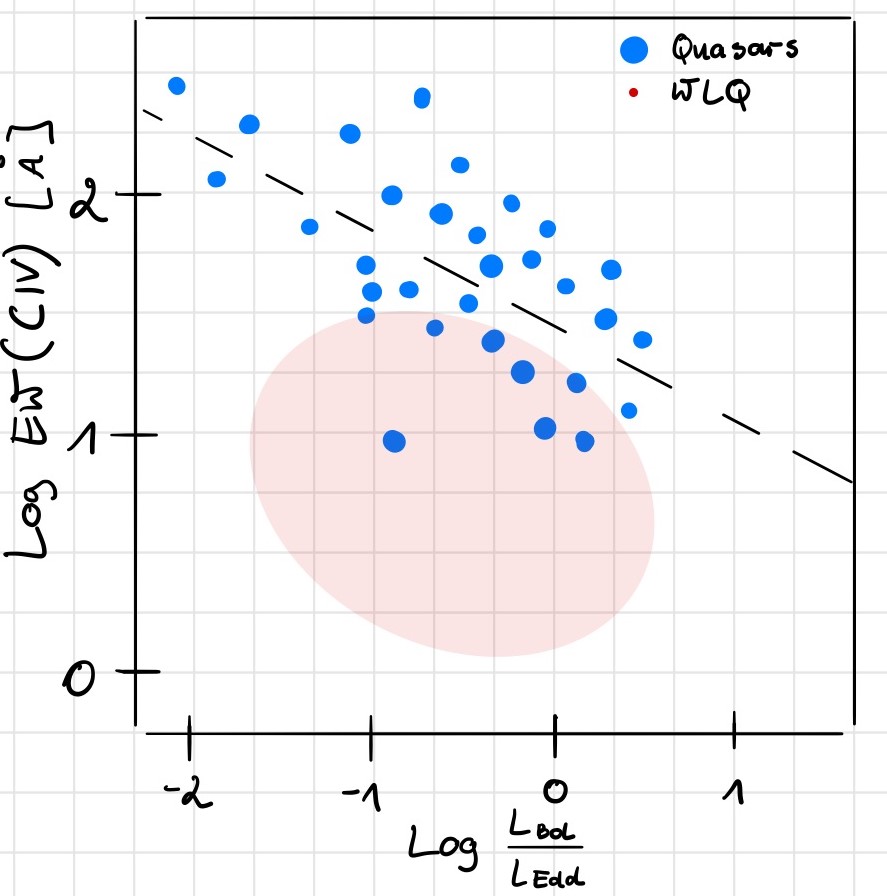}
  \end{center}
\caption[{Baldwin effect for WLQs.}]{Baldwin effect. Blue dots, sample of normal Quasars. Red area expected range of WLQs.}
\label{fig:EW}
\end{figure}
The Baldwin effect is the relation between the luminosity of AGNs and emission-lines in the electromagnetic spectra. It is an important feature in the description of quasars and AGNs because the value of EW is related to \dotm\ and/or spin of BH. The results from the first task will be used to determine the Baldwin effect for the broader WLQs sample. I will determine the Baldwin effect by a using broader sample of WLQs. Previous attempts have been done; however, I will use corrected \dotm. I will include new properties in the model description such as: spectral hardening factor, the effect of self-irradiation, and the effect of limb-darkening. I am aiming to see an anti-correlation between EW of lines e.g (\CIV) and $L/L_{Edd}$ relation in WLQ (Figure \ref{fig:EW}). However, due to the weakness of \CIV\ line, I expect to observe shifted anti-correlation which is different than for normal quasars (red area in Figure \ref{fig:EW}). 

Additionally, I will explore the influence of radiative efficiency on this in a broader WLQs sample. I will inspect the wide range of $\eta$ parameter, which influences the $L/L_{Edd}$ and thus the spin of BH. As a result, the shape of SED will be more constrained due to the strong dependency on $L/L_{Edd}$. In order to constraint/establish the nature of WLQs well, more research is needed. I would expand the WLQs sample and use the recent X-ray development in the WLQs matter.

{\bibliographystyle{thesis}}

{\bibliography{apj}}
\end{document}